\documentclass[11pt]{article}

\usepackage{xr}
\makeatletter
\newcommand*{\addFileDependency}[1]{
  \typeout{(#1)}
  \@addtofilelist{#1}
  \IfFileExists{#1}{}{\typeout{No file #1.}}
}
\makeatother

\usepackage{paper_style}

\usepackage{authblk}
\makeatother

\usepackage{calc}

\usepackage{accents}
\newcommand{\dbtilde}[1]{\accentset{\approx}{#1}}

\title{Bayesian Inference  for  Evidence Accumulation Models with Regressors}

\author[1,4]{Viet-Hung Dao}
\author[2]{David Gunawan}
\author[1]{Robert Kohn}
\author[3]{Minh-Ngoc Tran}
\author[5]{Guy E. Hawkins}
\author[5]{Scott D. Brown}
\affil[1]{Australian School of Business, University of New South Wales, Sydney, Australia}
\affil[2]{School of Mathematics and Applied Statistics, University of Wollongong}
\affil[3]{Discipline of Business Analytics, University of Sydney Business School }
\affil[4]{Department Of Mathematics, Thuyloi University, Vietnam}
\affil[5]{School of Psychological Sciences, University of Newcastle, Australia}

\def\b1{\boldsymbol{1}}
\def\b0{\boldsymbol{0}}
\def\ba{\boldsymbol{a}}

\def\bX{\boldsymbol{X}}

\def\by{\boldsymbol{y}}

\DeclareMathOperator{\diag}{diag}

\DeclareMathOperator{\logdet}{logdet}
\DeclareMathOperator{\tr}{trace}

\DeclareMathOperator{\vect}{vec}

\def\bPsi{\boldsymbol{\Psi}}
\def\bpsi{\boldsymbol{\psi}}
\def\sig2{\sigma^2}

\newcommand{\btheta}{\boldsymbol{\theta}}
\newcommand{\bthetatilde}{\boldsymbol{\tilde{\theta}}}

\newcommand{\brho}{\boldsymbol{\rho}}

\newcommand{\bmu}{\boldsymbol{\mu}}
\newcommand{\bomg}{\boldsymbol{\omega}}
\newcommand{\bldeta}{\boldsymbol{\eta}}
\newcommand{\bxi}{\boldsymbol{\xi}}
\newcommand{\Da}{D_{\alpha}}

\newcommand{\balph}{\boldsymbol{\alpha}}
\newcommand{\balphj}{\boldsymbol{\alpha}_j}

\newcommand{\balphJ}{\boldsymbol{\alpha}_{1:J}}
\newcommand{\balphJk}{\boldsymbol{\alpha}^{\boldsymbol{k}}_{1:J}}

\newcommand{\balphJR}{\boldsymbol{\alpha}^{1:R}_{1:J}}

\newcommand{\alphtilde}{\tilde{\alpha}}
\newcommand{\balphtildeJ}{\tilde{\boldsymbol{\alpha}}_{1:J}}

\newcommand{\bmualph}{\boldsymbol{\mu_{\alpha}}}

\newcommand{\bSigalph}{\boldsymbol{\Sigma_{\alpha}}}
\newcommand{\bSigalphinv}{\boldsymbol{\Sigma_{\alpha}^{-1}}}

\newcommand{\bSig}{\boldsymbol{\Sigma}}

\def\bk{\boldsymbol{k}} 
\def\ka{k_{\alpha}} 
\def\va{v_{\alpha}}

	\def\rt{\textrm{rt}}
	
	\def\minRT{\min \text{RT}}

	\newcommand{\modelpara}{\boldsymbol{\theta}}
	
	\def\muv{\mu_v}
	
	\def\sv{s_v}

	\def\muz{\mu_{z}}
	
	\def\sz{s_{z}}

	\def\Ltau{L_{\tau}}
	\def\Ltauj{L_{\tau_j}}
	\def\mutau{\mu_{\tau}}
	\def\stau{s_{\tau}}
	\def\mutauj{\mu_{\tau_j}}
	\def\stauj{s_{\tau_j}}
	
	\def\wfptpdf{\textrm{WFPT}(c,t_d|v,a,z)}

	\def\dfpdflowersmall{\textrm{DDM}^s(\textrm{``lower''},\rt)}
	\def\dfpdflowerlarge{\textrm{DDM}^l(\textrm{``lower''},\rt)}

	\def\fsmall{f^s(t_d|v,a,z)}
	\def\flarge{f^l(t_d|v,a,z)}

	\newcommand{\gsmall}{g^s(t_d|\modelpara)} 
	\newcommand{\glarge}{g^l(t_d|\modelpara)} 

	\def\Wkl{k\sin\left( \dfrac{\pi k z}{a}\right) \exp\left( -\dfrac{\pi^2k^2t_d}{2a^2}\right) } 
	\def\sumWkl{ \sum\limits_{k=1}^{\infty} \Wkl} 

	\def\Wks{\left(\dfrac{z}{a} +2k\right) \exp \left( -\dfrac{(z+2ka)^2}{2t_d}\right)} 
	\def\sumWks{ \sum\limits_{k=-\infty}^{\infty} \Wks} 

\newcommand{\lb}{\mathcal{L}(\boldsymbol{\lambda})}

\newcommand{\lbest}{ \widehat{\mathcal{L}(\blamb)} }
\newcommand{\gradlb}{\nabla_{\boldsymbol{\lambda}}\mathcal{L}(\boldsymbol{\lambda})}

\newcommand{\gradlbest}{\widehat{\nabla_{\boldsymbol{\lambda}}\mathcal{L}(\boldsymbol{\lambda})}}

\newcommand{\beps}{\boldsymbol{\epsilon}}

\newcommand{\blamb}{\boldsymbol{\lambda}}
\def\bmulamb{\boldsymbol{\mu}_{\boldsymbol{\lambda}}}

\def\Blamb{B_{\boldsymbol{\lambda}}}
\def\Dlamb{D_{\boldsymbol{\lambda}}}
\def\dlamb{d_{\boldsymbol{\lambda}}}
\def\bd{\boldsymbol{d}}

\def\bmulambalpha{\boldsymbol{\mu}_{1}}
\def\bmulambmu{\boldsymbol{\mu}_{2}}
\def\bmulambbeta{\boldsymbol{\mu}_{3}}
\def\bmulambhyper{\boldsymbol{\mu}_{4}}

\newcommand{\qvb}{q_{\boldsymbol{\lambda}}(\boldsymbol{\theta})}

\newcommand{\joint}{p(\boldsymbol{y},\boldsymbol{\theta})}

\newcommand{\bv}{\boldsymbol{v}}

\newcommand{\uone}{\dfrac{b-A-(t-\tau)v^c}{A}}
\newcommand{\utwo}{\dfrac{b-(t-\tau)v^c}{A}}

\newcommand{\xzero}{\Phi(\overline{\omega}_1-\overline{\omega}_2)}
\newcommand{\xone}{\phi(\overline{\omega}_1-\overline{\omega}_2)}
\newcommand{\xtwo}{\phi'(\overline{\omega}_1-\overline{\omega}_2)}

\newcommand{\zzero}{\Phi(\overline{\omega}_1)}
\newcommand{\zone}{\phi(\overline{\omega}_1)}
\newcommand{\ztwo}{\phi'(\overline{\omega}_1)}

\newcommand{\wone}{\overline{\omega}_1}
\newcommand{\wtwo}{\overline{\omega}_2}

\newcommand{\kb}{k_{\bbeta}} 

\newcommand{\bbeta}{\boldsymbol{\beta}} 
\newcommand{\betavec}{\boldsymbol{\beta}_{\text{vec}}} 
\newcommand{\mumu}{\boldsymbol{\mu}_{\boldsymbol{\mu}}} 
\newcommand{\Sigmu}{\boldsymbol{\Sigma}_{\boldsymbol{\mu}}} 
\newcommand{\mubeta}{\boldsymbol{\mu}_{\boldsymbol{\beta}}} 
\newcommand{\Sbeta}{\boldsymbol{\Sigma}_{\boldsymbol{\beta}}} 

\def\malpha{m_j(\balph_j|\by_j,\bX_j,\bbeta,\bmualph,\bSigalph)} 
\def\mbeta{m(\betavec|\by,\balphJ,\bX)} 

\newcommand{\Szero}{\boldsymbol{\Sigma}_0} 

\newcommand{\qvbthetaalpha}{q_{\boldsymbol{\lambda}}(\boldsymbol{\theta},\boldsymbol{\alpha}_{1:J})}
    
\newcommand{\predRTij}{\widetilde{RT}^{(t)}_{ij}}
\newcommand{\predREij}{\widetilde{RE}^{(t)}_{ij}}
\newcommand{\predRTjse}{\widetilde{RT}^{(t;j)}_{(s,e)}}

\newcommand{\medse}{\bar{M}^{(t)}_{s,e}}
\newcommand{\medsej}{M^{(t;j)}_{s,e}}

\newcommand{\accuracyse}{\bar{P}^{(t)}_{s,e}}
\newcommand{\accuracysej}{p^{(t;j)}_{s,e}}
	\def\muv{v}
	
	\def\muz{z}
	
	\def\mutau{\tau}
	\def\mutauj{\tau_j}
	
\begin{document}

{\setstretch{.8}
\maketitle


\begin{abstract}
Evidence accumulation models (EAMs) are an important class of cognitive models used to analyze both response time and response choice data recorded from decision-making tasks. Developments in estimation procedures have helped EAMs become important both in basic scientific applications and solution-focussed applied work. Hierarchical Bayesian estimation frameworks for the  linear ballistic accumulator model (LBA) and the diffusion decision model (DDM) have been widely used, but still suffer from some key limitations, particularly for large sample sizes, for models with many parameters, and when linking decision-relevant covariates to model parameters. We extend upon previous work with methods for estimating the LBA and DDM in hierarchical Bayesian frameworks that include random effects which are correlated between people, and include regression-model links between decision-relevant covariates and model parameters. Our methods work equally well in cases where the covariates are measured once per person (e.g., personality traits or psychological tests) or once per decision (e.g., neural or physiological data). We provide methods for exact Bayesian inference, using particle-based MCMC, and also approximate methods based on variational Bayesian (VB) inference. The VB methods are sufficiently fast and efficient that they can address large-scale estimation problems, such as with very large data sets. We evaluate the performance of these methods in applications to data from three existing experiments. Detailed algorithmic implementations and code are freely available for all methods.

\noindent

\end{abstract}
}



\section{Introduction}

Evidence accumulation models (EAMs) are a class of cognitive models that describe simple decisions \citep[for a review, see][]{ratcliff2016diffusion}. EAMs were initially developed with basic perceptual decisions in mind \citep{stone1960models,Laming1968} but their domain of explanation has steadily expanded to cover decisions about memories \citep[e.g.,][]{Ratcliff1978,evans2018modeling}, words \citep[e.g.,][]{wagenmakers2008diffusion}, and categorization \citep[e.g.,][]{PhiliastidesEtAl2006,nosofsky1997exemplar-based}, among others. Recent work has extended EAMs to describe much more complex decisions than they were originally developed for, for example, in consumer choices \citep[e.g.,][]{hawkins2014integrating}, clinical groups \citep[e.g.,][]{weigard2014diffusion,ratcliff2010individual,wall2021identifying}, and patient preferences in health care \citep[e.g.,][]{jones2015using}. As well as extending the domain of explanation of EAMs, the last decade has seen a concerted effort to more deeply integrate EAMs with theoretical advances in other areas of psychological research. One way in which this is accomplished is by including theoretically-motivated measured data as regressors, or covariates, to help understand their relationships with EAM parameters \citep[for example applications, see][]{van2017confirmatory,wiecki2013hddm,evans2017need}. 

The basic premise of an EAM is that decisions between competing alternative outcomes are made by accumulating evidence in favour of each possible response. The accumulation continues until a pre-defined threshold level of evidence is exceeded, after which the response corresponding to the evidence is executed. While all EAMs share this basic structure, they differ in the details of the accumulation process, the way different response options are represented, and many other details. For an overview of the rich variety of EAMs, see \cite{donkin2018response}. We restrict our analyses to two commonly used EAMs: the diffusion decision model \citep[sometimes abbreviated ``DDM'';][]{Ratcliff1978} and the linear ballistic accumulator \citep[LBA;][]{brown2008simplest}. The DDM models choices between two alternatives by assuming that evidence for and against the choices accumulates continuously over time, as a Wiener diffusion process. This process occurs between two absorbing boundaries that correspond to the two response options. On the other hand, the LBA is an accumulator-based model, which represents a choice between multiple alternatives as a race between multiple accumulators \citep[usually, one for each response, but see also][]{van2019accumulating}. 

Recently, there have been rapid advances in methods for estimating EAMs from data 
\citep{vandekerckhove2011hierarchical,wiecki2013hddm,gunawan2020new,dao2022efficient,turner2013method}. These  have improved the way in which EAMs (and other cognitive models) can accommodate more realistic and plausible assumptions about psychology.
%
%
%
%
%
%
%

The LBA makes some important simplifying assumptions about the evidence accumulation processes which permit closed-form expressions for the likelihood of the model. 
Efficient Bayesian methods based on particle Markov chain Monte-Carlo (MCMC) are available
for estimating the LBA model \citep[see, e.g.,][]{gunawan2020new,tran2021robustly}. These methods allow researchers to more adequately capture psychologically reasonable behaviour, such as the correlated nature of person-to-person variability represented by random effects. The simplicity of the analytic expressions in the LBA also makes it possible to link the model with other models of relevant data \cite[so called ``joint modelling'',][]{turner2016more,turner2013bayesian}. Joint modelling is successful in uncovering some links between neural and behavioral data with the LBA model. However, the estimation methods which have been used in this line of research do not easily lend themselves to a standard approach that has become very popular. A more standard approach is to link measured covariates with model parameters via regression. For example, researchers often want to know which model parameters are best related to neural data measured on each decision trial, or personality variables measured at the per-subject level. This kind of question can be easily investigated using the regression approach which is made easy for the DDM using the package developed by \cite{wiecki2013hddm}, but is not readily available for the LBA.

The likelihood function for the DDM is less tractable than for the LBA, and also much more computationally expensive. Because of this, many applications treat individual participants independently (i.e., non-hierarchical models, without random effects). The most commonly used estimation methods for estimating non-hierarchical DDMs are maximum likelihood \citep{ratcliff2002estimating} and the three binning methods: the chi-square method \citep{ratcliff2002estimating}, the multinomial likelihood ratio $\chi^2$ (also $G^2$ method), and the quantile maximum likelihood method \citep{heathcote2002quantile}. For recent reviews, see \citet{ratcliff2015individual,alexandrowicz2020comparing}. 

\citet{vandekerckhove2011hierarchical} and \citet{wiecki2013hddm} develop hierarchical structures for the DDM with corresponding Bayesian estimation methods. To overcome the difficulties imposed by the model's tractability, those methods adopt simplified diffusion models, either by removing some elements in the base model \citep{vandekerckhove2011hierarchical} or constraining some parameters to be fixed across all subjects \citep{wiecki2013hddm}. Both methods also  assume that each subject's parameters (the random effects) are uncorrelated a priori. This is implausible because the random effects represent cognitive processes which are likely to be highly related, for example, participants with faster processing speeds (drift rates) are also likely to have faster perception and response execution times (non-decision time) and unlikely to have excessive decision caution. Exact Bayesian methods, such as MCMC, for estimating the full DDM with correlated random effects are not readily available, and are also likely to be computationally prohibitive for many applications with a large number of subjects. An alternative approach is to speed up the estimation method by using variational Bayes (VB) inference. \citet{galdo2019variational} develop such an approach for the nonhierarchical LBA model using mean field VB. For the hierarchical LBA, \cite{dao2022efficient} develop a fixed form VB which speeds up the computation by a factor of 5-10 compared to MCMC.


\subsection{Our Objectives}

Our article  significantly advances the state of the art for estimating both of the most widely-used evidence accumulation models, the LBA and the DDM. Building on recent statistical developments, we develop new inference methods for both models. 
This includes the following advances beyond previous approaches: 
\begin{itemize}
    \item Exact Bayesian estimation, via particle MCMC, for the ``full'' DDM. This model includes all psychologically important elements which have been identified as necessary to account for benchmark data patterns \citep[see:][]{Ratcliff1978,ratcliff1998modeling,ratcliff2002estimating}. In addition, we extend the group-level model to include subject-to-subject correlations in random effects.
    \item Computationally efficient approximate Bayesian inference for the DDM. This follows recent developments for the LBA model \citep{dao2022efficient}. We adapt cutting edge variational Bayesian methods to overcome the substantial computational burdens of the full DDM. 
    \item 
    Regression approaches to link covariate data with random effects, for both the LBA and DDM. We extend earlier approaches to support these analyses using the LBA and using the full version of the DDM, and to the variational Bayesian methods, allowing application in larger data sets that would otherwise be computationally intractable. We provide methods and examples for covariates which are measured at both the decision-by-decision level (e.g., trial level neural data) and the person level (e.g., personality or demographic data). 
\end{itemize}

Our methodological advances make it possible to address psychological research problems that were not previously feasible. For example, in a detailed analysis below, we use the DDM and LBA to analyze data from more than one thousand participants, and link those data with a large set of person-level covariates. Existing estimation algorithms did not allow for this kind of analysis. Even if they had, using the full DDM for data of that magnitude would almost certainly have been practically infeasible, because of the computational burden. The variational Bayesian methods we develop make this analysis not just possible, but within the reach of most researchers, without incurring extensive computing costs. An important element in making this possible is also the contribution we make in finding automatic and high-quality starting points from which to initialize the  optimization algorithms used in VB. The code and data are made freely available at \url{https://github.com/Henry-Dao/RegEAMs}.

The rest of the paper is organized as follows. Section \ref{sec:EAMs} provides detailed specifications and notation for the LBA and DDM models, including the structures assumed for their random effects and covariates. Section \ref{sec:Bayesian-methods} discusses efficient exact (particle MCMC) and approximate (VB) Bayesian estimation methods. Section \ref{sec:simulation} evaluates the performance of the proposed estimation methods using simulated data with known ground truth. Section \ref{sec:reg-eams-real-data} provides detailed, real-world analysis examples using three data sets, including data from the Human Connectome Project \citep[HCP:][]{van-essen2013wu-minn} in which more than 1,000 subjects each made 
120 decisions in a memory task (the $n$-back paradigm). Section \ref{conclusions} concludes. The paper has an online supplement that contains some further empirical and technical results.

\section{EAMs}\label{sec:EAMs}

\subsection{The Linear Ballistic Accumulator (LBA) Model}\label{sec:LBA}

\begin{figure}
	\begin{center}
		\includegraphics[scale= 0.6]{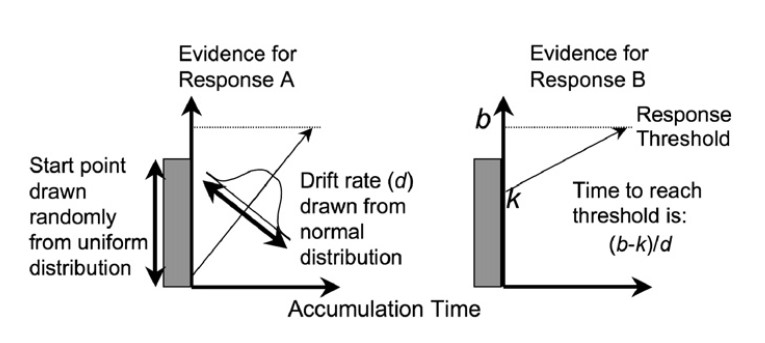}\\
	\end{center}
	\caption[LBA model]{An illustration of the LBA model for a binary choice with two evidence accumulators, one for ``Response A'' (left panel) and one for ``Response B'' (right panel). Evidence accumulates for each response until one reaches a threshold ($b$). The speed of evidence accumulation (drift rate $d$) and starting points ($k$) are random from decision to decision and between accumulators. From ``The simplest complete model of choice response time: Linear ballistic accumulation", by S.D. Brown and A. Heathcote, 2008, \emph{Cognitive Psychology}, 57, pp. 153-178. Copyright [2008] by Elsevier. Reprinted with permission.}\label{fig: LBA_picture}
\end{figure}

The LBA model of \citet{brown2008simplest} represents a choice between $C$ $( = 2, 3, ...)$ alternatives using $C$ different evidence accumulators, one for each response choice \citep[but see also][for more complex arrangements of accumulators]{van2019accumulating}. On any decision (a ``trial''), each evidence accumulator starts with evidence $k$ that increases at a speed given by the ``drift rate'' $d$ (see Figure \ref{fig: LBA_picture}). Accumulation continues until a response threshold $b$ is reached. The first accumulator to reach the threshold determines the response, and the time taken to reach the threshold is the response time (RT), plus some extra time for non-decision processes, $\tau$. This captures the time taken in decision-making before and after the evidence accumulation process, for example, encoding the evidence from a stimulus and executing a response. 

To explain the observed variability of human decision-making data, the model assumes that the starting points for the evidence accumulators ($k$) are randomly drawn from independent uniform distributions on the interval $[0,A]$, and the drift rates ($d$) are drawn from independent normal distributions with means $v_1, v_2, \dots, v_C$ for the different response accumulators. For convenience, a common and fixed variance $s^2=1$ is often assumed for all accumulators. There are other ways to obtain identifiability; for example, by the constraint on the sum of the drift rates \citep[see ][for more details]{DonkinPBR2009}. The likelihood function of the LBA model is available in closed form; see \citet{terry2015generalising} for details, including for non-normal distributions of drift rates.

Modern applications of the LBA model use a hierarchical Bayesian specification. Our article considers the hierarchical LBA model proposed by \citet{gunawan2020new} which allows the individual level parameters to vary from person to person as random effects. The random effects are assumed to be normally distributed, after log-transformation, with a covariance matrix that takes into account the correlations between the random effects. 

\subsection{The Diffusion Decision Model (DDM)}\label{sec:DDM}

\begin{figure}
	\includegraphics[scale=0.4]{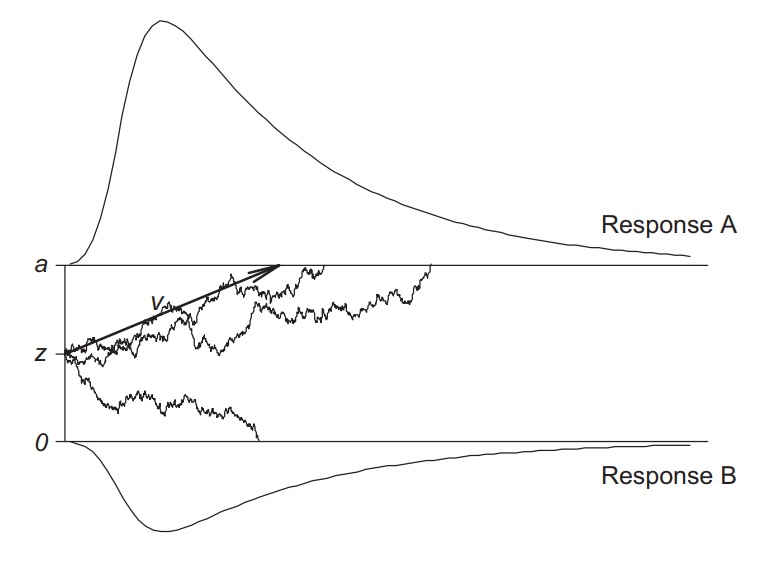}
	\caption[A Diffusion Process]{A simple diffusion decision model
		with constant drift $v$ starting at $z$. The process terminates as soon as it reaches one of the thresholds, at $a$ or $0$. Used with permission from \emph{Experimental Psychology} 2013; Vol. 60(6):385–402 ©2013 Hogrefe Publishing \url{www.hogrefe.com}\hspace{0.2cm} \url{https://doi.org/10.1027/1618-3169/a000218.} }\label{fig:simple_DDM}
\end{figure}

Diffusion decision models \citep[DDMs,][]{Ratcliff1978}  were first introduced in the psychology literature in the late 1970's as computational models to study and understand human decision-making. They have been successfully applied in many domains, for example, to study sleep deprivation, individual differences in IQ, and working memory; see \citet{ratcliff2016diffusion} for a recent review. Figure \ref{fig:simple_DDM} illustrates the decision-making process proposed in the DDM. Suppose a decision-maker chooses between responses A and B. A DDM assumes that information accumulates randomly and continuously over time (the time dimension is represented on the horizontal axis). The decision process starts with an initial amount of evidence denoted by $0<z<a$. Evidence is then accumulated randomly via Brownian motion with drift $v$ (i.e., a Wiener process). Accumulation continues until it reaches one of two absorbing boundaries, one for each response. For convenience, the criterion for one response is set at 0 (response B in this case) and the other is set at some value $a>z$. We call $a$ the boundary separation parameter. As evidence accumulates randomly from moment to moment, different sample paths are observed each time. Figure \ref{fig:simple_DDM} shows three such observed sample paths. The shape of the RT distributions is displayed on top for response A and upside down for response B. Just as for the LBA, the DDM also includes non-decision time denoted by $\tau$. 

The diffusion model described so far is sometimes known as the ``simple'' diffusion model; see \cite{stone1960models}. Modern applications often use the so-called ``full'' diffusion model, which includes trial-to-trial variability in three parameters: the start point of evidence accumulation, $z$; the drift rate, $v$; and the non-decision time, $\tau$. Variability in these parameters has been shown to be necessary to explain benchmark patterns in data, such as the relative speed of correct and incorrect decisions \citep{RatcliffRouder1998,RatcliffTuerlinckx2002}. The full DDM is characterised by seven parameters: drift-rate mean $\muv$, drift-rate standard deviation $\sv$, boundary separation $a$, starting point mean $\muz$ and width $\sz$, and non-decision time mean $\mutau$ and width $\stau$. The structure of the diffusion model imposes some constraints on its parameters -- for example, the separation between boundaries must be greater than the width of the uniform distribution of start points, $a>s_z$, and the start point of evidence accumulation must not be outside the response boundaries, $0.5s_z < z < a-0.5s_z$. We apply appropriate transformations of the raw parameters to ensure that all can be freely estimated on the full real line such that they all lie on the real line (see Section \ref{supp:sec-HDDMtransformation} of the online supplement for full details).

\subsection{Regression on Covariates}\label{sec:reg-eams-model-specification}

This section discusses the hierarchical version of the LBA and DDM, and how covariates can be linked to the random effects by the regression approach. We consider cases in which the  covariates are measured either at the subject or trial level (e.g., demographic data or neural measurements, respectively). The inclusion of covariate regression helps researchers to understand the effects of potential factors of influence on different elements of the decision process (e.g., eye movements, pupil diameter, skin conductivity, scalp electrical potential, personality measurements, and environmental variables). Including useful covariates in a model can also increase predictive power and reduce unexplained error variability. 

We first describe the hierarchical framework assumed for random effects, and then the inclusion of covariates. The hierarchical framework is the same one as proposed by \citet{gunawan2020new}, which assumes a multivariate normal distribution for the random effects which have been transformed to have support on the entire real line. Covariates influence each subject's random effect vector ($\balph_{j}$) through a linear combination of the covariate measurements for that subject and a matrix of coefficients ($\bbeta$). The coefficients are assumed to be fixed across subjects. The assumption of a fixed effect for coefficients is simpler than used in other approaches. For example, \cite{wiecki2013hddm} include random effects for regression coefficients, which is psychologically plausible: the relationship between, say, a scalp EEG measurement and a model parameter may plausibly differ between subjects, due to differences in scalp or cortex topography, or even mundane differences in the placement of an electrode cap. Our approach is more restrictive in its assumptions about regression coefficients because this allows us greater freedom in assumptions about the base cognitive model. For example, we are able to implement the ``full'' diffusion model with covariates, without assuming fixed effects for the model's trial-to-trial variability parameters \citep[as in][]{wiecki2013hddm}. Our approach also permits us to simultaneously estimate the effect of a set of covariates on all elements of the random effects vector in one step, which is not feasible with existing approaches.

Formally, let $X_{ij}^\top = [x_{ij;1},\dots,x_{ij;d}]$ be a vector consisting of $d$ covariates measured for each subject $j$  and trial $i$. Denote by $\bX_j$ the design matrix containing all trial-level covariates associated with subject $j$. Let $\balph_{j}$ be the random effects of subject $j$ and $\bbeta$ be a $\Da\times d$ matrix of regression coefficients which are assumed to be constant across participants (fixed effects), where $\Da$ is the dimension of the random effects. Elements of $\bbeta$ specify the effect that each covariate (indexed by column) has on each model parameter (indexed by row). The full model specification is then:
\begin{enumerate}
	\item The conditional density: 
	\begin{equation}\label{eq:reg-eams-conditional-density}
	    \begin{array}{l}
	    p(\by|\balphJ,\bbeta,\bX) = \prod\limits_{j=1}^{J}\prod\limits_{i=1}^{n_j} p(y_{ij}|\widetilde{\balph}_{ij}), \\
        \widetilde{\balph}_{ij} = \balph_j + \bbeta X_{ij}, \textrm{ for } j = 1,\dots,J,\, i = 1,\dots,n_j,
    \end{array}
	\end{equation}
	where $p(y_{ij}|\widetilde{\balph}_{ij})$ is the density of either the LBA or DDM, as appropriate.
	\item A multivariate normal distribution for the random effects
	\begin{equation*}
		\balph_{j}|\bmualph,\bSigalph \stackrel{ind.}{\sim} N(\bmualph,\bSigalph ).
	\end{equation*}
	Note that this normal distribution is over the random effects which have been transformed to the full real line. Section \ref{supp:sec-HDDMtransformation} of the online supplement gives further details of the transformations.
	\item Priors for the model parameters 
	\begin{enumerate}
	    \item A normal prior for the fixed effect coefficients $\betavec \sim N(\mubeta,\Sbeta),\textrm{ where }\betavec:= \textrm{vec}(\bbeta).$\footnote{$\vect(\bbeta)$ is a column vector obtained by stacking the columns of matrix $\bbeta$ under each other from left to right.}  In the simulation and real data examples, we set $\mubeta = \boldsymbol{0}$ and $\Sbeta = 9\boldsymbol{I}$, so that the prior can cover most of the plausible values. 
	    \item Group-level mean $\bmualph \sim N(\mumu,\Sigmu).$ The hyperparameters $\mumu$ and $\Sigmu$ are set to $\mumu = \boldsymbol{0}$ and $\Sigmu = 3\boldsymbol{I}$ so that the prior can cover most of the plausible values.
	    \item Group-level covariance matrix $\bSigalph$. Two priors are considered:
	    \begin{enumerate}
	        \item When the number of subjects is small (say, less than 30), we recommend using a relatively informative prior: $\bSigalph \sim \textrm{IW} \left(\ka,\bPsi\right).$ This prior is used in the real data example discussed in Section~\ref{sec:reg-eams-real-data} with the hyperparameters $\ka=20$ and $\bPsi = \boldsymbol{I}$. The notation $\textrm{IW}(a, A)$ means
            an inverse Wishart distribution with degrees of freedom $a$ and scale matrix $A$.
            \item For general use, we follow \citet{gunawan2020new} and adopt a marginally noninformative prior, proposed by \citet{huang2013simple} 
	        \begin{equation*}
        		\begin{array}{l}
        			\bSigalph|a_1,\dots,a_{\Da} \sim \textrm{IW} \left(\Da +1,\bPsi\right),\bPsi=4\textrm{diag}\left(\dfrac{1}{a_1},\dots,\dfrac{1}{a_{\Da}}\right),\\
        			a_1,\dots,a_{\Da} \sim \textrm{IG} \left( \dfrac{1}{2},1\right).
        		\end{array}
        	\end{equation*}
        	The notation $\textrm{IG}(a, b)$ means an inverse Gamma distribution with shape parameter $a$ and scale parameter $b$.
	        Denote $\ba = (a_1,\dots,a_{D_{\alpha}})$. The marginally noninformative prior is our default, and is used in all examples, unless stated otherwise.
	    \end{enumerate}
	\end{enumerate}
\end{enumerate}

Following many earlier developments \citep[starting with][]{ratcliff2002estimating}, when working with real data we augment the EAMs with a mixture process representing a small amount of ``contaminant'' responses. This assumption helps considerably with the stability of the estimation methods in the face of outliers in the data. The mixture is between the EAM density $p$ and a fixed density $p_0$
\begin{equation}\label{eq:reg_mixture}
    \tilde{p}(y_{ij}|\widetilde{\balph}_{ij}) = (1-\overline{\omega})p(y_{ij}|\widetilde{\balph}_{ij}) + \overline{\omega} p_0(y_{ij}),
\end{equation}
where $\overline{\omega}$ is a very small weight, $y_{ij}=(c_{ij},  \textrm{RT}_{ij})$, and $c_{ij}$ and $\textrm{RT}_{ij}$ are the response choice and the response time of individual $j$ at trial $i$, respectively. In the analyses below, we set $\overline{\omega} = 0.0001$. The psychological interpretation of Equation \eqref{eq:reg_mixture} is that the data are generated from a mixture of two processes, one from the EAM $p(y_{ij}|\widetilde{\balph}_{ij})$ and a rare (1 in 10,000) sample from the fixed distribution $p_0(y_{ij})$. The densities of response choice $c_{ij}$ and response time $\textrm{RT}_{ij}$ in the joint density $p_0(y_{ij})$ are assumed to be independent. The density of response choice $c_{ij}$ is assumed to follow uniform distribution over $C$ alternatives and the density of response times $\textrm{RT}_{ij}$ is assumed to follow a uniform distribution over the interval $(0,2)$, which matches experimental limitations set in the data collection procedures for the studies we consider here. 

\section{Bayesian Estimation Methods \label{sec:Bayesian-methods}}
We develop efficient exact Bayesian methods, based on particle MCMC, and very fast approximate Bayesian methods, using variational Bayesian inference. We extend both of these methods to the LBA and DDM models with the full random effects and covariate structures introduced in Section \ref{sec:reg-eams-model-specification}.

\subsection{Particle Metropolis within Gibbs (PMwG) Sampler}\label{sec:Bayesian-methods-PMwG}

\citet{gunawan2020new} propose an exact PMwG sampler for estimating hierarchical LBA models. We extend that method to include the DDM and also covariate regression. PMwG defines a target distribution on an augmented space that includes all the model parameters and random effects, called particles, and has the joint posterior density of the parameters and random effects as a marginal density. Extension to the DDM is complicated by the fact that many of the DDM's parameters have bounded and parameter-dependent support (for example, the start point $z$ must be more than zero but less than the theshold separation, $a$). This makes PMwG's assumption of a multivariate Gaussian distribution for random effects inappropriate. We develop parameter transformations for the DDM to address this problem.  Extensions to covariate regression are more subtle. All details can be found in Section~\ref{supp:PMwGalgorithm} of the online supplement.

\subsection{Variational Bayesian Estimation}\label{sec:vb-hddm-reg-eam}
Variational Bayes (VB) is used for estimating complex and high-dimensional statistical models when estimation using MCMC is too costly;
for reviews, see, e.g., \citet{blei2017variational,ormerod2010explaining}. VB can lead to substantial increases in computational speed (even x10 or x100), as well as improvements in parallelisability. The computational gain is practically important, especially for computationally costly models like the DDM. Exact Bayesian methods based on MCMC are often not practical for data sets of even moderate size for the hierarchical DDM. For this reason, faster VB approaches open up a new range of possibilities for analysis.

The efficiency of VB is achieved by approximating the posterior distribution of the parameters and random effects, $p(\btheta,\balphJ|\by)$, using a simpler and more tractable distribution, $q$. The approximating distribution $q$ is parameterised by a vector of variational parameters $\blamb$, and so the estimation problem is reduced to optimising over $\blamb$ to maximise the similarity between the approximating distribution $q$ and the target posterior. This optimization problem can be very difficult because the dimension of the vector $\blamb$ can be very large (e.g., with several entries for each subject, and so hundreds or thousands of elements in all). 

This particular VB method is known in the literature as ``fixed-form'' VB as the functional form of the approximating distribution $q$ is fixed a priori. The discrepancy between the approximate and the target distribution is measured using the Kullback-Leibler (KL) divergence
$$KL(\qvbthetaalpha||p(\btheta,\balphJ|\by)) := E_{\qvbthetaalpha}\left[ \log\dfrac{\qvbthetaalpha}{p(\btheta,\balphJ|\by)}\right];$$
the notation $E_f$ denotes the expectation with respect to the density $f$. The best approximation is obtained by minimizing $KL(\qvbthetaalpha||p(\btheta,\balphJ|\by))$, or equivalently, maximizing the so-called ``lower bound'' \citep{blei2017variational}
$$ \lb:= E_{\qvbthetaalpha}\left[\log p(\btheta,\balphJ,\by)  - \log\qvbthetaalpha\right]. $$

Usually, we cannot calculate $\lb$ and its gradient vector $\gradlb$ analytically (that is the case for all analyses considered in this paper). Instead, we have access to unbiased estimators of $\lb$ and $\gradlb$ and therefore can use stochastic gradient ascent methods \citep{robbins1951stochastic} to optimise $\lb$. Like all optimization algorithms, we start from some initial value $\blamb^{(0)}$ and then update recursively:
\begin{equation}
	\label{eqn:SGD}
	\blamb^{(t+1)} = \blamb^{(t)} + \boldsymbol{\rho}_t\odot\widehat{\nabla_{\blamb}\mathcal{L}(\blamb^{(t)})},
\end{equation}
where $\brho_t$ is a vector of step sizes or learning rates, $\odot$ denotes the element-wise product of two vectors, and $\gradlbest$ is an unbiased estimate of the gradient $\gradlb$. In practice, the success of the algorithm crucially depends on the choice of the learning rates $\brho_t$ and the efficiency of the unbiased estimator $\gradlbest$. We use ADAM adaptive learning rate \citep{kingma2014adam} and the so-called ``reparametrization trick'' to obtain $\gradlbest$. Section \ref{reparameterisationtrick} of the online supplement gives further details.

The massive increase in computational efficiency of VB over MCMC comes with some well-known drawbacks. First, VB produces an approximation to the posterior distribution, not an (asymptotically) exact posterior simulation, as in MCMC. The quality of the approximation depends on the choice of the predetermined class of densities $\qvbthetaalpha$. The variational densities should be selected to balance the flexibility of the variational approximation and the computational cost. Second, like all optimization-based methods, VB is subject to problems in search, such as getting stuck in local optima. For this reason, the initial values for the search ($\blamb^{(0)}$) are important factors that contribute to its performance. Third, many empirical studies have shown that even though VB can lead to quite accurate approximations of the mean of the posterior distribution, but often the variance is underestimated \citep{blei2017variational,giordano2018covariances}.

\subsubsection{VB for Evidence Accumulation Models Including Covariates}\label{subsec:reg-eam-bayes-methods}

This section presents a VB algorithm for EAMs that incorporates covariates. Additionally, we propose a further simplified and more efficient VB variant, named VBL, designed to handle big data (i.e., VB for Large data sets). 

\citet{dao2022efficient}  propose the VB method for estimating hierarchical LBA models without covariates. 
The posterior density of the parameters and random effects can be factored as
\begin{equation*}
    p(\balph_{1:J},\bmualph,\bSigalph,\ba|\by) = p(\balph_{1:J},\bmualph,\ba|\by)p(\bSigalph|\balph_{1:J},\bmualph,\ba,\by).
\end{equation*}
Assuming the conjugate prior for $\bSigalph$, \citeauthor{dao2022efficient} show that the conditional density \\$p(\bSigalph|\balph_{1:J},\bmualph,\ba,\by)$ is the density of $\textrm{IW}(\bSigalph|\nu,\bPsi')$
with $\nu = 2\Da + J + 1$ and $\bPsi' = \sum_{j=1}^J (\balph_j - \bmualph)(\balph_j - \bmualph)^\top + 4\diag\left(a^{-1}_1,\dots,a^{-1}_{\Da}\right)$. 

Defining $\tilde{\btheta} = (\balph_{1:J},\bmualph,\log \ba)$, \citeauthor{dao2022efficient} propose a hybrid Gaussian VB which simplifies estimation through approximating the (large) covariance matrix for $\tilde{\btheta}$ using a factor structure. With this assumption, the approximating distribution has the form
\begin{equation}\label{eq:old-vb}
    q_{\blamb}(\tilde{\btheta},\bSigalph) = N(\tilde{\btheta}|\bmu,BB^\top + D^2) \textrm{IW}(\bSigalph|\nu,\bPsi'),
\end{equation}
where $B$ is a $p\times r$ matrix, $p$ is the number of components in $\tilde{\btheta}$, $r$ is the number of factors, and $D^2$ denotes a diagonal matrix having the main diagonal given by vector $\bd^2:= (d_1^2,\dots,d_p^2)$. We call the variational approximation in Equation \eqref{eq:old-vb} ``VB''. In addition, when the number of subjects is small, informative priors can be used for the group-level covariance matrix to regularize and stabilize the optimisation. Detailed discussion on VB and its extension can be found in Section \ref{supp:VB} of the online supplement.

The variational approximation in~\eqref{eq:old-vb} becomes less efficient for data sets with a large number of subjects, because the estimated number of random effects grows linearly with the number of subjects, and the size of the covariance matrix of $\tilde{\btheta}$ grows with the square of the number of subjects. For efficient estimation in large data sets, we propose a restricted variant called VBL. VBL assumes that the posterior distribution of the random effects is independent across subjects. It has the form
\begin{equation}\label{eq:new-vb}
    q_{\blamb}(\dbtilde{\btheta},\balph_{1:J},\bSigalph) = N(\dbtilde{\btheta}|\bmu_{J + 1},B_{J + 1}B_{J + 1}^\top + D_{J + 1}^2) \textrm{IW}(\bSigalph|\nu,\bPsi') \prod\limits_{j = 1}^J N(\balph_j|\bmu_j,B_jB_j^\top + D_j^2) ,
\end{equation}
where $\dbtilde{\btheta} = (\bmualph,\log \ba)$ if the model does not include covariates, and $\dbtilde{\btheta} = (\bbeta,\bmualph,\log \ba)$ if the model includes covariates. 
The simplified assumption of independence removes the need to estimate a large covariance matrix. In Equation \eqref{eq:old-vb}, all the correlations in the random effects and parameters $\dbtilde{\btheta}$ are captured by the covariance matrix $BB^\top + D^2$. In Equation~\eqref{eq:new-vb}, these correlations are captured by blocks of smaller covariance matrices. The restricted VBL approach still captures the correlation of random effects \textit{within} subjects by the smaller matrices $B_jB_j^\top + D_j^2,j = 1,\dots, J + 1$. 

The simplified assumptions of VBL greatly speed up computation, especially when the number of subjects is large. For example, in Section~\ref{sec:hcp-data} we analyse a data set consisting of $1,066$ subjects. In this case, VB would model the dependence of all parameters and random effects, requiring a covariance matrix of size $18,000\times 18,000$. On the other hand, VBL models the dependence between random effects within subjects, requiring covariance matrices of size only around $20 \times 20$.  

Unless stated otherwise, we use $40$ factors (the number of factors, $r$, is precisely the number of columns of $B$) and use $10$ Monte Carlo samples to estimate the lower bounds and the gradients of the lower bounds. We use $2$ factors for the $B_1,\dots,B_J$ matrices and $10$ factors for $B_{J+1}$. We only use a single Monte Carlo sample for VBL for estimating the lower bounds. Section \ref{supp:sec-gradients-reg-eams} of the online supplement gives a more detailed discussion of VB and VBL, and Section \ref{supp:sec-VB-HDDM} discusses the implementation details for the DDM. 


\subsection{VB Initialization}\label{sec:vb-initialization}

Like all iterative search algorithms, the efficiency and stability of VB depends on good initialization -- reasonable start points for the search. While selecting plausible values using domain knowledge may suffice for simple models, it is insufficient for the complex models under consideration. We propose two initialization methods for the variational parameters for VB and VBL. For VB, the variational parameters $\blamb$ consist of the mean $\bmu_{\lambda}$, and the matrices $B$ and $D$, and for VBL, the variational parameters $\blamb$ consist of the means $(\bmu_{\lambda,1},\dots,\bmu_{\lambda,J+1})$ and the matrices  $B_1,\dots,B_{J + 1}, D_1,\dots,D_{J+1}$. 

Two initialization methods are considered for the mean components $\bmu_{\lambda}$ for VB and $(\bmu_{\lambda,1},\dots,\bmu_{\lambda,J+1})$ for VBL. The first is based on maximum a posteriori (MAP) estimation, and the second is based on running a small number of Markov chain iterations, using PMwG. In all the experiments considered in this paper, the elements in these matrices $B,B_1,\dots,B_{J + 1},D, D_1,\dots,D_{J+1}$ are initialized as $0.01$.  These values ensure that the optimization algorithm is stable because generated values will be close to the mean values in the first few initial iterations. 



\subsubsection{MAP Initialization for VB and VBL}
To initialize the VB and VBL parameters using maximum a posteriori estimation (MAP), we first find MAP estimates of the $\balph_{1:J}$ and $\bbeta$ by fitting the non-hierarchical EAMs independently for each subject. The steps are: 
\begin{enumerate}
    \item Data from  subject $j$,  for $j=1,\dots,J$, we consider a non-hierarchical EAM, 
    \begin{align*}
        &p(\by_j|\balph_j,\bX_j,\bbeta^{(j)}) = \prod\limits_{i=1}^{n_j}p(y_{ij}|\widetilde{\balph}_{ij}),\\
        &\widetilde{\balph}_{ij} = \balph_j + x^\top_{ij}\bbeta^{(j)},\\
        &\betavec^{(j)} \sim N(\boldsymbol{0},10I),\quad \balphj \sim N(\boldsymbol{0},10I).
    \end{align*}
    We maximize the density $p(\by_j|\balph_j,\bX_j,\bbeta^{(j)})p(\balph_j)p(\bbeta^{(j)})$ with respect to $\balph_{j}$ and $\bbeta^{(j)}$ to obtain the MAP estimates denoted by $\widehat{\balph}_j$ and $\widehat{\bbeta}^{(j)}$. To initialize the maximization procedure, a set of plausible values for $\balphj$ is selected, and the elements of $\bbeta^{(j)}$ are set to zero. Note that the priors in this step are used only to regularize the optimization, hence, they are different from the priors in the model specification in Section~\ref{sec:reg-eams-model-specification}. 
    \item For VB, we first partition $\bmu_{\lambda}$ into four blocks, $\bmu_{\lambda} = (\widetilde{\bmu}_1,\widetilde{\bmu}_2,\widetilde{\bmu}_3,\widetilde{\bmu}_4)$. The blocks correspond to the four different types of parameters or random effects in the models defined in Section \ref{subsec:reg-eam-bayes-methods}. That is, $\widetilde{\bmu}_1$, $\widetilde{\bmu}_2$, $\widetilde{\bmu}_3$, and $\widetilde{\bmu}_4$ are the means of $\balphJ$, $\bmualph$, $\bbeta$, and $\log \ba$, respectively. Recall from Section~\ref{sec:reg-eams-model-specification} that $\ba = (a_1,\dots,a_{D_{\alpha}})$ are hyperparameters coming from the marginally noninformative prior for the group-level covariance matrix. 
    
    The initial values for $\widetilde{\bmu}_1$, $\widetilde{\bmu}_2$, and $\widetilde{\bmu}_3$ are obtained from the MAP estimates obtained from Step 1. 
    We set the initial values for $\widetilde{\bmu}_1 = (\widehat{\balph}_1,\dots,\widehat{\balph}_{J})$. The initial values of the elements of $\widetilde{\bmu}_2$ and $\widetilde{\bmu}_3$ are obtained by averaging over the elements of $\widehat{\balph}_j$ and $\widehat{\bbeta}^{(j)}$, respectively, over the $J$ subjects. The initial values for the elements of $\widetilde{\bmu}_4$ are set to zero. 

    \item For VBL, we set the initial values for $\bmu_{\lambda,j}=\widehat{\balph}_j$ for $j=1,...,J$. We then partition $\bmu_{\lambda,J+1}=(\boldsymbol{\mu}_{\lambda,J+1,1},\boldsymbol{\mu}_{\lambda,J+1,2},\boldsymbol{\mu}_{\lambda,J+1,3})$, where $\boldsymbol{\mu}_{\lambda,J+1,1}$, $\boldsymbol{\mu}_{\lambda,J+1,2}$, and $\boldsymbol{\mu}_{\lambda,J+1,3}$ are the means of $\bmualph$, $\bbeta$, and $\log \ba$, respectively.
    The initial values of the elements of $\boldsymbol{\mu}_{\lambda,J+1,1}$ and $\boldsymbol{\mu}_{\lambda,J+1,2}$ are obtained by averaging over the elements of $\widehat{\balph}_j$ and $\widehat{\bbeta}^{(j)}$, respectively, over the $J$ subjects. The initial values for the elements of $\boldsymbol{\mu}_{\lambda,J+1,3}$ are set to zero. 

    
\end{enumerate}
There are two special cases where we cannot obtain the initial values for $\bbeta$ using the MAP initialisation method. The first case is when the data does not involve covariates, the variational means $\bmulamb$ consists of $(\widetilde{\bmu}_1,\widetilde{\bmu}_2,\widetilde{\bmu}_4)$ for VB and $\bmu_{\lambda,J+1}$ consists of $(\boldsymbol{\mu}_{\lambda,J+1,1}, \boldsymbol{\mu}_{\lambda,J+1,3})$ for VBL. The second case is when the data consists of subject-level (as opposed to trial-level) covariates that are fixed across trials (section~\ref{sec:hcp-data}). For this case, we use data for subject $j$ to fit a non-hierarchical EAM
\begin{align*}
    &p(\by_j|\balph_j,\bX_j,\bbeta^{(j)}) = \prod\limits_{i=1}^{n_j}p(y_{ij}|\balph_{j}),\\
    &\balphj \sim N(\boldsymbol{0},10I),
\end{align*}
The initial values for $\widetilde{\bmu}_3$ in VB and $\boldsymbol{\mu}_{\lambda,J+1,3}$ for VBL are set to zero.




\subsubsection{PMwG Initialization}

In this approach, the PMwG sampler is applied to the full hierarchical EAM for a small number of iterations. The average of the last draws of the chain is used as the initial value for $\bmu_{\lambda}$. The advantage of PMwG compared to MAP is that it provides initial values of all the mean components of $\bmu_{\lambda}$ for VB and $(\bmu_{\lambda,1},\dots,\bmu_{\lambda,J+1})$ for VBL. The disadvantage of the PMwG initialization method is that it is computationally expensive when the number of subjects is large. In all examples considered in this paper, we run PMwG for $200$ iterations and use the average of the last $100$ draws as the initial value.




\section{Simulation study}\label{sec:simulation}

This section studies the performance of PMwG and VB methods for estimating hierarchical DDMs using simulated data. The data are simulated from a model mimicking the conditions inspired by Experiment 1 of \citet{wagenmakers2008diffusion} discussed in Section \ref{sec:hddm-real-data}. We don't replicate the same analysis for the hierarchical LBA here as it was documented in \cite{dao2022efficient}. 

We generated two data sets, one with $J=12$ subjects and one with $J=50$ subjects. Each subject made $1,200$ decisions split equally between two experimental conditions ($600$ trials for each condition). Trials in each condition were split between four types of stimulus (high frequency words, low frequency words, very low frequency words, and non-words) with frequencies: $100$, $100$, $100$ and $300$. The generated data have nonzero correlations between the random effects which reflects both the improved specification of the hierarchical DDM as well as plausible psychological assumptions about individual differences. The group-level parameters were set to match those estimated for the real data.

Three  sampling stages are employed in PMwG. In the initial stage, the first $500$ iterates are discarded as burn-in; then the next $1,500$ iterates are used in the adaptation stage to construct the efficient proposal densities for the random effects for the final sampling stage; finally, a total of $14,000$ ($J=12$) and $8,000$ ($J=50$) MCMC posterior draws are obtained in the sampling stage; see \citet{gunawan2020new} for further details.

The performance of the PMwG sampler is assessed based on two criteria: the efficiency of the sampler and the recovery of the data-generating (``true'') parameter values. The efficiency of the sampler is measured by the integrated autocorrelation time (IACT, see 
Section \ref{supp:IACT} of the online supplement), shown in Table \ref{tab:HDDM_PMwG_mimic_Lexical} in Section \ref{additionalresultssimstudy} of the online supplement for each parameter and both simulated data sets ($J=12$ and $J=50$). Overall, the estimated IACTs are quite small, except for the tenth component, indicating that the sampler is efficient. The tenth component is the parameter for the variability of the starting point of evidence accumulation, $\log (\sz)$, which has previously been found to be difficult to estimate \citep{boehm2018estimating,ratcliff2015individual,wiecki2013hddm}. As expected, the PMwG sampler also performs significantly better when there are more subjects in the simulated data.

\begin{figure}
    \centering
    \includegraphics[scale=0.24]{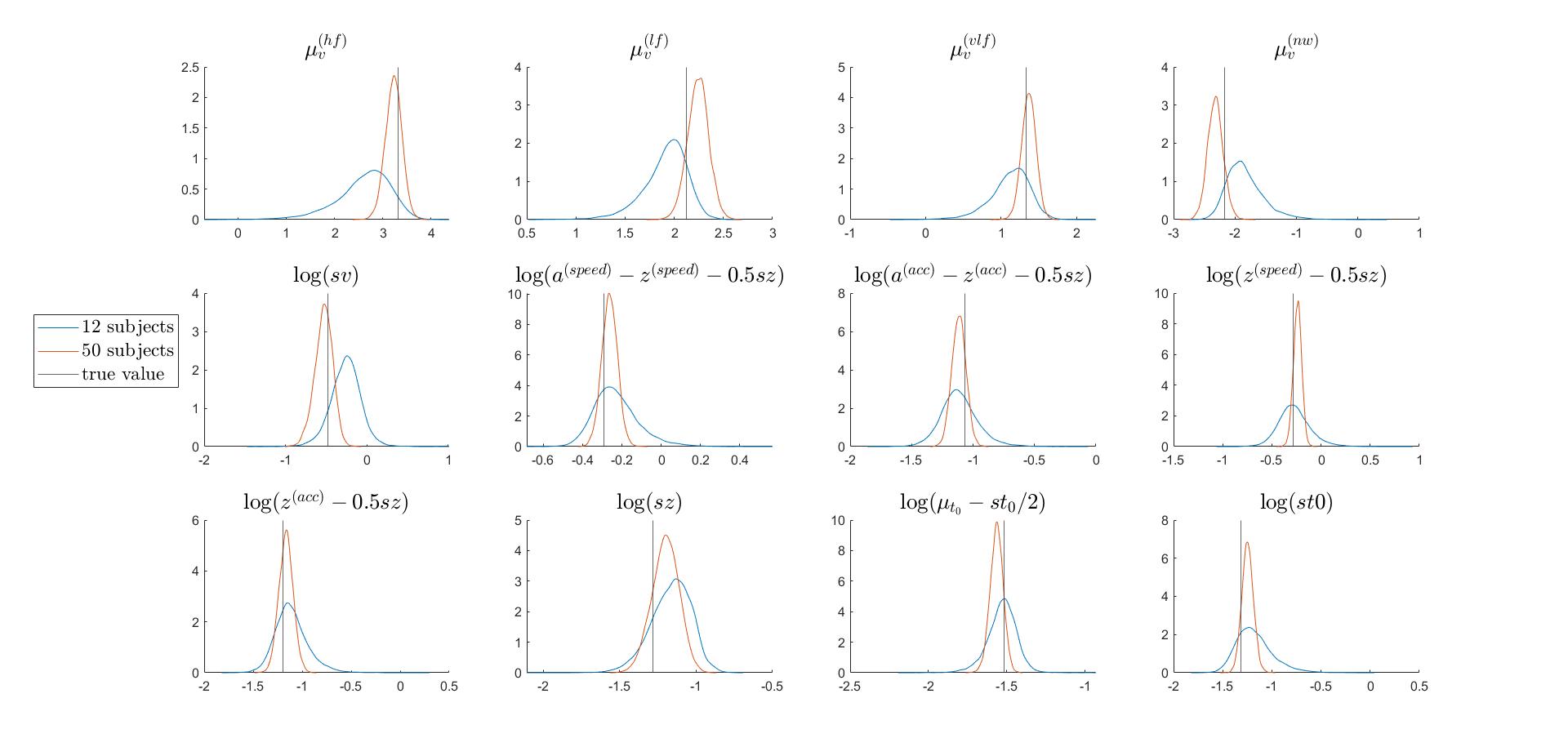}
    \caption[DDM - Simulation Study. Kernel density estimates of the group-level means.]{Kernel density estimates of the group-level mean $\bmualph$ estimated using the PMwG method in the first simulation setting (12 subjects, blue color) and the second setting (50 subjects, red color). The vertical lines show the true values.}
    \label{fig:hddm_mimic_Lexical_density_mean}
\end{figure}

In terms of parameter recovery, Figure~\ref{fig:hddm_mimic_Lexical_density_mean} shows the posterior densities estimated by PMwG for the group-level means $\bmualph$ together with vertical lines showing the data-generating (true) parameter values. Figure \ref{fig:hddm_mimic_Lexical_density_var} in section \ref{additionalresultssimstudy} of the online supplement shows corresponding details for the posterior densities of the variances (diagonal of $\bSigalph$). For all parameters in the figure, the mode of the posterior distribution was close to the data-generating value, indicating good parameter recovery.

We now study the performance of the variational Bayes algorithm, VB, with the covariance matrix modelled using 40 factors (refer to Section \ref{subsec:reg-eam-bayes-methods}) for estimating the DDM. The VB initial values were obtained by using the MAP strategy. The performance of VB is assessed based on its accuracy and computational cost, treating the posterior distributions estimated using PMwG as the ground truth. 

\begin{figure}
    \centering

    \hspace*{-2cm}  \includegraphics[scale=0.22]{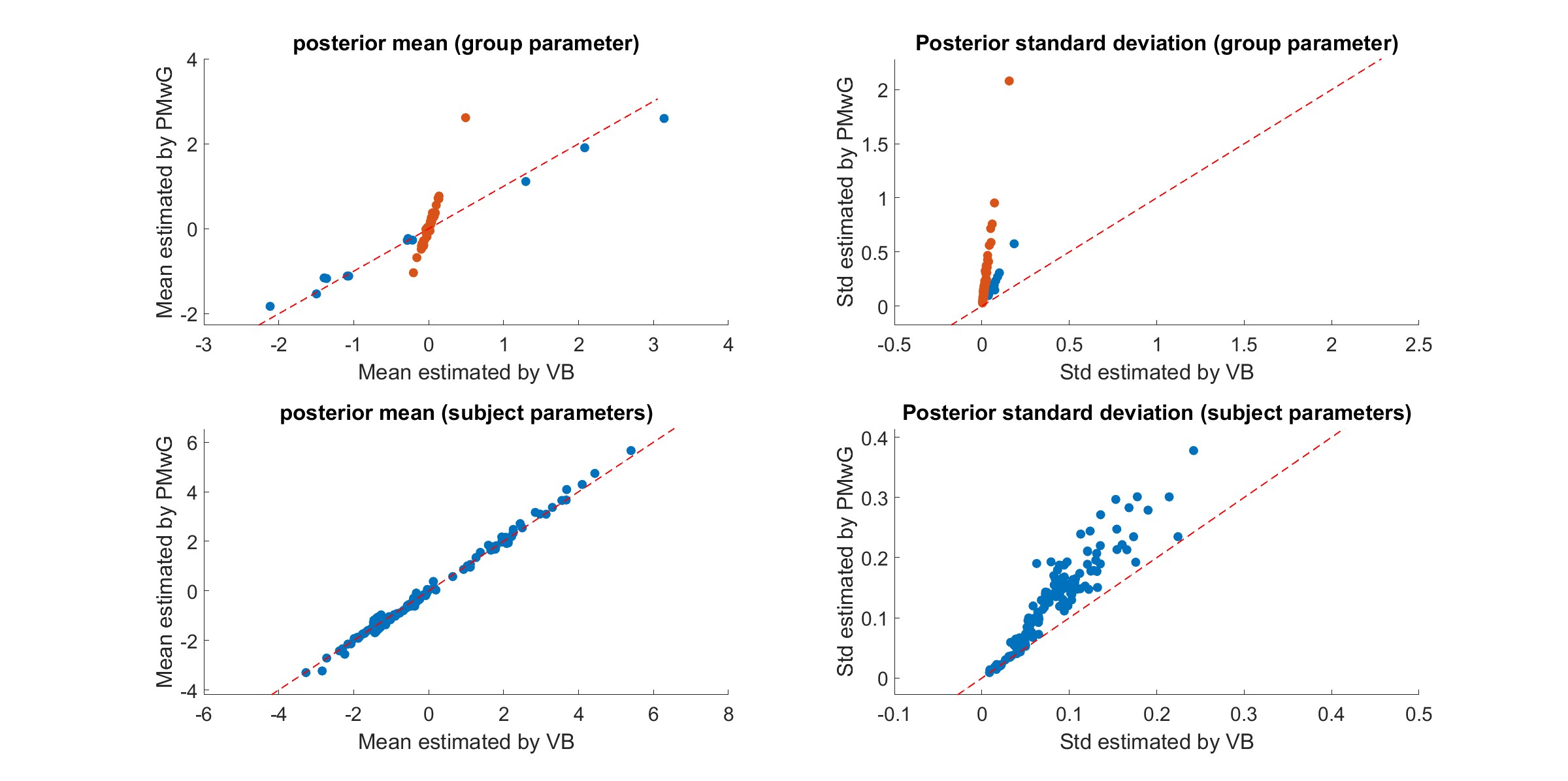}     
    \caption[DDM - Simulation Study (12 subjects). Comparing the posterior moments estimated using PMwG against using VB.]{Comparing the means and standard deviations of the marginal posterior distributions estimated by VB (horizontal axis) against the exact values calculated using PMwG (vertical axis), for the simulation study with $J=12$ subjects. The top panels give the means and standard deviations of the group-level parameters. The bottom panels show the means and standard deviations of the subject-level random effects. Notice that all the parameters are plotted on the transformed scale. Additionally, the red dots appearing in the top left and top right panels indicate the post mean and posterior standard deviation of the group-level covariance matrix.}
    \label{fig:hddm_simdata_medium_vb_pmwg_moment}
\end{figure}

\begin{figure}
    \centering
 
    \hspace*{-2cm} \includegraphics[scale=0.22]{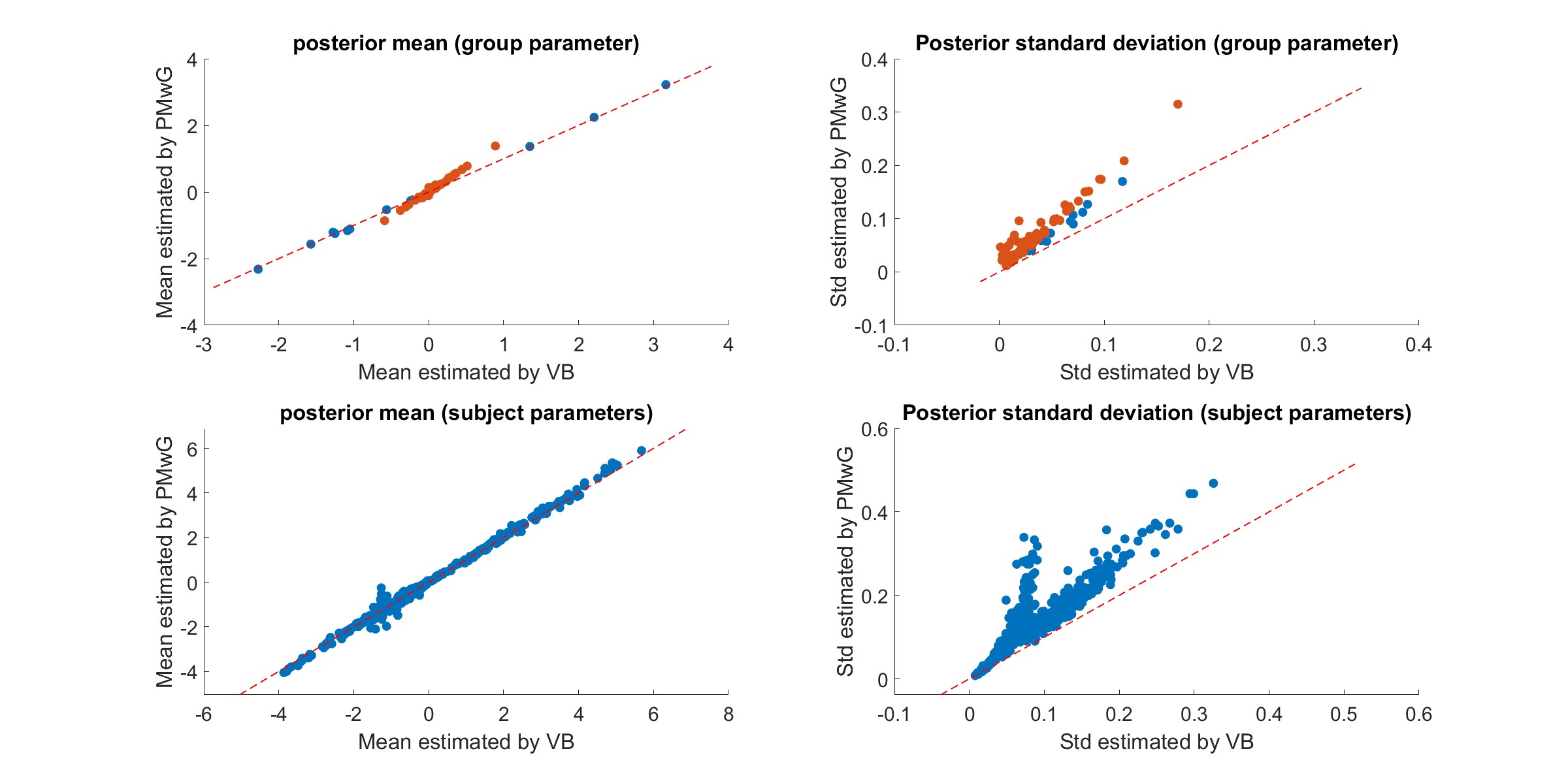}    
    \caption[DDM - Simulation Study (50 subjects). Comparing the posterior moments estimated using PMwG against using VB .]{Comparing the means and standard deviations of the marginal posterior distributions estimated by VB (horizontal axis) against the exact values calculated using PMwG (vertical axis) for the simulation study with $J=50$ subjects. The top panels give the means and standard deviations of the group-level parameters. The bottom panels show the means and standard deviations of the subject parameters. Additionally, the red dots appearing in the top left and top right panels indicate the posterior mean and posterior standard deviation of the group-level covariance matrix.}
    \label{fig:hddm_simdata_large_vb_pmwg_moment}
\end{figure}

Figures \ref{fig:hddm_simdata_medium_vb_pmwg_moment} and \ref{fig:hddm_simdata_large_vb_pmwg_moment} extend the comparison of VB estimates against the ground truth of all parameters and  random effects for individual subjects. The figures plot the posterior means and standard deviations for group-level parameters and random effects estimated using VB against those estimated using PMwG. The results indicate that the VB approximations provide accurate estimates of the posterior means for both group-level means and random effects (as shown by the blue dots), although the VB estimates of posterior variances tend to be underestimated. However, it should be noted that in the small-sample setting (as seen in Figure~\ref{fig:hddm_simdata_medium_vb_pmwg_moment}), the VB method does not perform as well in approximating the group-level covariance matrix compared to other parameters (red dots). This is due to the fact that accurate estimation of the covariance matrix requires a sufficient number of observations (i.e., participants). As demonstrated in Figure~\ref{fig:hddm_simdata_large_vb_pmwg_moment}, the approximation of the covariance matrix improves significantly as the number of observed subjects increases.

Figure~\ref{fig:hddm_simdata_medium_vb_pmwg_density} in Section \ref{additionalresultssimstudy} of the online supplement shows the posterior marginal densities of the group-level mean parameters estimated using PMwG and VB methods for the first simulated data (12 subjects). The figure shows that VB can approximate the posterior mean of the drift rate and the non-decision time parameters well. However, the posterior variances are  underestimated and the starting point variability parameters are harder to approximate. Figure \ref{fig:hddm_simdata_large_vb_pmwg_density} in section \ref{additionalresultssimstudy} of the online supplement shows a similar plot for the simulation study with 50 subjects.


The primary advantage of the VB approach is its computational efficiency. Table \ref{tab:HDDM_mimic_Lexical_comparison} shows that VB is around 5 times faster than PMwG for data with 12 subjects, and around 15 times faster for data with 50 subjects. For this comparison, it is important to note that PMwG was implemented using parallel programming, and run on a high-performance computer cluster with 48 CPU cores. In contrast, VB can run on a standard desktop computer, which makes it more accessible and cheaper in terms of computational resources. The comparison in Table \ref{tab:HDDM_mimic_Lexical_comparison} compares computational efficiency in ``real time'' -- when different computational resources are available. The gain in efficiency for VB would be even more marked if the two algorithms were compared on equivalent capacity machines (i.e., both 8-core machines, or both 48-core machines).

\begin{table}[H]
    \centering
    \begin{tabular}{c|c|c||c|c}
    \multirow{3}{*}{\textbf{Method}}  & \multicolumn{2}{c||}{\textbf{Running time }} & \multicolumn{2}{c}{\textbf{Number of}} \\
    & \multicolumn{2}{c||}{\textbf{(hour:minute)}}  & \multicolumn{2}{c}{\textbf{cpu-cores}} \\
    \cline{2-5}
    & $J=12$ & $J=50$ & $J=12$ & $J=50$\\
    \hline
    \hline
    PMwG & 31:04 & 49:52 & 48 & 48 \\
    \hline
    VB  & 1:17 & 7:52 &  8 & 8 \\
     \hline
    \end{tabular}
    \caption[DDM - Simulation Study. Comparing running  time  between VB and PMwG.]{A comparison between PMwG and VB in terms of running time and computational resources in the simulation study under two settings: 12 subjects ($J=12$) and 50 subjects ($J=50$).}
    \label{tab:HDDM_mimic_Lexical_comparison}
\end{table}



We do not present direct comparisons in computational efficiency between our proposed methods and the HDDM package \citep{wiecki2013hddm}. Such comparisons are not informative, because the two packages implement sufficiently different model structures to make comparisons meaningless. There are three important model simplifications imposed in the HDDM package. First, all the variability parameters $\sv,\sz,\stau$ are fixed across subjects. Second, HDDM constrains random effects that depend on experimental conditions ($a^{(speed)},a^{(acc.)}$, or $\muz^{(speed)},\muz^{(acc.)}$) and random effects that depend on stimuli ($\muv^{(hf)},\muv^{(lf)},\muv^{(vlf)},\muv^{(nw)}$) to have identical group-level variances. Third, HDDM assumes independent (uncorrelated) priors for the random effects. 

Apart from these model-based differences, we have implemented our proposed PMwG and VB approaches in Matlab. Due to the differences in modeling assumptions and software implementations, a detailed comparison between HDDM and the PMwG and VB approaches is not feasible and is beyond the scope of the current paper. Instead, we focus on presenting the PMwG and VB methods as alternative approaches to estimate more complex DDMs and provide a thorough evaluation of their performance and accuracy using simulation studies and real data analyses. 



\section{Real Data Analysis}\label{sec:reg-eams-real-data}

Below, we analyze data from three very different experiments, chosen to highlight strengths and limitations of the various approaches for model estimation. The first experiment had a moderate number of subjects (17) who each completed a large number of trials (around $1,700$) in a lexical decision task \citep{wagenmakers2008diffusion}. Analysis of this experiment demonstrates that our new methods for applying the DDM in hierarchical settings with PMwG and VB are practical, efficient, and produce outcomes that align with earlier analyses using the LBA model \citep{gunawan2020new} and also simpler estimation methods for the DDM \citep{wagenmakers2008diffusion}. The second experiment has a small number of subjects (9) complete a moderate number of trials (around 600) in a mental rotation decision-making task \citep{provost2013two}. During the experiment, scalp electrical potentials were recorded for every trial. Previous analyses focused on linking these EEG covariates with drift rate parameters of the LBA model \citep{van2017confirmatory}. We extend these earlier analyses to the more complete LBA model specification, and then also extend to the corresponding analysis using the DDM. The third experiment uses data from the Human Connectome Project \citep[HCP:][]{van-essen2013wu-minn} in which more than 1,000 subjects made just 120 decisions each in a memory task (the $n$-back paradigm). A key feature of the HCP is the measurement of interesting covariates at the person-by-person level, including personality measurements, demographic data, medical histories, structural brain measurements, and even genetic data. Data from the 2-back memory task have previously been analyzed using the LBA model and compared with genetic overlap data \citep[twin status;][]{evans2018modeling}, but analysis of larger data sets including covariates has not previously been within reach for the LBA model, and certainly not for the DDM.


\subsection{Data Set 1: Lexical Decision}\label{sec:hddm-real-data}
This example uses data from Experiment $1$ reported by \citet{wagenmakers2008diffusion} to demonstrate the new estimation methods we have developed for the DDM, using both PMwG and VB. The data consist of 17 participants who were asked to determine whether letter strings were valid English words or non-words, under two experimental conditions: speed and accuracy emphases. There were four kinds of letter strings randomly mixed from trial to trial: high frequency words, low frequency words, very low frequency words and non-words. For the DDM, we adopt the model specification considered by \citeauthor{wagenmakers2008diffusion}, and embed this within a hierarchical structure with correlated random effects. Different mean drift rates are estimated for different stimulus classes, denoted by $\muv^{(hf)}$ (high frequency), $\muv^{(lf)}$ (low frequency), $\muv^{(vlf)}$ (very low frequency), and $\muv^{(nw)}$ (non-words). The boundary separation $a$ and the starting point $\muz$ are different across the speed-emphasis and accuracy-emphasis conditions. In total, each subject has 12 parameters,
$$ (\muv^{(hf)},\muv^{(lf)},\muv^{(vlf)},\muv^{(nw)},\sv,a^{(speed)},a^{(acc.)},\muz^{(speed)},\muz^{(acc.)},\sz,\mutau,\stau). $$
As usual, we transform the subject parameters to the full real line (using the transformations in Section~\ref{supp:sec-hddm-simulation-transform} of the online supplement). We denote the transformed vector of parameters by $\balph$.

For exact Bayesian estimation, we ran a PMwG algorithm for $14,000$ iterations with the first $500$ iterations discarded as burn-in. Table \ref{tab:PMwG_HDDM_Lexical} shows the posterior means, the posterior standard deviations and the IACT estimates of the group-level parameters. The IACTs of these parameters are small, indicating that the PMwG sampler is efficient. The estimation results are consistent with the ones found in \citet{wagenmakers2008diffusion}, i.e., higher frequency words correspond to higher positive drift rates ($\muv^{(hf)}>\muv^{(lf)}>\muv^{(vlf)})$ and the drift rate is negative for nonwords ($\muv^{(nw)}$). Under the speed condition, the boundary separation is around half of what it in the accuracy emphasis condition ($a^{(speed)}=0.93$ compared with $a^{(accuracy)}=1.78$). In the speed emphasis condition, the amount of variability in the starting point is also proportionally larger relative to boundary separation than in the accuracy condition, so the error responses tend to be faster than correct responses and vice versa across conditions.

\begin{table}
    \centering
    \begin{tabular}{l|r|c || l|r|c || l|c|c}
    & \multicolumn{1}{|c|}{Estimate} & IACT &   & \multicolumn{1}{|c|}{Estimate} & IACT &   &  \multicolumn{1}{|c|}{Estimate}  & IACT \\
     \hline
$\bmualph_{1}$&3.31 (0.34)&2.56&$\muv^{(hf)}$&3.31 (0.34)&2.56&$\bSigalph_{1}$&1.44 (0.79)&3.22\\
$\bmualph_{2}$&2.12 (0.22)&2.42&$\muv^{(lf)}$&2.12 (0.22)&2.42&$\bSigalph_{2}$&0.62 (0.33)&3.06\\
$\bmualph_{3}$&1.33 (0.18)&2.25&$\muv^{(vlf)}$&1.33 (0.18)&2.25&$\bSigalph_{3}$&0.41 (0.21)&2.67\\
$\bmualph_{4}$&-2.17 (0.24)&2.40&$\muv^{(nw)}$&-2.17 (0.24)&2.40&$\bSigalph_{4}$&0.72 (0.38)&3.03\\
$\bmualph_{5}$&-0.48 (0.18)&3.48&$\sv$&0.63 (0.11)&3.27&$\bSigalph_{5}$&0.38 (0.21)&4.73\\
$\bmualph_{6}$&-1.07 (0.10)&2.45&$a^{(speed)}$&0.93 (0.05)&1.83&$\bSigalph_{6}$&0.13 (0.06)&2.69\\
$\bmualph_{7}$&-0.29 (0.08)&2.08&$a^{(acc.)}$&1.78 (0.11)&1.96&$\bSigalph_{7}$&0.09 (0.04)&1.96\\
$\bmualph_{8}$&-1.19 (0.13)&2.40&$\muz^{(speed)}$&0.45 (0.03)&1.84&$\bSigalph_{8}$&0.20 (0.09)&3.01\\
$\bmualph_{9}$&-0.28 (0.09)&1.87&$\muz^{(acc.)}$&0.90 (0.06)&1.85&$\bSigalph_{9}$&0.11 (0.04)&1.65\\
$\bmualph_{10}$&-1.28 (0.13)&4.87&$\sz$&0.28 (0.04)&4.85&$\bSigalph_{10}$&0.15 (0.08)&5.81\\
$\bmualph_{11}$&-1.52 (0.08)&2.06&$\mutau$&0.36 (0.01)&2.23&$\bSigalph_{11}$&0.08 (0.04)&2.24\\
$\bmualph_{12}$&-1.31 (0.12)&2.52&$\stau$&0.27 (0.04)&2.69&$\bSigalph_{12}$&0.19 (0.10)&2.91\\
     
    \end{tabular}
    \caption[DDM - Real data. PMwG estimation results.]{Experiment 1 by \citet{wagenmakers2008diffusion}. Posterior means with posterior standard deviations (in brackets) and the IACT estimates for the group-level parameters. The estimates are calculated based on the PMwG draws. The first three columns refer to the transformed group-level parameters and the next three columns refer to group-level parameters back-transformed to the raw parameter scale.}
    \label{tab:PMwG_HDDM_Lexical}
\end{table}

\begin{figure}
    \centering
    
    \hspace*{-0cm} \includegraphics[scale=0.4]{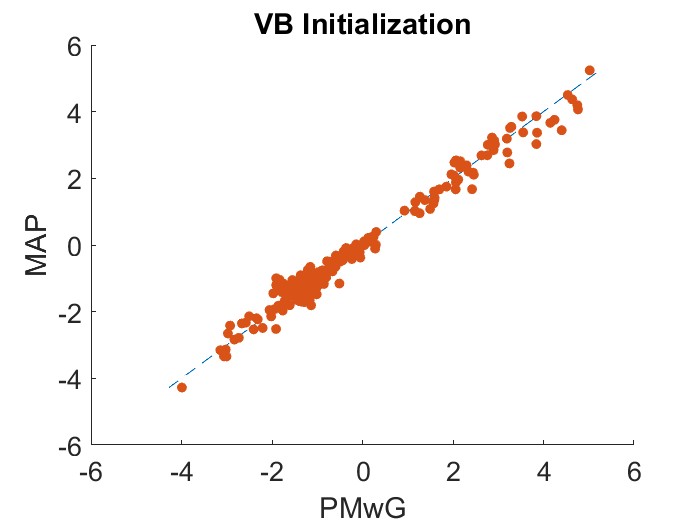}
    \caption[DDM - Real data. Comparing the initial values for VB.]{Experiment 1 by \citet{wagenmakers2008diffusion}. Comparing the initial values for VB obtained by PMwG method (horizontal axis) against the values calculated using MAP (vertical axis).}
    \label{fig:hddm_Lexical_scatterplot_map_vs_pmwg}
\end{figure}

We repeated the same analysis but this time using a VB approximation, for computational efficiency. We compare the MAP and PMwG initialisation methods discussed in Section \ref{sec:vb-initialization}. The two methods provided very similar start points for the VB search, as shown in Figure  \ref{fig:hddm_Lexical_scatterplot_map_vs_pmwg}. The MAP initialization method does not generate initial values for the hyperparameters $\log \ba$, and so we set them to zero. The computing time for the MAP method was only $8.6$ minutes compared to the PMwG initialization method's $327$. In what follows, we report results from VB initialized using the MAP method.

\begin{figure}
    \centering
    \hspace*{-2cm} \includegraphics[scale=0.22]{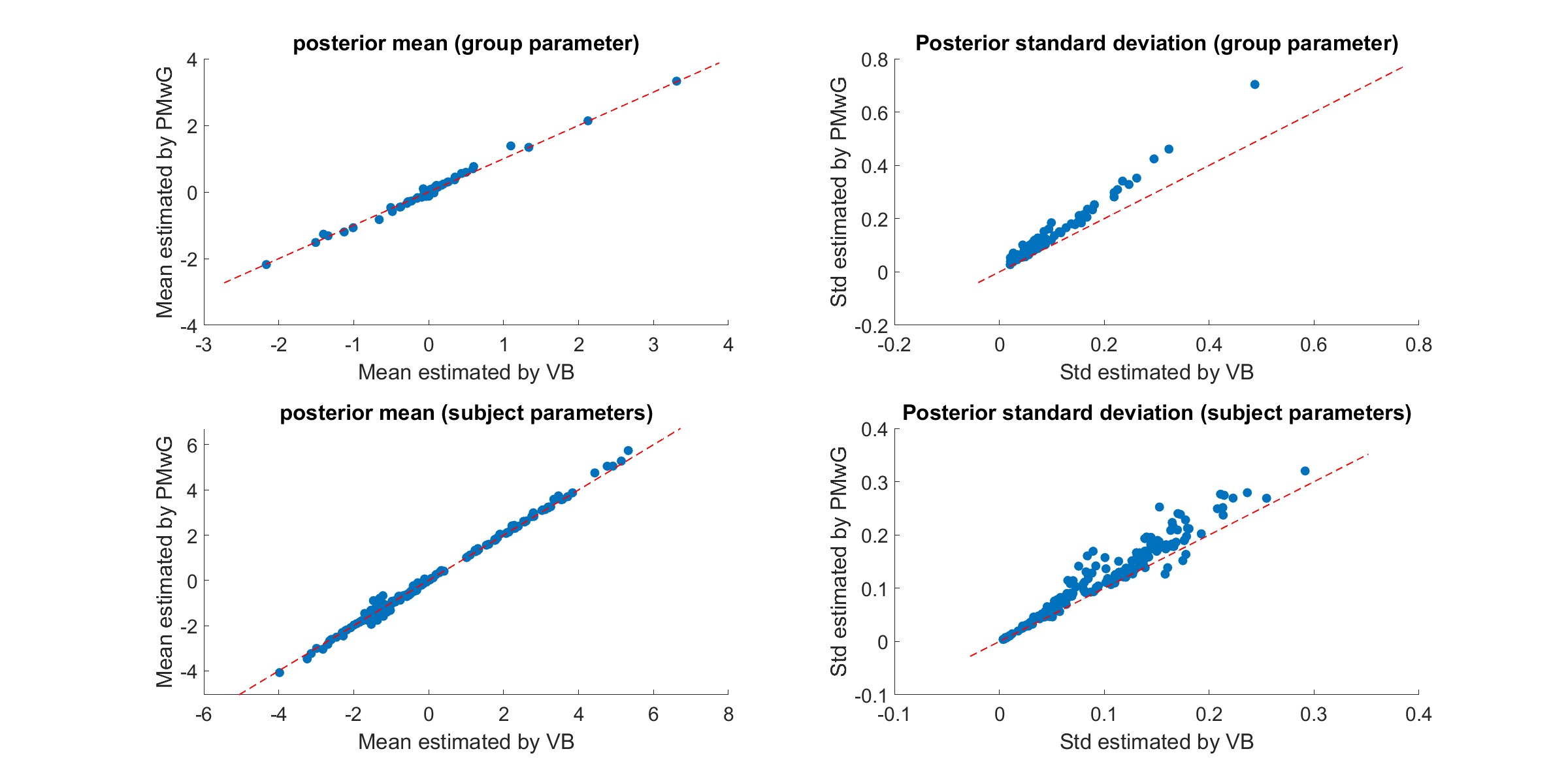}
    \caption[DDM - Real data. Comparing the posterior moments estimated using PMwG against using VB.]{Experiment 1 by \citet{wagenmakers2008diffusion}. Comparing the means and standard deviations of the marginal posterior distributions estimated by VB (horizontal axis) against the exact values calculated using PMwG (vertical axis). The top panels show the means and standard deviations of the group-level parameters. The bottom panels show the means and standard deviations of the subject parameters.}
    \label{fig:hddm_Lexical_vb_pmwg_moment2}
\end{figure}

Estimation results from the VB approximation were comparable to those obtained from the exact MCMC approach using PMwG. Figure~\ref{fig:hddm_Lexical_vb_pmwg_moment2} compares the mean and standard deviation of the posterior distribution between the VB and PMwG, for both group-level and subject-level parameters. Posterior means (left hand panels) are estimated very accurately by VB. As is typical for VB methods, posterior standard deviations (right hand panels) are underestimated, but the difference is not extreme. Estimation using VB took 6 minutes, which was nearly 10x faster than PMwG (52 minutes), even though VB used fewer computational resources (8 cpu-cores for VB; 48 cpu-cores for PMwG).

\subsection{Data Set 2: Mental Rotation with Neural Covariates}\label{sec:mental-rotation}

This section applies the PMwG and VB methods to analyse the mental rotation experiment reported by \citet{provost2013two}. In this experiment, two images of the same object were shown on the screen, side by side. The left image was fixed across trials, and only the right one was changed by flipping and/or rotating. There were five possible rotation angles: $0^{\circ}, 45^{\circ}, 90^{\circ}, 135^{\circ},$ and $180^{\circ}$; which results in ten conditions when crossed with the flipping manipulation. Participants were tasked with deciding whether the right hand image was a flipped (``mirror'') or not flipped (``same'') version of the left hand image. Nine participants each performed approximately 600 trials. Previous analyses of these data by \cite{van2017confirmatory} have linked LBA drift rates on each trial with scalp electrical activity (EEG). We extend previous analyses to use both LBA and DDM models. We also adopt a more plausible and complete specification for random effects, and explore more possibilities for links between the EEG covariates and the random effects. 

\subsubsection{LBA Model}

The maximally flexible LBA model specification for the mental rotation experiment could have up to $20$ random effects for each subject for the drift rates (five rotation angles crossed with the same and mirror stimulus types, each for two different response accumulators). We adopt a more parsimonious approach by forcing the drift rate to change linearly with the  rotation angle, which reduces the model to just 6 drift rate parameters. Let $v_{ij;s}$ and $v_{ij;m}$, respectively, be the drift rates corresponding to the ``same'' and ``mirror'' accumulators in trial $i$ for subject $j$. Let $E_{ij} \in \{ 0,45,90,135,180 \}$ represent the rotation angle of the stimulus for trial $i$ and subject $j$. 

The covariates in this experiment were scalp electrode potentials measured via EEG, and recorded hundreds or thousands of times within the period of each decision. Following \cite{van2017confirmatory}, we represent the covariates using the mean amplitude recorded in each of eight time windows following stimulus onset. We denote these values $x_{ij}$, for the neural covariates for subject $j$ at trial $i$. Finally, let $\bbeta$ be the vector of covariate coefficients (fixed across subjects and trials), which link neural covariates with drift rates. Then, 
\begin{align*}
	v_{ij;s} &= \left\{
	\begin{array}{l l}
		v^{s}_{j;s} + E_{ij}\times v_{j;c} + \bbeta^\top x_{ij},  & \textrm{ if } S_{ij} = \text{``same''},\\  
		v^{m}_{j;s} + E_{ij}\times v_{j;e} + \bbeta^\top x_{ij},  & \textrm{ if } S_{ij} = \textrm{``mirror''},\\
	\end{array}\right.\\
	v_{ij;m} &= \left\{
	\begin{array}{l l}
		v^{s}_{j;m} + E_{ij}\times v_{j;e} + \bbeta^\top x_{ij},  & \textrm{ if } S_{ij} = \text{``same''},\\  
		v^{m}_{j;m} + E_{ij}\times v_{j;c} + \bbeta^\top x_{ij},  & \textrm{ if } S_{ij} = \textrm{``mirror''},\\
	\end{array}\right.\\
	\end{align*}
where $E_{ij} \in \{ 0,45,90,135,180 \}$ represents different experimental conditions (rotation angles). All other random effects are assumed to be the same across accumulators. The non-decision time is allowed to vary across rotation angles via the following equation
\begin{equation*}
    \tau_{ij} = \tau_0 + E_{ij}\tau.
\end{equation*}
Under the specified LBA model, each subject has 10 random effects
\begin{equation}
	\bldeta = (b,A, v^{s}_{s}, v^{m}_{s},  v^{s}_{m}, v^{m}_{m},  v_c, v_{e},\tau_0,\tau).
\end{equation}
To capture the constraints on these parameters, such as strict positivity for $\tau$, we apply the usual log transformations.



\subsubsection{Diffusion Model}
This mental rotation experiment has not previously been modelled using the DDM, and so we specify parameters for the DDM by following as closely as possible the LBA specification. We identify the upper boundary of the diffusion process with the ``same'' response, and the lower boundary with the ``mirror'' response. Once again, we constrain drift rates to follow linear functions of the stimulus rotation angle, $E_{ij}$, and neural covariates, $x_{ij}$:

\begin{equation}\label{eq:erp-hddm-linking-equation}
	v_{ij} = \left\{
	\begin{array}{l l}
		v_j^0 + kv_j + \bbeta^\top x_{ij}, & \textrm{ if } S_{ij} = \text{``same'' and }E_{ij} = k,\\  
		-v_j^0 - kv_j + \bbeta^\top x_{ij}, & \textrm{ if } S_{ij} = \textrm{``mirror'' and }E_{ij} = k,\\
	\end{array}\right.
\end{equation}
where $k \in \{ 0,45,90,135,180 \}$ representing various rotation angles. The non-decision time is allowed to vary across rotation angles via the equation
\begin{equation*}
    \tau_{ij} = \tau_0 + k\tau,\textrm{ if } E_{ij} = k, \textrm{ where } k \in \{ 0,45,90,135,180 \}.
\end{equation*}
The nine random effects for each subject are
\begin{equation}
	\bldeta = (v^{0},v,\sv,a,z,\sz,\tau_0,\tau,\stau).
\end{equation}
To enforce the constraints on these parameters, we apply corresponding transformations as for the lexical decision experiment (see Section~\ref{supp:sec-regddm-real-data-transform} of the online supplement). The neural covariates, $x_{ij}$, and the fixed coefficients which link them to the drift rates, $\bbeta$, are specified in the same way as for the LBA model above.

\subsubsection{Prior Distributions}
For both LBA and DDM, we used the multivariate Gaussian priors described in section \ref{subsec:reg-eam-bayes-methods} for the group-level mean $\bmualph$ and the fixed coefficients $\bbeta$. For the group-level covariance matrix $\bSigalph$, instead of using the marginally noninformative prior described in Section \ref{subsec:reg-eam-bayes-methods}, we adopted an inverse Wishart prior with 20 degrees of freedom and an identity scale matrix. This prior is more informative, which is appropriate because the very small sample in this experiment (only 9 participants) is too few to stably estimate subject-to-subject correlations.

\subsubsection{Estimation Results} 
\begin{figure}
    \centering
    
    \hspace*{-0.5cm} \includegraphics[scale=0.4]{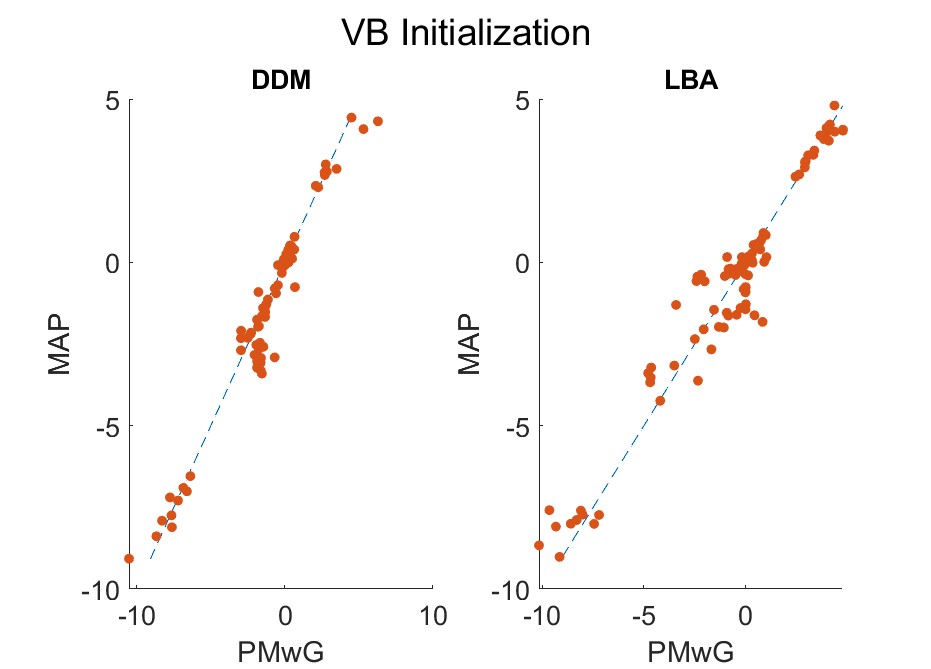}
    \caption[DDM - Real data. Comparing the initial values for VB.]{Mental rotation experiment by \cite{provost2013two}. Comparing the initial values for VB search obtained by PMwG method (horizontal axis) against the values calculated using MAP (vertical axis) for DDM (left panel) and LBA (right panel).}
    \label{fig:hddm_mentalrotation_scatterplot_map_vs_pmwg}
\end{figure}

We estimated both the LBA and DDM using PMwG and VB methods. For PMwG, we ran $25,000$ iterations with the first $500$ iterations discarded as burn-in. For VB, initialization using both MAP and PMwG (with 200 iterations) resulted in very similar starting values, as shown in Figure~\ref{fig:hddm_mentalrotation_scatterplot_map_vs_pmwg}. As before, MAP initialization was faster than PMwG: 0.1 seconds vs. 2 minutes for the LBA model, and 5.4 minutes vs. 98 minutes for the DDM. We conclude that the MAP initialization method provides initial values that are very close to those of PMwG while being much faster. This suggests a scalability advantage for MAP initilization, which we explore with a large data set in Section \ref{sec:hcp-data}. For the rest of this section, we report VB results obtained using MAP initialization.

\begin{figure}
	\centering
	\hspace*{-2cm}\includegraphics[scale=0.2]{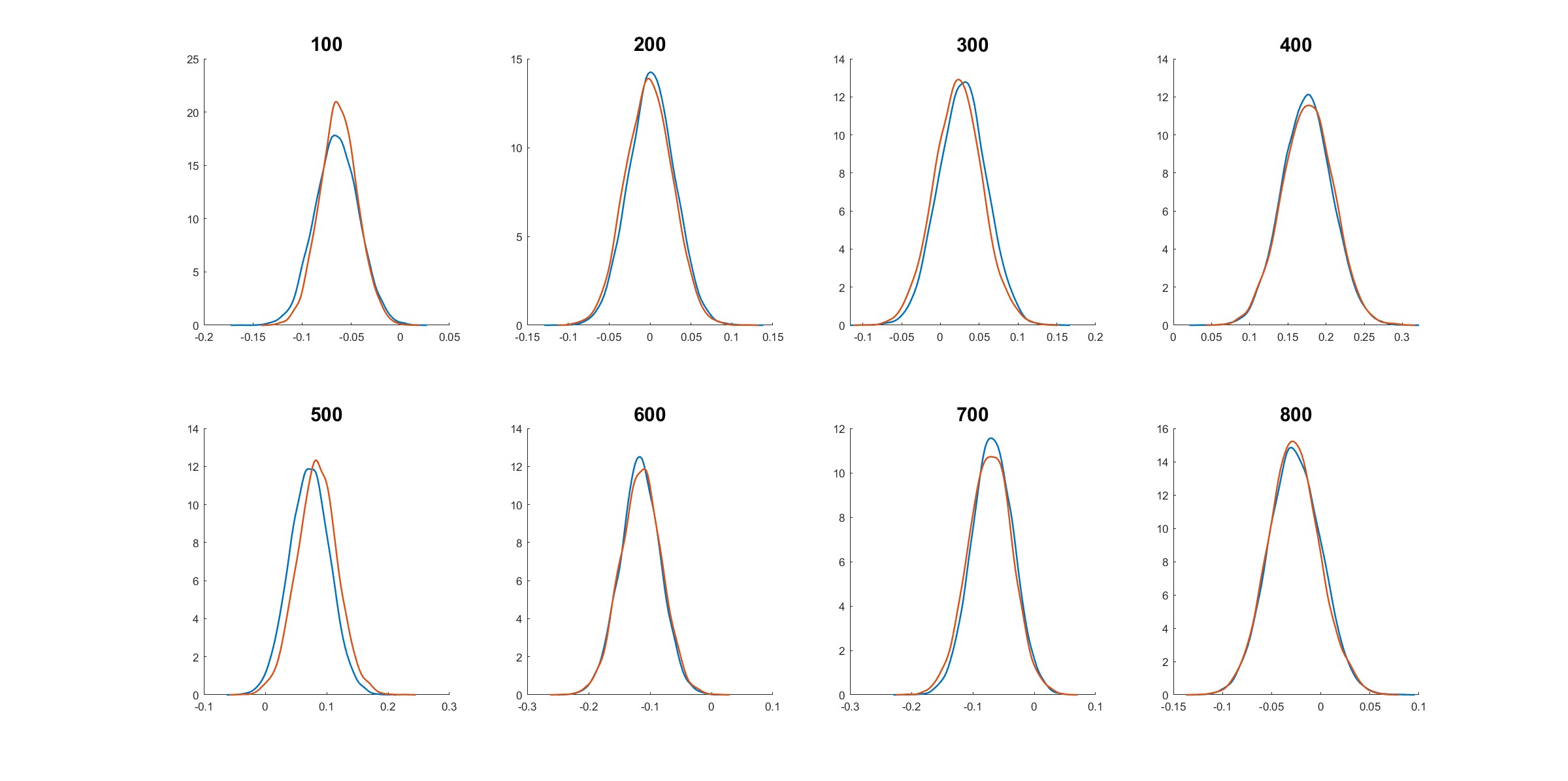}
	\caption[Mental Rotation. Comparing the posterior densities estimated using PMwG against using VB.]{
    Mental rotation experiment by \cite{provost2013two}. Kernel density estimates of marginal posterior densities of the coefficients of ERP epochs ($100,200,\dots,800$) from the LBA model, estimated using PMwG (blue) and VB (red) methods.}
    \label{fig:reglba_mental_rotation_vb_pmwg_density_beta}
\end{figure}

\begin{figure}
	\centering
	\hspace*{-2cm}\includegraphics[scale=0.2]{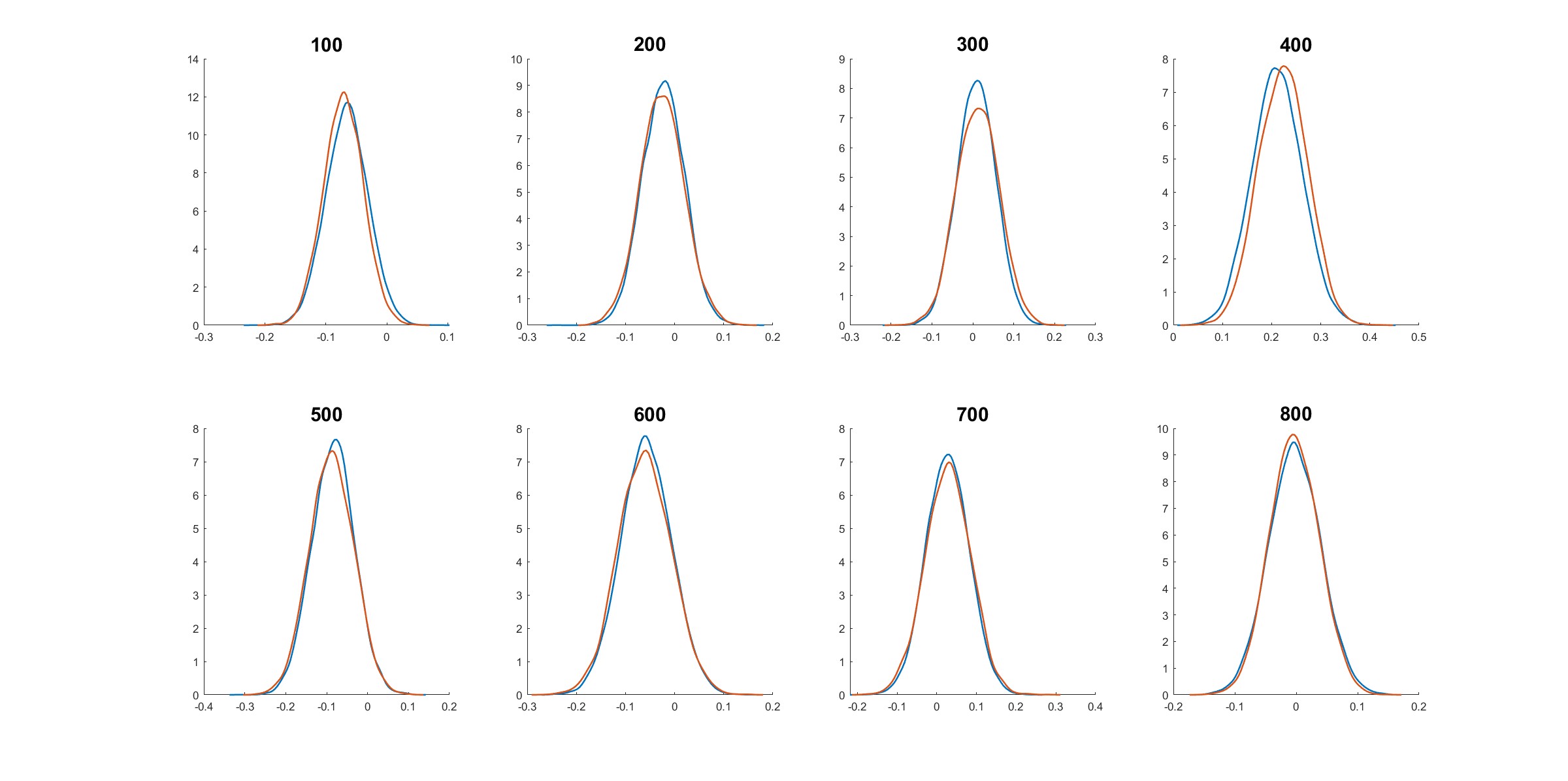}
	\caption[DDM - Mental Rotation. Comparing the posterior densities estimated using PMwG against using VB.]{
    Mental rotation experiment by \cite{provost2013two}. Kernel density estimates of marginal posterior densities of the coefficients of ERP epochs ($100,200,\dots,800$) from the DDM model, estimated using PMwG (blue) and VB (red) methods.}
    \label{fig:reghddm_mental_rotation_vb_pmwg_density_beta}
\end{figure}

The key interest in analyzing this experiment is in the coefficients $\bbeta$, which describe how scalp EEG measurements are associated with changes in decision information (drift rates). Figures~ \ref{fig:reglba_mental_rotation_vb_pmwg_density_beta} and \ref{fig:reghddm_mental_rotation_vb_pmwg_density_beta} compare posterior distributions for the eight components of $\bbeta$ (one for each time period in the EEG measurement window) between estimates using the asymptotically exact PMwG algorithm and the VB approximation. VB captures these fixed effects (at least marginally) very precisely for both models. For other parameters, VB can capture also the posterior means relatively well; see Section~\ref{supp:sec-further-results} for further details. In addition, VB estimation was around 10x faster than PMwG for the LBA model (7 seconds compared with 75 seconds) and around 5x faster for the DDM (6 minutes compared with 31 minutes).

\begin{figure}
		\centering
		
	\hspace*{-0.5cm}\includegraphics[scale=0.2]{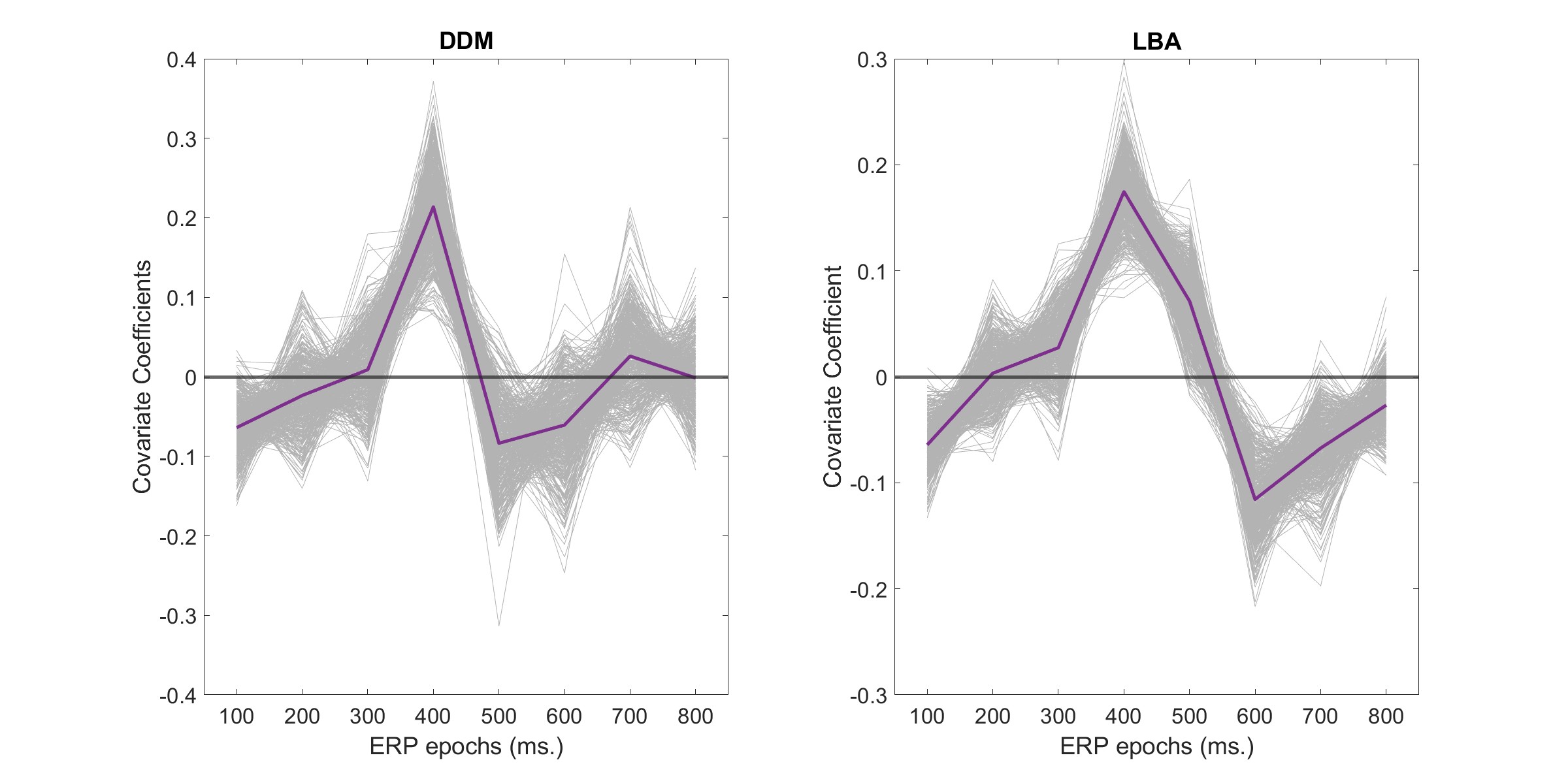}	\caption{Mental rotation experiment by \cite{provost2013two}. Posterior samples, obtained by PMwG method, of the coefficients $\bbeta$ from the DDM (left panel) and LBA (right panel). In both panels, each posterior sample is represented by a grey line and the purple line segments show the averages.}
		\label{fig:line-plot-beta}
	\end{figure}

Figure~\ref{fig:line-plot-beta}	represents the posterior distribution (estimated using PMwG) of the coefficients $\bbeta$  in a manner similar to that used by \cite{van2017confirmatory}, which emphasizes change over time rather than comparison between VB and PMwG estimation methods. The thick lines show how the coefficient which links EEG measurements to drift rate changes with time during the course of each decision trial. Note that the conclusions to be drawn from this analysis are quite consistent between LBA and DDM versions. For example, both show a positive link around the fourth time period (400-500 msec after stimulus onset), and a smaller negative link soon after.

\subsubsection{Goodness of Fit} 

To assess the goodness of fit of the estimated models, we follow \citet{van2017confirmatory} and consider two summary statistics: the median response time and the mean proportion of correct responses. These summary statistics are calculated per condition and averaged over participants. The estimated models are used to simulate posterior predictive data which are summarised in the same way. Section \ref{sec:simulate-pred-data} of the online supplement details the algorithm to generate posterior predictive data.

\begin{figure}
		\centering
		
          \hspace*{-1cm}\includegraphics[scale=0.2]{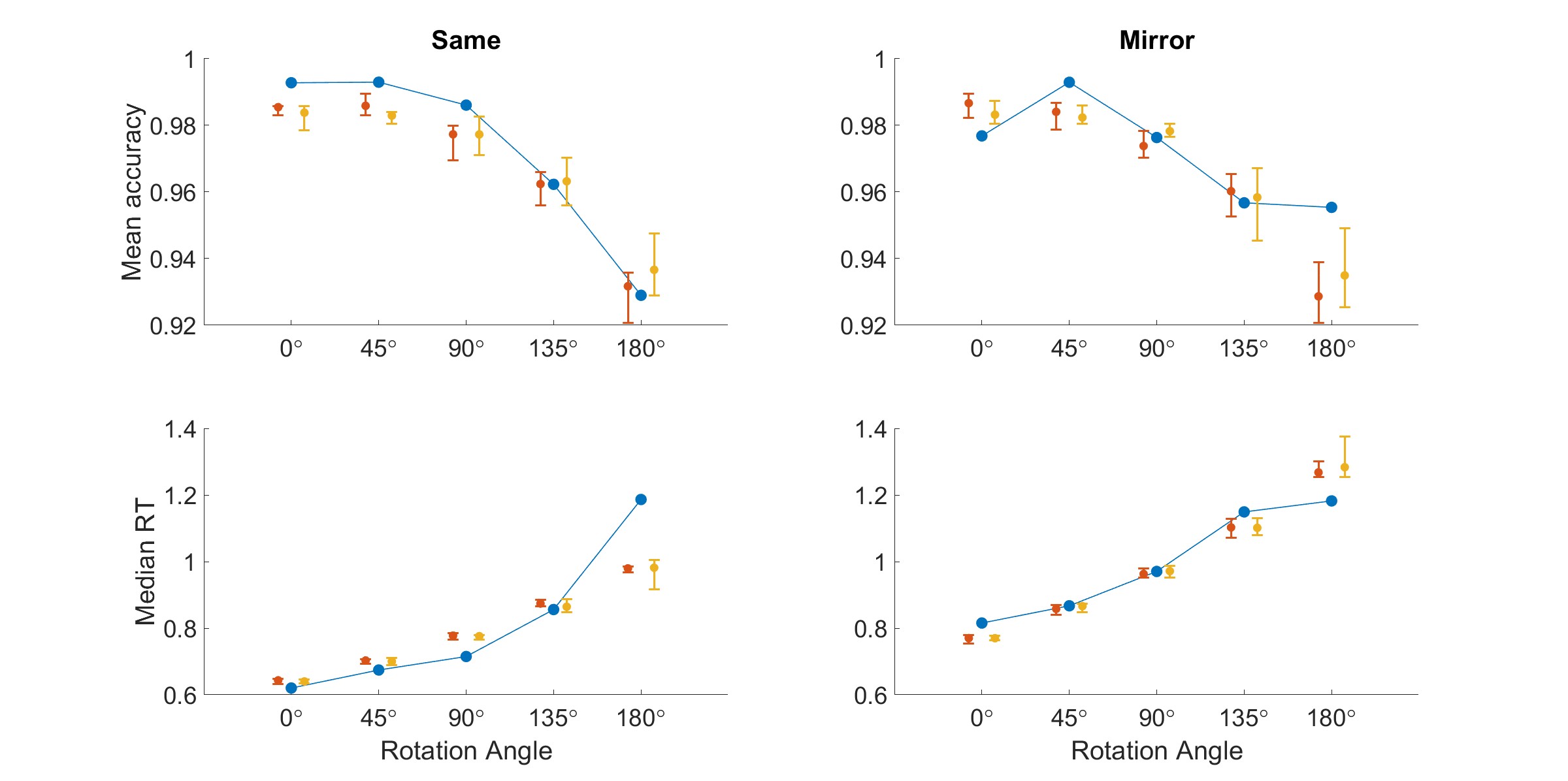}
		\caption{Mental rotation experiment by \cite{provost2013two}. Comparing the posterior predictive summary statistics generated via PMWG and VB, against observed data for the LBA model. In all the panels, the vertical bars represent the posterior predictive distribution obtained by PMwG (red) and VB (yellow). In each vertical bar, the upper and lower tails together with the dot in between represent the 25th, 75th, and 50th  percentiles of the posterior predictive distribution, respectively. The top panels and the bottom panels show the posterior predictive mean accuracy and median RT, respectively, under 5 rotation angle conditions (from left to right): $0^{\circ},45^{\circ},90^{\circ},135^{\circ},$ and $180^{\circ}$.  The blue dots are the corresponding statistics evaluated from the observed data.}
		\label{fig:hlba-boxplot-mcmc-vs-vb}
	\end{figure}

\begin{figure}
		\centering
		
          \hspace*{-1cm}\includegraphics[scale=0.2]{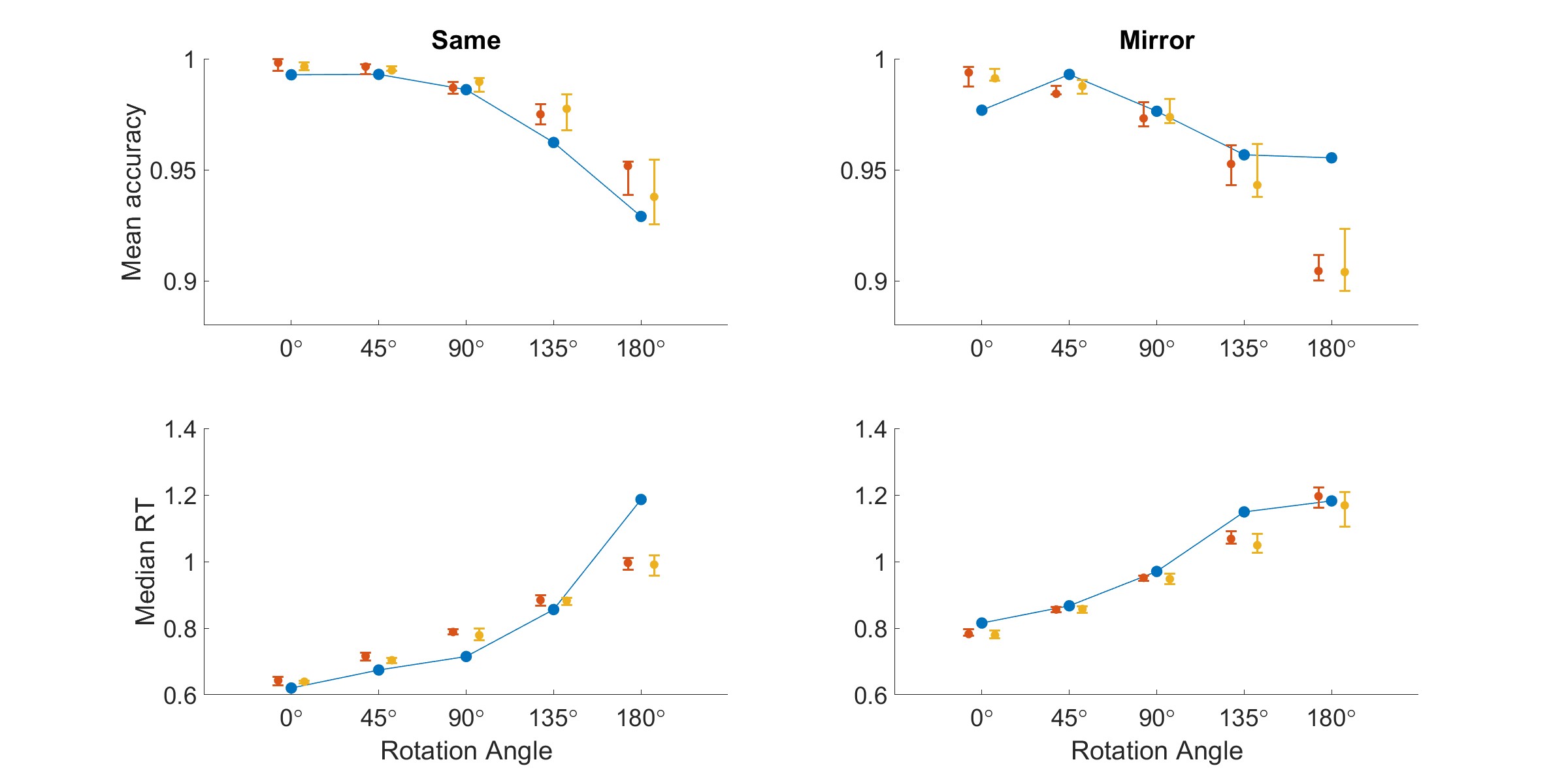}
		\caption{Mental rotation experiment by \cite{provost2013two}. Comparing the posterior predictive summary statistics from DDM, generated using PMWG and VB methods, against observed data. In all the panels, the vertical bars represent the posterior predictive distribution obtained by PMwG (red) and VB (yellow). In each vertical bar, the upper and lower tails together with the dot in between represent the 25th, 75th, and 50th  percentiles of the posterior predictive distribution, respectively. The top panels and the bottom panels show the posterior predictive mean accuracy and median RT, respectively, under 5 rotation angle conditions (from left to right): $0^{\circ},45^{\circ},90^{\circ},135^{\circ},$ and $180^{\circ}$. The square dots are the corresponding statistics evaluated from the observed data.}
		\label{fig:hddm-boxplot-mcmc-vs-vb}
	\end{figure}

Figure~\ref{fig:hlba-boxplot-mcmc-vs-vb} compares the posterior predictive data simulated from the LBA model estimated using PMwG (red bars) and VB (yellow bars) with the observed data (blue dots). The red and yellow bars are similar, indicating that the prediction obtained from VB is as accurate as PMwG. The predictions obtained from VB and PMwG are both close to the observed data, with exceptions in conditions which match those observed by \cite{van2017confirmatory}. For example, the model underestimates response times in the largest rotation condition for ``same'' stimuli, but overestimates response times in the same condition for ``mirror'' stimuli. Figure~\ref{fig:hddm-boxplot-mcmc-vs-vb} makes the corresponding comparison, but for the DDM. Once again, the VB and PMwG estimates are very similar, and both are close to the observed data.

\subsection{Data Set 3: A Large-N Memory Task with Person-Level Covariates}\label{sec:hcp-data}



Key strengths of the VB approaches we have developed are efficiency and the corresponding potential to practically address larger problems. To evaluate this potential, we consider data from the Human Connectome Project \citep[HCP:][]{van-essen2013wu-minn}, with more than 1,000 participants, and many interesting covariates measured at the person-by-person level, such as personality traits and neuropsychological measurements. In a data set this large, exact Bayesian methods based on MCMC can become impractical. For example, even using the efficient particle-based approach of PMwG and using a 48-cpu-core workstation, estimating the LBA model on the HCP data took more than two days. Estimating the DDM using corresponding MCMC methods was simply not practical; extrapolating from earlier results, we estimate running time would have been around one month. In contrast, VB methods produced good estimates from this data set in around two hours for the LBA model and two days for the DDM.

The Human Connectome Project had participants perform many experimental tasks; we investigate data from the 0-back and 2-back memory tasks. For each of around 150 trials, participants were shown a picture and had to decide whether the picture was the same as the picture shown 2 trials ago (in the 2-back condition) or the same as a fixed target picture (in the 0-back condition). There were three kinds of trial, defined by the picture matches: ``target'' trials, in which the picture really did match; ``non-target'' trials, in which it did not match; and ``lure'' trials, in which the picture did not match the target, but did match some other closely related pictures.

Participants in the HCP also provided data on hundreds of person-level covariates, including MRI structural scans, detailed demographic data, and psychological survey results. To demonstrate how our estimation methods can extend to person-level covariates, we extracted eight measurements for each person in the sample:
 \begin{itemize}
     \item Mini-Mental-State-Exam (``MMSE''). A survey administered in clinics and hospitals to quickly screen people for signs of cognitive impairment \citep{creavin2016mini}.
     \item The Pittsburgh sleep quality index (``PSQI''), which measures sleep quality and quantity \citep{smyth1999pittsburgh}.
     \item NIH Toolbox Picture Sequence Memory Test (``Sequencing''), which measures aspects of comprehension and memory \citep{baron1986mechanical}.
     \item The Wisconsin Card Sort (``Card Sort''), used to measure reasoning and executive function \citep{heaton1993wisconsin}.
     \item NIH Toolbox Picture Vocabulary test (``Vocab''), which measures comprehension with reduced influence from English language ability \citep{dunn1965peabody}.
     \item NIH Toolbox Pattern Completion Processing Speed task (``Speed''). A simple reaction time task to measure how quickly people make simple decisions \citep{carlozzi2015nih}.
     \item Fluid Intelligence (``Fluid''), measured using the Penn Progressive Matrices \citep{moore2015psychometric}.
     \item Crystalized Intelligence (``Crystal'') composite score from the NIH Cognitive Toolbox \citep{heaton2014reliability}.
 \end{itemize}

We used age-adjusted scores for all measures, except for the MMSE and PSQI (HCP data did not include age adjustments for those). This set of eight covariates was chosen from the very large set measured in the HCP because they satisfy the purpose of illustrating the estimation method with regression on person-level measurements. Our estimation methods support the use of much larger sets of covariates, where necessary. The particular covariates used above have plausible psychological links with the components of the decision models. For example, processing speed has been shown to be predictive of general intelligence \citep{salthouse1996processing-speed} and has been linked with drift rates in EAMs \citep{dutilh2017test,RatcliffEtAl2008WPR}. We investigated a quite general set of possible links between covariates and model parameters. For both DDM and LBA, we included regression components linking each of the eight covariates with eight different model parameters: one for response caution and three for drift rates, in both the 0-back and 2-back conditions.



\subsubsection{LBA Model}

We specified an LBA model for the HCP data following previous work \citep{evans2018modeling}. Let $v^{target}_{ij}$ and $v^{nontarget}_{ij}$ be the drift rates corresponding to the ``target'' and ``nontarget'' accumulators in trial $i$ of subject $j$, respectively. For each parameter, we use superscripts $(0)$ and $(2)$ to indicate the 0-back and 2-back conditions, respectively. In the 0-back condition, the drift rates $v^{target}_{ij}$ and $v^{nontarget}_{ij}$ vary according to the  stimuli and they are modelled as follows. For stimuli that are ``lures'', we have:
\begin{align}
	v^{target}_{ij} &= v^{(0)}_{j} - v^{(0)}_{j;lure} + X^\top_{j}\bbeta^{(0)}_{target,lure} , \label{eq:hcp-regddm-link-eq1}\\
	v^{nontarget}_{ij} &= v^{(0)}_{j} + v^{(0)}_{j;lure} + X^\top_{j}\bbeta^{(0)}_{nontarget,lure}. \label{eq:hcp-regddm-link-eq2}
\end{align}
Similarly, the drift rates for the ``target'' are:  
\begin{align}
	v^{target}_{ij} &= v^{(0)}_{j} - v^{(0)}_{j;target} + X^\top_{j}\bbeta^{(0)}_{target,target} , \label{eq:hcp-regddm-link-eq3}\\
	v^{nontarget}_{ij} &= v^{(0)}_{j} + v^{(0)}_{j;target} + X^\top_{j}\bbeta^{(0)}_{nontarget,target}, \label{eq:hcp-regddm-link-eq4}
\end{align}
and for ``nontarget'' stimuli are:
\begin{align}
	v^{target}_{ij} &= v^{(0)}_{j} - v^{(0)}_{j;nontarget} + X^\top_{j}\bbeta^{(0)}_{target,nontarget} , \label{eq:hcp-regddm-link-eq5}\\
	v^{nontarget}_{ij} &= v^{(0)}_{j} + v^{(0)}_{j;lure} + X^\top_{j}\bbeta^{(0)}_{nontarget,nontarget}. \label{eq:hcp-regddm-link-eq6}
\end{align}
The drift rates for the 2-back condition are modelled similarly. Following previous work \citep[e.g., in ageing][]{thapar2003diffusion} we explore links between the covariates and both the drift rate parameters (as above) and the response threshold parameter. Let $c=b-A$, the distance between the upper limit of the starting points for evidence accumulation and the response threshold. This parameterization helps enforce the constraint that the threshold is above the starting point ($b>A$). For the 0-back block, we assume
\begin{equation}\label{eq:hcp-regddm-link-eq3}
    \log (c_{ij}) = \log\left( c^{(0)}_j \right) + X^\top_{j}\bbeta^{(0)}_{c},
\end{equation}
The model for $c$ for the 2-back block is defined similarly. Altogether, each subject has $14$ random effects
\begin{equation}
	\bldeta = \left(c^{(0)},A^{(0)}, v^{(0)}, v^{(0)}_{lure},  v^{(0)}_{target},  v^{(0)}_{nontarget},  \tau^{(0)},  c^{(2)},A^{(2)}, v^{(2)}, v^{(2)}_{lure},  v^{(2)}_{target},  v^{(2)}_{nontarget},  \tau^{(2)}\right).
\end{equation}
To capture the constraints on these parameters, we apply the usual log transformations for the random effects.
The coefficient matrix $\bbeta$ has 8 columns corresponding to 8 covariates and 14 rows (12 rows to model the drift rates and 2 rows corresponding to the response threshold parameter).

\subsubsection{Diffusion Model}
We specified a DDM for these data following closely the LBA specification above. We allowed all random effects to vary across 0-back and 2-back conditions. We also assume that only the drift rate parameter changes according to three experimental conditions: ``lure'', ``target'', and ``nontarget''. This leads to a model with $18$ random effects for each subject: 
\begin{align*}
    \bldeta = &\left(v^{(0)}_{lure},  v^{(0)}_{target},  v^{(0)}_{nontarget}, \sv^{(0)},a^{(0)},z^{(0)},\sz^{(0)}, \tau^{(0)}, \stau^{(0)},\right. \\
    & \left. v^{(2)}_{lure},  v^{(2)}_{target},  v^{(2)}_{nontarget},  \sv^{(2)}, a^{(2)}, z^{(2)}, \sz^{(2)}, \tau^{(2)}, \stau^{(2)}\right).
\end{align*}

We once again linked the drift rates and the response caution (boundary separation) parameters with the covariates. Because we identified the upper response boundary with the ``target'' response, drift rates for stimuli in non-target and lure conditions are negative, and the following regression equations respect this. If we let $v_{ij}$ be the drift rate on trial $i$ for subject $j$, then for the 0-back block:
\begin{equation}\label{eq:hcp-hddm-linking-equation}
	\begin{array}{l l}
	\log\left(-v_{ij}\right) = \log \left(-v^{(0)}_{lure} \right) + X^\top_{j}\bbeta^{v^{(0)}}_{lure}, & \textrm{ if condition = ``lure''},\\  
	\log\left(v_{ij}\right) = \log \left(v^{(0)}_{target} \right) + X^\top_{j}\bbeta^{v^{(0)}}_{target}, & \textrm{ if condition = ``target''},\\  
	\log\left(-v_{ij}\right) = \log \left(-v^{(0)}_{nontarget} \right) + X^\top_{j}\bbeta^{v^{(0)}}_{nontarget}, & \textrm{ if condition = ``nontarget''}.\\  
	\end{array}
\end{equation}
The drift rates of 2-back blocks are modelled similarly. For the boundary separation parameter $a$, we use a regression equation which again respects the DDM's parameter constraints (e.g. that $a > z+0.5s_z$):

\begin{equation}\label{eq:hcp-hddm-linking-equation}
	\begin{array}{l l}
	\log\left(a-z-0.5\sz\right) = \log \left(a^{(0)} - z^{(0)} -0.5\sz^{(0)} \right) + X^\top_{j}\bbeta^{(0)}_{a}, & \textrm{ if block = ``0-back''},\\  
    \log\left(a-z-0.5\sz\right) = \log \left(a^{(2)} - z^{(2)} -0.5\sz^{(2)} \right) + X^\top_{j}\bbeta^{(2)}_{a}, & \textrm{ if block = ``2-back''}.\\  
	\end{array}
\end{equation}
To capture the constraints on these parameters, we apply the usual transformations, specified in detail in section~\ref{supp:sec-regddm-real-data-transform} of the online supplement. 
The coefficient matrix $\bbeta$ has 8 columns corresponding to 8 covariates and 8 rows (6 rows to model the drift rates and 2 rows corresponding to the boundary separation parameter). 

\subsubsection{Prior distributions}
The multivariate Gaussian priors described in Section \ref{subsec:reg-eam-bayes-methods} are used for the group-level mean $\bmualph$ and the fixed coefficients $\bbeta$. For the group-level covariance matrix $\bSigalph$, instead of using the marginally noninformative prior described in section \ref{subsec:reg-eam-bayes-methods}, we adopt an inverse Wishart prior with degrees of freedom equal to the number of random effects in the model, and an identity scale matrix. 


\subsubsection{Results} 


Because of the large size of the HCP data set, we carried out extensive checks of our estimation methods, which are reported in the supplementary material (Section~\ref{supp:sec-further-results}). In these checks, we first estimated the LBA model without including covariates, just in the now-standard hierarchical framework, using both exact inference via PMwG and approximate inference via VBL. This check confirmed that the posterior predictive data generated from the model under the two estimation methods were very similar, and that the posterior distributions themselves agreed closely in their means (the variances were underestimated by VB, as is typical). We could not perform the same check for the DDM, because estimation of the DDM using MCMC is impractical in large data sets. Instead, we used VBL to estimated a corresponding DDM, with standard hierarchical structure but no covariates. We then confirmed that the posterior predictive data generated from that model were consistent with those generated by the LBA.

\begin{figure}
    \hspace*{-1cm} \includegraphics[scale=0.20]{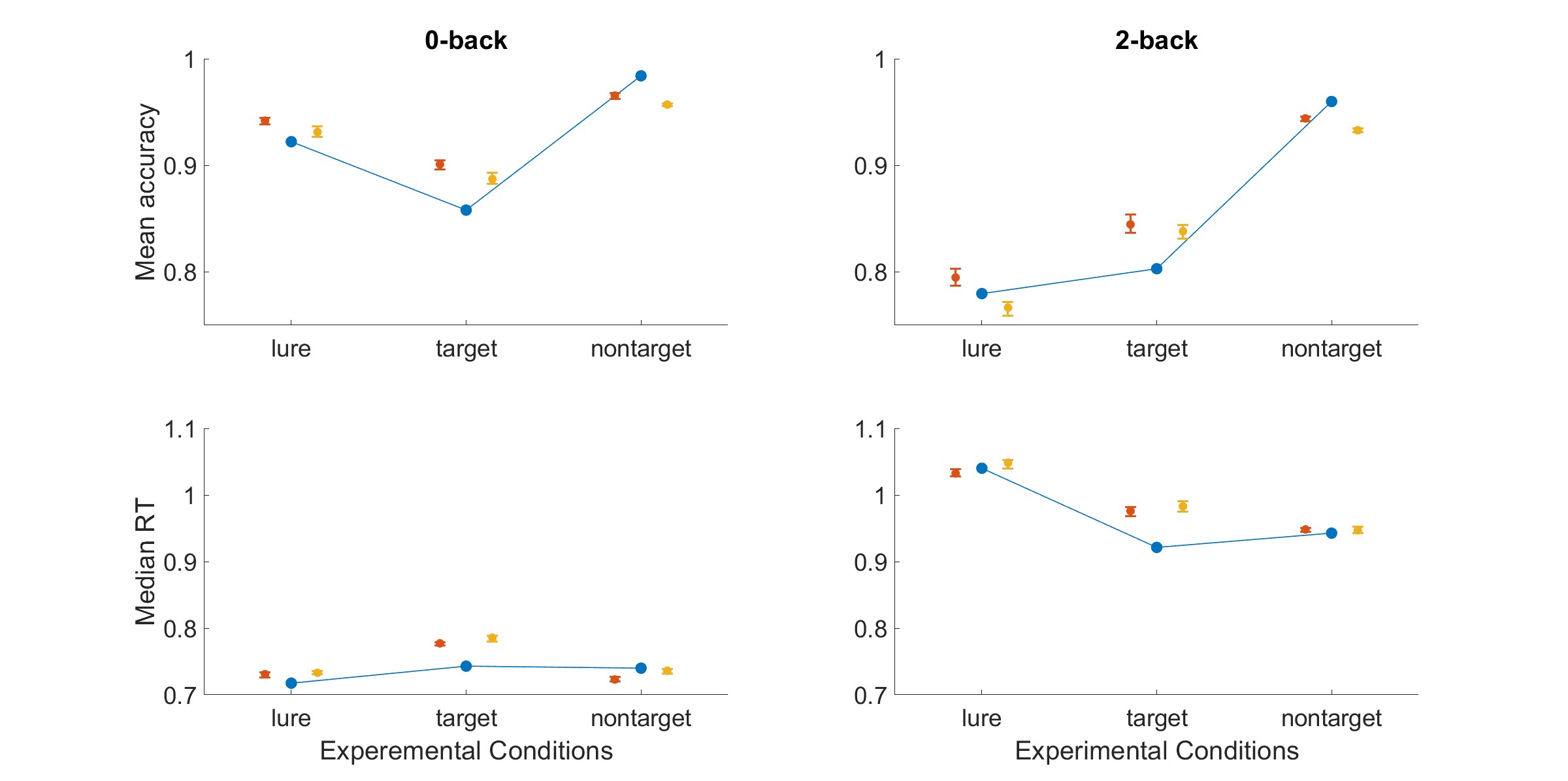}	
\caption{HCP 0-back and 2-back data. Comparing the posterior predictive summary statistics, simulated from LBA (represented by yellow vertical bars) and DDM (represented by red vertical bars) using the VBL method, against the observed data (represented by blue dots). Within each vertical bar, the upper and lower tails, along with the dot in between, represent the $2.5\%$, $97.5\%$, and $50\%$ quantiles of the predictive distribution, respectively. The top panels show the posterior predictive mean accuracy and the bottom panels show median RT. Left panels correspond to the 0-back conditions and the right panels correspond to the 2-back conditions.}
	\label{fig:hcp-data-lba-vs-hddm-with-covariates-boxplot-sumstats}	
\end{figure}

Given the reasonable results of VBL methods when estimating the models without covariates, we then moved on to estimate both models using VBL also including the covariates. Figure~\ref{fig:hcp-data-lba-vs-hddm-with-covariates-boxplot-sumstats} compares the posterior predictive data obtained from both models with the observed data. The two models produced very similar predictions, which is typical \citep{donkin2011diffusion}. Both models accounted well for the observed accuracy of decisions, but had a tendency to overestimate response times for ``target'' stimuli, by about 50 msec. on average.

\begin{figure}
	
    \hspace*{-1.2cm}\includegraphics[scale=0.2]{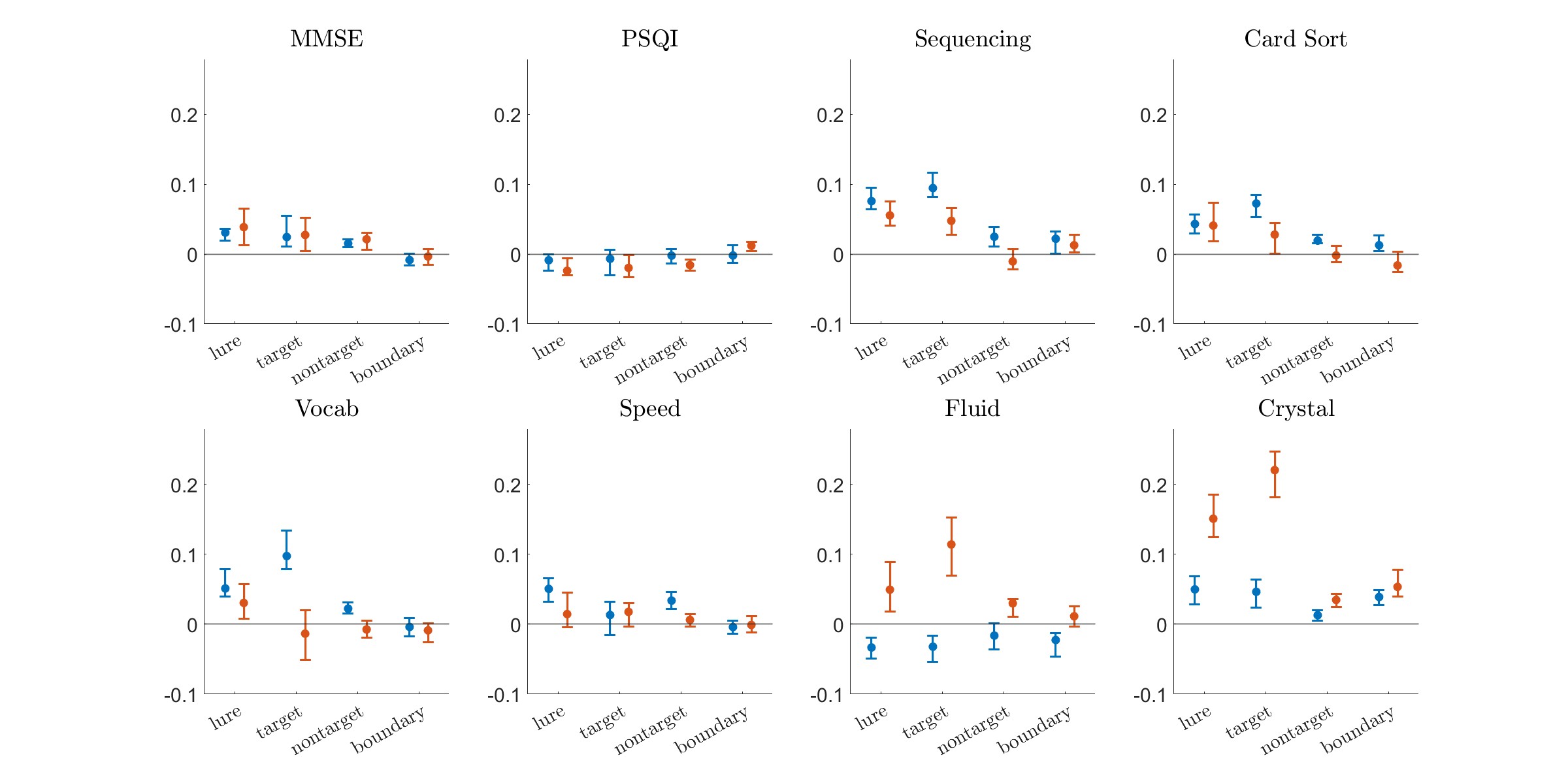}	
	\caption{HCP 0-back and 2-back data. Posterior distribution of the coefficients $\bbeta$ estimated using the VBL method for the DDM, with vertical bars representing the 0-back (blue) and 2-back (red) conditions. Within each vertical bar, the upper and lower tails, along with the dot in between, represent the $2.5\%$, $97.5\%$, and $50\%$ quantiles of the posterior distribution, respectively. Each panel corresponds to a covariate. For example, the top left panel shows the posterior distribution of all the coefficients of MMSE. The second top left panel displays the posterior distribution of all coefficients corresponding to PSQI, and so on and for other panels.}
	\label{fig:hcp-data-hddm-boxplot-beta-covariates}	
\end{figure}

Our approach to simultaneously estimating coefficients for many covariates and model parameters provides an integrated perspective on known and expected relationships in the literature, and has strong potential to highlight new relationships. The estimated coefficients are summarised in Figure~\ref{fig:hcp-data-hddm-boxplot-beta-covariates}, which displays the (marginal) posterior distributions of the coefficients of the covariates for the DDM. The LBA results were largely similar and for brevity we report them in Figure~\ref{fig:hcp-data-reglba-boxplot-beta-covariates} of the supplement. 

We first highlight some `sanity checks' -- analyses that lend confidence the estimated coefficients are consistent with reasonable intuition. For instance, there are positive coefficients for the MMSE covariate and the three drift rate parameters (target, nontarget, lure), indicating participants with lower cognitive impairment (higher MMSE score) tend to have better processing efficiency in the 0-back and 2-back tasks. For the same covariate, there is no relationship with boundary separation -- which is reasonable, given there is no a priori reason why degree of cognitive impairment may alter boundary separation, at least within the range of MMSE scores for the mostly-healthy HCP sample. Similarly, there are negative coefficients for the PSQI covariate and the three drift rate parameters, on average, indicating participants with greater sleep-related impairment (higher PSQI) tend to have poorer processing efficiency. Further, this relationship was magnified in the more challenging 2-back task compared to the 0-back task. While these coefficients are in the expected direction, the credible intervals include 0, for the most part. 

Our approach also identifies known relationships in the literature. For instance, the estimated coefficients identified the positive relationship between crystallised and fluid intelligence and drift rate \citep[e.g.,][]{lerche2020diffusion,van2011integrated}, and the magnitude of the relationship was stronger in the more challenging 2-back task. The coefficients also identify plausible relationships. For example, the 'speed' covariate, which is taken a simple reaction time task, was more strongly related to processing efficiency in the simpler 0-back task than the more cognitively demanding 2-back task. Finally, the coefficients provide insight into expected though not previously reported relationships, at least to our knowledge, between tasks in standard cognitive battery toolkits and parameters of a cognitive model. For instance, positive coefficients for the sequencing (comprehension and memory) and card sorting (reasoning and executive function) covariates and the set of drift rate parameters, indicating participants who perform better in those subtasks of the toolkit tend to process the 0-back and 2-back tasks more efficiently. Taken together, our approach is a powerful way to summarise relationships between components of processing in cognitive models, such as EAMs, and a diverse array of covariates, all within a psychologically plausible and single-step hierarchical model.

\section{Conclusions\label{conclusions}}

Evidence accumulation models have become an important tool in cognitive psychological research and in end-user-focussed applications. An important factor underpinning this success is the development of powerful methods for estimating the models from data. In the past decade or so, these methods have included included Bayesian estimation frameworks that have become important tools in allowing researchers to address the questions they are interested in, using the data they have available. In particular, hierarchical model frameworks with random effects for participants, and the inclusion -- by regression -- of information from covariates measured separately from the central decision process have been key drivers of impact 
\citep{vandekerckhove2011hierarchical,wiecki2013hddm,galdo2019variational,dao2022efficient,gunawan2020new}.

These more sophisticated estimation procedures have exposed new limitations which prevent researchers from applying the models even more widely. The limitations manifest either as impractical computation times (e.g., the models can easily require hours or days to estimate, even with substantial computing resources) or unreliable estimation outcomes (e.g., Markov chains which fail to converge, or mix well). The limitations are caused by many factors, but two primary contributors include the computational cost associated with the diffusion model's density functions and the ever-increasing size of the models, as more parameters are added to extend the models' accounts to include new covariates and conditions. These limitations force modellers to make simplifying assumptions for practical reasons, which can cause tension with scientific goals. For example, the computational cost of using the DDM is often reduced by simplifying the model's assumptions about trial-to-trial variation in parameters, reducing the DDM from the modern versions \citep[e.g.,]{RatcliffTuerlinckx2002,ratcliff2016diffusion} to much older versions \citep[often the ``simple'' random walk of][]{stone1960models} which have well-known limitations as scientific theories. Similarly, modellers often make simplifying structural assumptions which have questionable scientific plausibility, such as assuming a priori uncorrelated random effects across people \citep[e.g.,][]{turner2013method}. The value of these simplifying assumptions is clear, though; for example, with sufficient simplification, EAMs can be estimated from data using off-the-shelf, general-purpose Bayesian sampling software such as JAGS \citep{plummer2003jags} or Stan \citep{carpenter2017stan}, for example as in applications reported by \cite{boehm2018using} and \cite{manning2021perceptual}. Such software has great advantages in ease of use, but is limited in the size of the models and data sets that can be addressed. 

Here, we have developed new methods and extended existing methods to push back even further the limitations described above. We have built on recent work using particle-based MCMC \citep{gunawan2020new} to provide efficient and robust exact Bayesian inference methods for both LBA and DDM, without making common simplifying assumptions: our methods work with the full DDM, and include hierarchical random effects structures with person-level correlations built in. For very large models or very large data sets, even these efficient MCMC methods can become impractically slow, and so we also developed approximate Bayesian methods based on variational inference. For all methods, we also include coherent frameworks for linking information measured in decision-relevant covariates with model parameters, including when the covariates are measured once per person or once per trial. For this to work efficiently at scale, we assumed fixed effects for the covariates (i.e., constant regression coefficients across people), which is another simplifying assumption that is not made by some other approaches to the problem \citep[e.g.,][]{jobst2020comparison,wiecki2013hddm}, and is something we are investigating in ongoing research.

Through analysis of three data sets, we demonstrated that our new methods produce robust and efficient estimates of model parameters and of the coefficients which link those parameters with decision-relevant covariates. Comparisons between exact (MCMC) and approximate (VB) inference revealed that the VB methods were often an order of magnitude faster, and that the central tendency of the posterior distributions was accurately recovered. Approximate inference often under-estimated the variance of the posterior distributions, which is a commmon finding for VB methods \citep{blei2017variational}. Two practically important contributions in our development of the VB methods are the investigation of different ways to initialize the starting points for optimization, and the investigation of a further simplified VB algorithm with potential to scale to very large estimation problems. The efficiency of VB makes it feasible for researchers to address interesting problems which are otherwise difficult, such as comparing performance across large sets of candidate EAMs \citep{dao2022efficient} or in large data sets, or with large numbers of covariates. The code and data are made freely available at \url{https://github.com/Henry-Dao/RegEAMs}.

\section{Acknowledgements} 
The research of Viet Hung Dao, David Gunawan, Minh-Ngoc Tran, Robert Kohn and Scott Brown was partially supported by Australian Research Council (ARC) Discovery Project (DP210103873). The research of Guy Hawkins and Scott Brown was partially supported by ARC Discovery Project (DP210100313). The funding sources had no role in the study design; in the analysis and interpretation of data; in the writing of the report; and in the decision to submit the article for publication. Declarations of interest: none. 

\appendix

\medskip


\title{Online Supplement for Bayesian Inference for Evidence Accumulation Models with Regressors}

\maketitle


\section{The Diffusion Decision model}\label{DDMdetail}
This section gives further details of the DDM density and its approximation. Section \ref{simpleDDMversion} considers a simple version of DDM; section \ref{sec:ddm_density} discusses the standard DDM; section \ref{supp:sec-HDDMtransformation} considers the required transformation for the random effects in the DDM; section \ref{HierarchicalDDMdetails} gives further details of the hierarchical DDM.

\subsection{A simple DDM}\label{simpleDDMversion}

\begin{figure}
	\centering
	\includegraphics[scale=0.4]{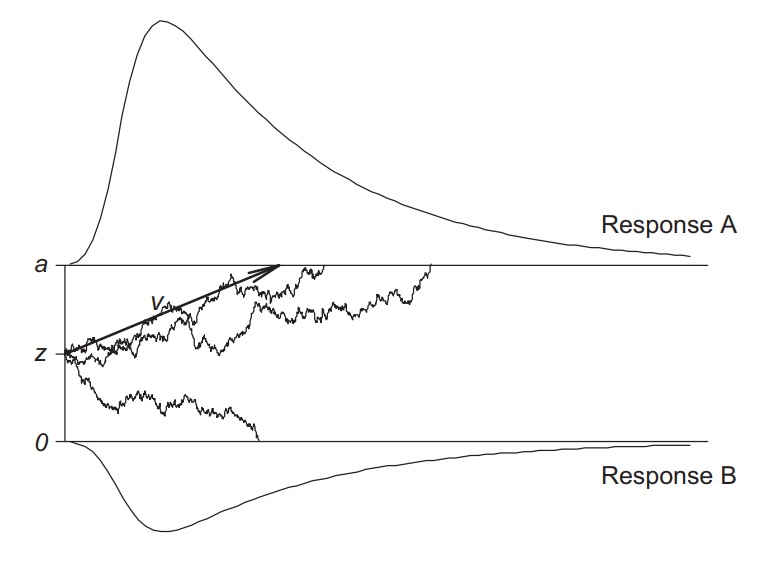}
	\caption[A Diffusion Model]{A simple diffusion decision model
		with constant drift $v$ starts at $z$ and terminates as soon as it reaches one
		of the thresholds at $a$ or $0$, respectively. The figure is used with permission from \emph{Experimental Psychology} 2013; Vol. 60(6):385–402
©2013 Hogrefe Publishing \url{www.hogrefe.com}\hspace{0.2cm} \url{https://doi.org/10.1027/1618-3169/a000218.}}\label{fig:wiener_process}
\end{figure}

The simplest diffusion model assumes that the information accumulates continuously by a Wiener diffusion process \citep{smith2000stochastic,smith2015introduction}. Let $X(t)$ be the ``state of evidence'' at the time $t$.   Figure~\ref{fig:wiener_process} shows that the process begins accumulating evidence at time $t=0$, with the initial evidence state $X(0)=z$ lying in the range $0<X(0)<a$. Information continues accumulating  at the drift rate $v$ until one of the two boundaries is reached. The decision depends on the boundary reached; i.e., if the process reaches the upper boundary first, response A is made. The sign of the drift rate $v$ (the drift direction) depends on the presented stimulus, i.e., when stimulus A is presented, the drift is positive and the accumulating evidence state tends to increase with time, making it more likely to terminate at the upper boundary and result in response A. Analogously, when stimulus B is presented, the drift rate is negative, and the process tends to end up at the lower boundary, resulting in response B. The drift rate's magnitude is determined by the stimulus discriminability. For example, in a lexical decision task, high frequency words correspond to high drift rates and low frequency words correspond to low drift rates. Hence, the drift rate $v$ is used to quantify ``ease of processing'' \citep{VossEtAl2004}. Large values of drift rate result in fast and accurate responses, whereas small drift rates produce slow and inaccurate decisions. 

The stochastic differential equation
\begin{equation}\label{eq:sde}
    dX(t) = vdt + sdW(t),
\end{equation}
defines the accumulation process, 
where $W(t)$ is  a Brownian motion (Wiener process), $v$ is the drift rate and $s$ is the 
standard deviation which
determines the variability in the sample paths of the process. 

The  diffusion model parameters are only identified up to a ratio; i.e., 
all the model parameters can be multiplied by a constant without affecting  the predictions.
Hence, to make the parameters estimable, one parameter needs to be fixed. 
It is common in practice to fix $s$ at $s=1$ \citep[see, e.g.,][and references therein]{DonkinPBR2009}. The other parameters are then expressed in units of infinitesimal standard deviation per unit time\footnote{\citet{Ratcliff1978} fixes $s=0.1$. The only difference is that the other parameters will all be 10 times smaller than the size of the corresponding parameters if $s=1$.}.

\subsubsection{The Wiener first-passage time density}
The observable model random variables  are the choices $c$ and the corresponding decision-time 
$t_d$ (the time it takes for the accumulator to reach a boundary). 
The joint density of the choice and the decision time is called the Wiener first-passage time 
density, denoted $\wfptpdf$. Let $f(t_d|v,a,z)$ be the WFPT density \footnote{To be precise, $f(t_d|v,a,z)$ is not a density in the strict sense as $\int\limits_0^{\infty}f(t_d|v,a,z)dt_d$ is often not equal to 1. For simplicity, following \citet{navarro2009fast,gondan2014even} and \citet{foster2021another}, we will refer to $f(t_d|v,a,z)$ as a density.}
describing the chance that the diffusion process is absorbed at time $t_d$ at the lower boundary.

Then, 
\begin{equation*}
    \wfptpdf = \begin{cases}
	f(t_d|v,a,z)& \text{ if } c = \text{``lower"},\\
	f(t_d|-v,a,a-z)& \text{ if } c = \text{``upper"}.
\end{cases}
\end{equation*}
\citet{feller1968introduction} provides two equivalent representations for the function $f(t_d|v,a,z)$: 
\begin{equation*}
    \flarge = \dfrac{\pi}{a^2}\exp\left( -v z-\dfrac{v^2t_d}{2}\right) \sum\limits_{k=1}^\infty k\sin \left (\dfrac{k\pi z}{a} \right)\exp \left( -\dfrac{k^2\pi^2t_d}{2a^2}\right),
\end{equation*}
and 
\begin{equation*}
    \fsmall = \dfrac{a}{\sqrt{2\pi t_d^3}} \exp\left( -v z-\dfrac{v^2t_d}{2}\right) \sum\limits_{k=-\infty}^\infty \left(\dfrac{z}{a}+2k\right)\exp \left( -\dfrac{(z+2ka)^2}{2t_d}\right).
\end{equation*}


The former is called the ``large-time'' representation and the latter is called the ``small-time'' representation. Their convergence rates depend on the value of the decision time $t_d$. The reason for referring to the two representations as large-time and small-time is that $\flarge$ converges at a faster rate for large values of $t_d$, whereas $\fsmall$ has a better rate of convergence with small values of $t_d$ \citep{navarro2009fast}.

\subsubsection{Approximating the Wiener first-passage time density:}
In practice, it is crucial to approximate the WFPT density and evaluate the approximation error.
For recent reviews on how to evaluate the density accurately and efficiently, see \citet{foster2021another} and references therein. Our work is based on the important paper by \citet{navarro2009fast} who propose an automatic rule to choose between the two representations and to truncate the series such that the approximation error is less than a predetermined accuracy level $\epsilon$. For mathematical convenience, we replace the starting point $z$ ($0<z<a$) with the relative starting point $\omega = z/a$ ($0<\omega<1$). Under the new parameterisation, 
the large-time representation density becomes
\begin{equation*}
    f^l(t_d|v,a,\omega) = \dfrac{\pi}{a^2}\exp\left( -va\omega-\dfrac{v^2t_d}{2}\right) \sum\limits_{k=1}^\infty k\sin \left (k\pi\omega \right)\exp \left( -\dfrac{k^2\pi^2t_d}{2a^2}\right).
\end{equation*}
When $v=0$ and $a=1$, 
\begin{equation*}
    f^l(t_d|0,1,\omega) = \pi \sum\limits_{k=1}^\infty k\sin \left (k\pi \omega \right)\exp \left( -\dfrac{k^2\pi^2t_d}{2}\right).
\end{equation*}
It is now straightforward to show that  
\begin{equation*}
    f^l\left(\dfrac{t_d}{a^2}
    \bigm\vert 0,1,\omega\right) = \pi \sum\limits_{k=1}^\infty k\sin \left (k\pi \omega \right)\exp \left( -\dfrac{k^2\pi^2t_d}{2a^2}\right),
\end{equation*}
and
\begin{equation*}
    f^l(t_d|v,a,\omega) = \dfrac{1}{a^2}\exp\left( -v a\omega-\dfrac{v^2t_d}{2}\right)f^l\left(\dfrac{t_d}{a^2}\left|\right.0,1,\omega\right).
\end{equation*}
Analogously, we can factorise the small-time density similarly, that is, 
\begin{equation*}
    f^s(t_d|v,a,\omega) = \dfrac{1}{a^2}\exp\left( -v a\omega-\dfrac{v^2t_d}{2}\right)f^s\left(\dfrac{t_d}{a^2}\left|\right.0,1,\omega\right),
\end{equation*}
where 
\begin{equation*}
    f^s\left(t_d\left|\right.0,1,\omega\right) = \dfrac{1}{\sqrt{2\pi t_d^3}}  \sum\limits_{k=-\infty}^\infty \left(\omega+2k\right)\exp \left( -\dfrac{(\omega+2k)^2}{2t_d}\right).
\end{equation*}
In general, both the small-time and large-time densities, denoted by the function $f(t_d|v,a,z)$, can be factorized as
\begin{equation*}
    f(t_d|v,a,\omega) = \dfrac{1}{a^2}\exp\left( -v a\omega-\dfrac{v^2t_d}{2}\right)f\left(\dfrac{t_d}{a^2}\left|\right.0,1,\omega\right).
\end{equation*}

The problem of approximating $f(t_d|v,a,\omega)$, a density with three parameters, now becomes a problem of estimating a density with a single parameter $f(t_d|0,1,\omega)$.

\citeauthor{navarro2009fast} suggest the following  rule to choose a suitable representation and to truncate the infinite series.  
\begin{equation*}
    f(t_d|0,1,\omega) \approx \begin{cases}
		\dfrac{1}{\sqrt{2\pi t_d^3}} \sum\limits_{k=-\floor*{(\kappa-1)/2}}^{\ceil*{(\kappa-1)/2}} \left(\omega + 2k\right)\exp \left( -\dfrac{(\omega+2k)^2}{2t_d}\right) & \textrm{ if } \lambda(t_d)<0\\
		& \\
		\pi \sum\limits_{k=1}^\kappa k\exp \left( -\dfrac{k^2\pi^2t_d}{2}\right) \sin \left(k\pi\omega \right) & \textrm{ if } \lambda(t_d)\geq 0,
	\end{cases}
\end{equation*}
where 
	\[ \lambda(t_d)=2+\sqrt{-2t_d\log(2\epsilon\sqrt{2\pi t_d})}-\sqrt{\dfrac{-2\log(\pi t_d\epsilon)}{\pi^2 t_d}}, \]
and $\epsilon$ is a predefined level of accuracy. The number of terms needed in the truncated series is denoted by $\kappa$ and is determined by the following rule:
\begin{itemize}
    \item large-time representation: $\kappa=\ceil*{\max \left\{\sqrt{\dfrac{-2\log(\pi t_d \epsilon)}{\pi^2 t_d}},\dfrac{1}{\pi\sqrt{t_d}} \right\} },$ \\
    \item small-time representation: $\kappa= \ceil*{\max\left\{ 2+\sqrt{-2t_d\log(2\epsilon\sqrt{2\pi t_d})},1+\sqrt{t_d}\right\} },$
\end{itemize}
where $\ceil*{\cdot}$ and $\floor*{.}$ are the ceiling and floor functions, respectively.

\textbf{Remark:} In the simple DDM, we do not  observe  the decision-time $t_d,$ but the response time $RT = t_d + \tau.$ Hence, the simple DDM density is the WFPT density with $t_d = RT - \tau.$

\subsection{The standard diffusion decision model}\label{sec:ddm_density}

The drawback with the simple model is that it cannot capture many important RT distributions observed from empirical data. To address this, \citet{Ratcliff1978} extends the model by assuming that the drift rate $v$, the starting point $z$ and the non-decision time $\tau$ can change from trial to trial. More specifically, at each trial, the value of the drift rate $v$ is assumed to be drawn from a Normal distribution with mean $\muv$ and variance $\sv^2$. Independently, the starting point $z$ and the non-decision time $\tau$ are drawn from uniform distributions centred at $\muz$ and $\mutau$, respectively. The width of the uniform distributions are $\sz$ (of the starting point) and $\stau$ (of the non-decision time). With these assumptions, the density of the standard diffusion model is defined as:
\begin{equation*}
    \textrm{DDM}(c,RT) = \dfrac{1}{\sz\stau}\int\limits_{\muz - \frac{\sz}{2}}^{\muz + \frac{\sz}{2}}\int\limits_{\mutau - \frac{\stau}{2}}^{\mutau + \frac{\stau}{2}}\int\limits_{-\infty}^{\infty} \wfptpdf N(v|\muv,\sv^2) dv  d\tau dz,
\end{equation*}
with $t_d = RT -\tau.$ 
Similarly  to
the WFPT density, the standard diffusion density
has two equivalent representations.  Denote the large-time and the small-time representations of the standard diffusion density by $\textrm{DDM}^{l}(c,RT)$ and $\textrm{DDM}^{s}(c,RT)$, respectively. Then, the densities are:
\begin{align*}
\textrm{DDM}^{l}(\textrm{"lower"},RT) &=\dfrac{1}{\sz\stau}
\int\limits_{\muz - \frac{\sz}{2}}^{\muz + \frac{\sz}{2}}\int\limits_{\mutau - \frac{\stau}{2}}^{\mutau + \frac{\stau}{2}}\int\limits_{-\infty}^{\infty}\dfrac{\pi}{a^2}\exp\left( -v z- \dfrac{v^2t_d}{2}\right)\sum\limits_{k=1}^\infty k\exp \left( -\dfrac{k^2\pi^2 t_d}{2a^2}\right)\\
&\times\sin \left(\dfrac{k\pi z}{a}\right)N(v|\muv,\sv^2)dv d \tau dz \\   
&= \dfrac{1}{\sz\stau}\int\limits_{\muz - \frac{\sz}{2}}^{\muz + \dfrac{\sz}{2}}\int\limits_{\mutau - \dfrac{\stau}{2}}^{\mutau + \frac{\stau}{2}} \dfrac{\pi}{a^2\sqrt{1+t_d\sv^2}} \exp\left(-\dfrac{\muv^2 t_d + 2\muv z - z^2\sv^2}{2(t_d\sv^2+1)} \right)\\
&\times \sumWkl d \tau dz .
\end{align*}

\begin{align*}
\textrm{DDM}^{s}(\textrm{"lower"},RT)&=
\dfrac{1}{\sz\stau}
\int\limits_{\muz - \frac{\sz}{2}}^{\muz + \frac{\sz}{2}}\int\limits_{\mutau - \frac{\stau}{2}}^{\mutau + \frac{\stau}{2}}\int\limits_{-\infty}^{\infty}  \dfrac{at_d^{-3/2}}{\sqrt{2\pi}} \exp\left( -v z- \dfrac{v^2t_d}{2}\right)\\ 
&\times \sum\limits_{k=-\infty}^\infty \left(\dfrac{z}{a} +2k\right) \exp \left( -\dfrac{(z+2k a)^2}{2t_d}\right) N(v|\muv,\sv^2)dv d \tau dz \\   
&= \dfrac{1}{\sz\stau}\int\limits_{\muz - \frac{\sz}{2}}^{\muz + \frac{\sz}{2}}\int\limits_{\mutau - \frac{\stau}{2}}^{\mutau + \frac{\stau}{2}} \dfrac{at_d^{-3/2}}{\sqrt{2\pi(1+t_d\sv^2)}} \exp\left(-\dfrac{\muv^2}{2\sv^2}+ \dfrac{(\muv - z\sv^2)^2}{\sv^2(1+t_d\sv^2)}\right)\\
&\times\sumWks d \tau dz.
\end{align*}

The integrals with respect to $z$ and $\tau$ are analytically intractable and must be approximated numerically using Gaussian quadrature. 

\subsection{Transformations}\label{supp:sec-HDDMtransformation}

The standard DDM has seven main 
parameters satisfying several constraints.
The variability parameters $\sv,\sz,$ and $\stau$ are positive. The boundary separation must be greater than the upper bound of the starting point, i.e., $a>\muz + 0.5\sz.$ The lower bound of the starting point and the non-decision time must be greater than zero, i.e., $\muz - 0.5\sz>0,$ and $\mutau-0.5\stau>0$. 
The full hierarchical structure considered here allows all the parameters to vary across subjects; hence, each subject has a vector of random effects, 
$$\bldeta = (\muv,\sv,a,\muz,\sz,\mutau,\stau).$$
The correlation between the subject parameters is captured by using a multivariate Gaussian with a full covariance matrix. To do so, we transform the subject parameters as follows, and denote the transformed parameters by 

$\balph := (\alpha_1,\alpha_2,\alpha_3,\alpha_4,\alpha_5,\alpha_6,\alpha_7),$
with
\begin{equation}\label{eq:hddm_transform1}
    \left\{ \begin{array}{ll}
      \alpha_1 &= \muv,	\alpha_2 = \log \sv,\\
	\alpha_3 &= \log(a - \muz - 0.5\sz), \alpha_4 = \log (\muz - 0.5\sz), \alpha_5 = \log(\sz),\\
	\alpha_6 &= \log(\mutau - 0.5\stau), \alpha_7 = \log \stau.\\
    \end{array}
    \right.
\end{equation}

\subsubsection{Hierarchical DDM - Experiment 1}\label{supp:sec-hddm-simulation-transform}
The transformations for the DDM in fitting the real data are discussed in Section~\ref{sec:hddm-real-data}. Each subject has the following parameters
$$ (\muv^{(hf)},\muv^{(lf)},\muv^{(vlf)},\muv^{(nw)},\sv,a^{(speed)},a^{(acc.)},\muz^{(speed)},\muz^{(acc.)},\sz,\mutau,\stau). $$

To capture the constraints on the DDM parameters, we apply the following transformations:
 $$\balph = (\alpha_1,\alpha_2,\alpha_3,\alpha_4,\alpha_5,\alpha_6,\alpha_7,\alpha_8,\alpha_9,\alpha_{10},\alpha_{11},\alpha_{12}),$$
with
\begin{equation*}
\left\{    \begin{array}{ll}
      \alpha_1 &= \muv^{(hf)}, \alpha_2 = \muv^{(lf)}, \alpha_3 = \muv^{(vlf)}, \alpha_4 = \muv^{(nw)}, \alpha_5 = \log \sv,\\
     \alpha_6 &= \log (a^{(speed)} - \muz^{(speed)} - 0.5\sz), \alpha_7 = \log (\muz^{(speed)} - 0.5\sz),\\
	\alpha_8 &= \log (a^{(acc.)} - \muz^{(acc.)} - 0.5\sz), \alpha_9 = \log (\muz^{(acc.)} - 0.5\sz), \\
	\alpha_{10} &= \log(\sz), \alpha_{11} = \log(\mutau - 0.5\stau), \alpha_{12} = \log \stau.
    \end{array}
\right.
\end{equation*}

\subsubsection{Simulation study for EAMs}\label{supp:sec-regeam-simulation-transform}

To capture the natural constraints of the random effects $\bldeta$ for simulation study discussed in section \ref{supp:sec-reg-eams-simulation} of the online supplement, we apply the following transformations. Denote the transformed random effects by $\balph$, we have
$$\balph = (\alpha_1,\alpha_2,\alpha_3,\alpha_4,\alpha_5,\alpha_6,\alpha_7,\alpha_8,\alpha_9),$$
with
\begin{equation}\label{eq:eeg-hddm-transform}
    \left\{ \begin{array}{ll}
      \alpha_1 &= \muv^{(h)},	\alpha_2 = \muv^{(m)},	\alpha_3 = \muv^{(e)},    \alpha_4 = \log \sv,\\
	\alpha_5 &= \log(a - \muz - 0.5\sz), \alpha_6 = \log (\muz - 0.5\sz), \alpha_7 = \log(\sz),\\
	\alpha_8 &= \log(\mutau - 0.5\stau), \alpha_9 = \log \stau.\\
    \end{array}
    \right.
\end{equation}

\subsubsection{EAMs - real data}
\paragraph{RegLBA}\label{supp:sec-reglba-real-data-transform}
We discuss the transformations applied for the RegLBA considered in Section~\ref{sec:mental-rotation}. The constraints on the random effects are: $b>A$ and all parameters must be positive. To capture these conditions, we apply the following transformations: 
\begin{equation}\label{eq:erp-lba-transform}
    \left\{ \begin{array}{ll}
      \alpha_1 &= \log( b - A),\\
	\alpha_2 &= \log(A),\\
	\alpha_3 &= v_s^s, \alpha_4 = v_m^s, \alpha_5 = v_s^m, \alpha_6 = v_m^m, \alpha_7 = v_c, \alpha_8 = v_e, \\
	\alpha_9 &= \log \tau.
    \end{array}
    \right.
\end{equation}
Denote the transformed random effects by $\balph = (\alpha_1,\alpha_2,\alpha_3,\alpha_4,\alpha_5,\alpha_6,\alpha_7,\alpha_8,\alpha_9).$

\paragraph{DDM}\label{supp:sec-regddm-real-data-transform}
We use the following transformations to enforce the constraints on the DDM parameters
in the DDM discussed in  Section~\ref{sec:mental-rotation}. 
Let $\balph = (\alpha_1,\alpha_2,\alpha_3,\alpha_4,\alpha_5,\alpha_6,\alpha_7,\alpha_8)$ denote the transformed parameters, with
\begin{equation}\label{eq:erp-hddm-transform}
    \left\{ \begin{array}{ll}
      \alpha_1 &= v^{0},	\alpha_2 = v,	\alpha_3 = \log \sv,\\
	\alpha_4 &= \log(a - z - 0.5\sz), \alpha_5 = \log (z - 0.5\sz), \alpha_6 = \log(\sz),\\
	\alpha_7 &= \log(\tau - 0.5\stau), \alpha_8 = \log \stau.\\
    \end{array}
    \right.
\end{equation}

\subsection{Hierarchical Diffusion Model} \label{HierarchicalDDMdetails}
This section gives further details on the hierarchical DDM. Denote by $p(y_{ij}|\balph_j)$ the density of an EAM; the HEAM is specified as follows.
\begin{enumerate}
	\item Conditional density: 
	\begin{equation*}\label{eq:heams-conditional-density}
	    \begin{array}{l}
	    p(\by|\balphJ) = \prod\limits_{j=1}^{J}p(\by_j|\balph_j),\\
        p(\by_j|\balph_j) =  \prod\limits_{i=1}^{n_j} p(y_{ij}|\balph_{j}), \textrm{ for } j = 1,\dots,J,\, i = 1,\dots,n_j.
    \end{array}
	\end{equation*}
	\item A multivariate normal distribution for the random effects
	\begin{equation*}
		\balph_j|\bmualph,\bSigalph \stackrel{ind.}{\sim} N(\bmualph,\bSigalph ).\label{eq:reg-eams-prior-random-effects}
	\end{equation*}
	\item Priors for the model parameters 
	\begin{enumerate}
	    \item Group-level mean $\bmualph \sim N(\mumu,\Sigmu).$ 
	    \item Group-level covariance $\bSigalph$. 
	    \begin{enumerate}
	        \item Informative prior $\bSigalph \sim \textrm{IW} \left(\ka,\bPsi\right),$ with the hyperparameters $\ka=20$ and $\bPsi = \boldsymbol{I}$.
	        \item Marginally noninformative prior by \citet{huang2013simple} 
	        \begin{equation*}
        		\begin{array}{l}
        			\bSigalph|a_1,\dots,a_{\Da} \sim \textrm{IW} \left(\Da +1,\bPsi\right),\bPsi=4\textrm{diag}\left(\dfrac{1}{a_1},\dots,\dfrac{1}{a_{\Da}}\right),\\
        			a_1,\dots,a_{\Da} \sim \textrm{IG} \left( \dfrac{1}{2},1\right).
        		\end{array}\label{eq:reg-eams-prior-group-level-covariance}
        	\end{equation*}
	    \end{enumerate}
	\end{enumerate}
\end{enumerate}

\subsubsection*{Remark 2}
The model specified in equation \eqref{eq:reg-eams-conditional-density} in section \ref{sec:reg-eams-model-specification} assumes the same transformations are applied to the random effects and the model parameters. However, in practice, this is not always the case. For example, Section~\ref{sec:reg-eams-real-data} considers a real example in which many different values of the drift rate can be modelled using just two parameters. This parsimonious parameterisation is crucial in  obtaining reliable estimates as the sample size is small. Hence, we will introduce several new notations so that the model specification is more flexible, and hence allows various parameterisations. 
First, let $\bomg$ and $\bpsi$ denote the set of model parameters with and without transformation, respectively. Similarly, we denote the natural and transformed random effects by $\bldeta$ and $\balph$. 

The linking equation is
\begin{equation}\label{eq:linking-flexible}
    \begin{array}{l}
        p(\by_j|\balph_j,\bbeta,\bX_j) =  \prod\limits_{i=1}^{n_j} p(y_{ij}|\bomg_{ij}),\\
        \bomg_{ij} = \bxi_j + \bbeta X_{ij}, \textrm{ for } j = 1,\dots,J,\, i = 1,\dots,n_j, 
    \end{array}
\end{equation}
where $\bxi_j$ denotes the random effects that take part in the linking equation. The transformation on the random effects is needed, as we want to use an unconstrained Gaussian distribution to capture the correlation of the effects. Notice that it is $\bxi_j$ and not $\balph_j$, that appears in the linking equation. The transformed model parameters $\bomg_{ij}$ are needed as the right side of the linking equation has an unrestricted range. Equation~\eqref{eq:linking-flexible} gives more modelling flexibility and allows parsimonious parameterisations such as the one in Section~\ref{sec:reg-eams-real-data}. 

\section{Variational Bayes}\label{supp:VB}
This section gives further details of the variational Bayes method. Subsection \ref{reparameterisationtrick} discusses variance reduction using the reparameterisation trick. Subsection \ref{learningrate} discusses the learning rates. Subsection \ref{supp:sec-gradients-reg-eams} gives details for estimating EAMs. Subsection \ref{supp:sec-VB-HDDM} discusses further details of the VB method for estimating a hierarchical DDM.

\subsection{The ``reparameterisation trick''}\label{reparameterisationtrick}
The performance of stochastic gradient ascent depends greatly on the variance of the noisy gradient estimate $\gradlbest$. Its performance can  be greatly improved by employing variance reduction methods. A popular variance reduction method  is the so-called ``reparameterisation trick'' \citep{kingma2013auto,rezende2014stochastic}. If we can write $\btheta\sim \qvb$ as $\btheta = u(\beps;\blamb)$, with $\beps \sim f_{\epsilon}$ not depending on $\blamb$, then the lower bound and its gradient can be written as the expectations
\begin{align}\label{eq: gradlb}
	\lb &= E_{f_{\epsilon}}\left[ \log p(\by,u(\beps;\blamb)) - \log q_{\blamb}(u(\beps;\blamb)) \right],\notag\\
	\gradlb &= E_{f_{\epsilon}}\bigg[ \nabla_{\blamb} u(\beps;\blamb) \left[\nabla_\theta \log \joint -  \nabla_\theta \log \qvb \right] \bigg].
\end{align}
By sampling $\beps\sim f_{\epsilon}$, it is straightforward to obtain the unbiased estimates of the lower bound and its gradient
\begin{align}\label{eq: gradlb_hat}
	\lbest& := \dfrac{1}{N}\sum\limits_{i=1}^N\left[ \log p(\by,u(\beps^{(i)};\blamb)) - \log q_{\blamb}(u(\beps^{(i)};\blamb)) \right],\notag\\
	\gradlbest& := \dfrac{1}{N}\sum\limits_{i=1}^N\bigg[ \nabla_{\blamb} u(\beps^{(i)};\blamb) \left(\nabla_\theta \log p(\by,\btheta^{(i)}) -  \nabla_\theta \log q_{\blamb}(\btheta^{(i)}) \right) \bigg],
\end{align}
with $\beps^{(i)} \sim f_{\epsilon}, i = 1,\dots,N$. We use $N=10$ in our applications.

\subsection{Learning rates and stopping rule}\label{learningrate}
\subsubsection{ADADELTA}
The elements of the vector $\blamb$ may need very different step sizes (learning rates) during the search, to account for scale or the geometry of the space. We set the step sizes adaptively using the ADADELTA method \citep{zeiler2012adadelta}, with different step sizes for each element of $\blamb$. At iteration $t+1$, the $i$th element $\lambda_i$ of $\blamb$ is updated as
$$ \lambda_i^{(t+1)}=\lambda_i^{(t)}+\Delta\lambda_i^{(t)}.$$
The step size $\Delta\lambda_i^{(t)}:=\rho_i^{(t)}g_{\lambda_i}^{(t)}$, where $g_{\lambda_i}^{(t)}$ denotes the $i$th component of $\widehat{\nabla_{\blamb}\mathcal{L}(\boldsymbol{\lambda}^{(t)})}$ and
$$ \rho_i^{(t)} := \dfrac{\sqrt{E(\Delta_{\lambda_i}^2)^{(t-1)} + \xi}}{\sqrt{E(g_{\lambda_i}^2)^{(t)} + \xi}},$$
where $\xi$ is a small positive constant, with
\begin{align*}
	E(\Delta_{\lambda_i}^2)^{(t)} &= v E(\Delta_{\lambda_i}^2)^{(t-1)} + (1-v)(\Delta_{\lambda_i}^{(t)})^2,\\
	E(g_{\lambda_i}^2)^{(t)} &= v E(g_{\lambda_i}^2)^{(t-1)} + (1-v)(g_{\lambda_i}^{(t)})^2.
\end{align*}
The ADADELTA default settings are $\xi =  10^{-6},v=0.95$ with initialisation $E(\Delta_{\lambda_i}^2)^{(0)} := E(g_{\lambda_i}^2)^{(0)}=0.$ However, in our experiments, we use $\xi=10^{-7}$ as we obtain better results.

\subsubsection{ADAM}

Another popular choice for selecting the learning rate is ADAM \citep{kingma2014adam}. Recall that the vector variational parameters used in the variational densities considered in this paper is $\blamb = (\bmulamb,\Blamb,\dlamb)$. Notice also that, for 
notation simplicity, in this section we shall omit the subscript indices, i.e., we use $\bmu$ instead of $\bmulamb$. We first explain the updating rule using ADAM method for the first component $\bmu$. ADAM requires several tuning parameters such as a stepsize $\alpha_{\mu}$, and two exponential decay rates for the moment estimates denoted by $\beta_1$ and $\beta_2$.\footnote{In the paper, we do use $\alpha$ and $\beta$ to denote the random and fixed effects when introducing the models. However, as we are purely discussing the general optimization algorithm without being specific about any model, the duplication in notation here should not cause any confusion.}
\begin{algorithm}
\caption{ADAM}\label{algo:adam}
\begin{enumerate}
    \item Initialisation: select initial values: $\bmu^{(0)}$ and set $m_0 = v_0 =0$.
    \item While not convergence do:
    \begin{itemize}
        \item[(1)] $ t \leftarrow t + 1$
        \item[(2)] Calculate $g_t \leftarrow \widehat{\nabla_{\bmu}\mathcal{L}(\boldsymbol{\lambda}^{(t)})}$
        \item[(3)] Update the biased first moment estimate
        \begin{equation*}
            m_t \leftarrow \beta_1 m_{t-1} + (1-\beta_1)g_t
        \end{equation*}
        \item[(4)] Correct the biased first moment estimate
        \begin{equation*}
             \hat{m}_t \leftarrow \dfrac{m_t}{1-\beta_1^t}
        \end{equation*}
         \item[(5)] Update the biased second moment estimate
        \begin{equation*}
            v_t \leftarrow \beta_2 v_{t-1} + (1-\beta_2)g^2_t
        \end{equation*}  
        \item[(6)] Correct the biased second moment estimate
        \begin{equation*}
            \hat{v}_t \leftarrow \dfrac{v_t}{1-\beta^t_2}
        \end{equation*}
        \item[(7)] Update parameters
        \begin{equation*}
            \bmu^{(t)} \leftarrow \bmu^{(t-1)} - \alpha_{\mu}\dfrac{ \hat{m}_t}{\sqrt{\hat{v}_t} + \epsilon}
        \end{equation*}
    \end{itemize}
\end{enumerate}
\end{algorithm}
The ADAM algorithm for other variational parameters $B$ and $\bd$ is performed in a similar manner. We fix $\beta_1 = 0.9, \beta_2 = 0.99,$ and $\epsilon = 10^{-8}$. For the stepsizes, we set the stepsize of $\bmu$ to be equal to $0.01$, and choose the same stepsize of $0.001$ for $B$ and $\bd$.

\subsubsection{Stopping rule}
A popular  criterion for the search algorithm is to stop when the moving average lower bound estimate $\overline{\textrm{LB}}_t =\dfrac{1}{m} \sum\limits_{i=t-m+1}^t \widehat{\mathcal{L}(\blamb^{(i)})} $ does not improve after $k$ consecutive iterations \citep{tran2017variational}. We find that the choice of $m=100$ and $k=50$ works well in most cases.

\subsection{Implementing VB for approximating the EAMs}\label{supp:sec-gradients-reg-eams}
In Section~\ref{subsec:reg-eam-bayes-methods} we introduce two VB classes. The first one, which we call VB 1, uses the hybrid Gaussian VB is of the form
\begin{equation*}
    q_{\blamb}(\btheta_1,\bSigalph) = N(\btheta_1|\bmulamb,\Blamb \Blamb^\top + \Dlamb^2) \textrm{IW}(\btheta_2|\nu,\bPsi'),
\end{equation*}
where $\btheta_1 = (\balph_{1:J},\bbeta,\bmualph,\log \ba), \nu = 2\Da + J + 1,$ and $\bPsi' = \sum_{j=1}^J (\balph_j - \bmualph)(\balph_j - \bmualph)^\top + 4\diag\left(a^{-1}_1,\dots,a^{-1}_{\Da}\right)$. The second class, called VB 2, is a special case of the first one in the sense that that the random effects $\balph_1,\dots,\balph_J$ are further assumed to be conditionally independent, i.e.,
\begin{equation*}
    q_{\blamb}(\btheta_1,\bSigalph) = N(\btheta_2|\bmu_{J + 1},B_{J + 1}B_{J + 1}^\top + D_{J + 1}^2) \textrm{IW}(\bSigalph|\nu,\bPsi') \prod\limits_{j = 1}^J N(\balph_j|\bmu_j,B_jB_j^\top + D_j^2),
\end{equation*}
where $\btheta_1 = (\balph_{1:J},\bpsi)$ with $\bpsi = (\bbeta,\bmualph,\log \ba)$. The best approximate distribution in terms of minimizing the KL divergence is found by maximizing the lower bound $\lb$ with respect to $\blamb$. Starting from some initial value $\blamb^{(0)}$, the values of for $\blamb$ are updated iteratively  as
\begin{equation}
	\label{eqn:SGD}
	\blamb^{(t+1)} = \blamb^{(t)} + \boldsymbol{\rho}_t\odot\widehat{\nabla_{\blamb}\mathcal{L}(\blamb^{(t)})},
\end{equation}
where $\brho_t$ is a vector of step sizes or learning rates, $\odot$ denotes the element-wise product of two vectors, and $\gradlbest$ is an unbiased estimate of the gradient of $\gradlb$. \citet{dao2022efficient} show the gradient of the lower bound with respect to the variational parameters is\footnote{Here $\btheta_2$ represents $\bSigalph$.}
\begin{align*}
	\gradlb &=  E_{\beps}\bigg[ \nabla_{\blamb} u(\beps;\blamb)\nabla_{\btheta_1} \log \bigg(\dfrac{p(\btheta_1,\by)}{q_{\blamb}(\btheta_1)}  \bigg) \bigg]\\
	&= E_{(\beps,\btheta_2)}\bigg[ \nabla_{\blamb} u(\beps;\blamb)\nabla_{\btheta_1}\log  \bigg(\dfrac{p(\btheta_1,\by)p(\btheta_2|\btheta_1,\by)}{q_{\blamb}(\btheta_1)p(\btheta_2|\btheta_1,\by)}  \bigg)  \bigg]\\
	&= E_{(\beps,\btheta_2)}\bigg[ \nabla_{\blamb} u(\beps;\blamb)\nabla_{\btheta_1}\log  \bigg(\dfrac{p(\btheta_1,\btheta_2,\by)}{q_{\blamb}(\btheta_1,\btheta_2)}  \bigg)  \bigg]\\
	&= E_{(\beps,\btheta_2)}\bigg[ \nabla_{\blamb} u(\beps;\blamb) \bigg(\nabla_{\btheta_1}\log p(y|\btheta_1,\btheta_2) + \nabla_{\btheta_1}\log p(\btheta_1,\btheta_2) \\
	&\quad\quad\qquad -\nabla_{\btheta_1}\log q_{\blamb}(\btheta_1,\btheta_2) - \nabla_{\btheta_1}\log p(\btheta_2|\btheta_1,\by) \bigg)  \bigg].
\end{align*}
To implement VB, we need to derive $\nabla_{\btheta_1} \log p(\by|\btheta_1,\btheta_2),  \nabla_{\btheta_1}\log p(\btheta_1,\btheta_2), \nabla_{\btheta_1}\log q_{\blamb}(\btheta_1)$ and $\nabla_{\btheta_1}\log p(\btheta_2|\btheta_1,\by)$. We note that, because
\[E_{(\beps,\btheta_2)}\bigg[  \nabla_{\btheta_1}\log p(\btheta_2|\btheta_1,\by) \bigg]=E_{\beps}\bigg[\int \nabla_{\btheta_1}p(\btheta_2|\btheta_1,\by) d\btheta_2\bigg]=0,\]
we can remove the term $\nabla_{\btheta_1}\log p(\btheta_2|\btheta_1,\by)$ from the calculation of $\gradlb$. However, this term also plays the role of a control variate and is useful in reducing the variance of the gradient estimate in finite sample sizes (recall we use $N=10$). We therefore include this term in all the computations reported in the paper.
The data-parameter joint density is
\begin{align*}
	p(\by,\balph_{1:J},\bbeta,\bmualph,\bSigalph,\log \ba) =&N(\bbeta|\mubeta,\Sbeta) N(\bmualph|\mumu,\Sigmu)\textrm{IW}(\bSigalph|\nu,\bPsi) \times \\ & \prod\limits_{d=1}^{\Da}\textrm{IG}(a_d|1/2,1)\left| J_{a_d\rightarrow\log a_d}\right|
	\prod\limits_{j=1}^J p(\by_j|\balph_j)N(\balph_j|\bmualph,\bSigalph)\\
\end{align*}
where $J_{a_d\rightarrow\log a_d} = a_d$ is the Jacobian of the transformation.

\subsubsection{Deriving $\nabla_{\btheta_1}\log q_{\blamb}(\btheta_1,\btheta_2)$}

\subsubsubsection*{VB 1}
\citet{ong2018gaussian} show that $\nabla_{\btheta_1}\log q_{\blamb}(\btheta_1) = \nabla_{\btheta_1}\log N(\btheta_1|\bmulamb,\Blamb \Blamb^\top + \Dlamb^2) =  -(\Blamb \Blamb^\top+\Dlamb^2)^{-1}(\btheta_1-\bmulamb)$.
We  then use the Woodbury formula \citep{Petersen2012matrix} to compute the inverse using 
$$ (\Blamb\Blamb^\top+\Dlamb^2)^{-1} =  \Dlamb^{-2} - \Dlamb^{-2}\Blamb(I+\Blamb^\top \Dlamb^{-2}\Blamb)^{-1}\Blamb^\top \Dlamb^{-2}.$$
This is computationally efficient because it only requires finding the inverses of the diagonal matrix $\Dlamb$  and of $(I+\Blamb^\top \Dlamb^{-2}\Blamb)$, which is a much smaller $r\times r$ matrix.
\subsubsubsection*{VB 2}
\begin{equation*}
    q_{\blamb}(\btheta_1) = N(\btheta_2|\bmu_{J + 1},B_{J + 1}B_{J + 1}^\top + D_{J + 1}^2) \prod\limits_{j = 1}^J N(\balph_j|\bmu_j,B_jB_j^\top + D_j^2),
\end{equation*}
Similar to VB 1, we have
\begin{equation*}
    \nabla_{\balph_j}\log q_{\blamb}(\btheta_1) = \nabla_{\balph_j}\log N(\balph_j|\bmu_j,B_jB_j^\top + D_j^2) = -(B_j B_j^\top+D_j^2)^{-1}(\balph_j-\bmu_j), \text{ for }j = 1,\dots,J;
\end{equation*}
and 
\begin{equation*}
    \nabla_{\bpsi}\log q_{\blamb}(\btheta_1) = \nabla_{\bpsi}\log N(\bpsi|\bmu_{J+1},B_{J+1} B_{J+1}^\top + D_{J+1}^2) = -(B_{J+1} B_{J+1}^\top+D_{J+1}^2)^{-1}(\bpsi-\bmu_{J+1}).
\end{equation*}
Then, $\nabla_{\btheta_1}\log q_{\blamb}(\btheta_1) = \left( \nabla_{\balph_1}\log q_{\blamb}(\btheta_1) ,\dots, \nabla_{\balph_J}\log q_{\blamb}(\btheta_1) , \nabla_{\bpsi}\log q_{\blamb}(\btheta_1)\right)$
\subsubsection{Deriving $\nabla_{\btheta_1} \log p(\by|\btheta_1,\btheta_2)$}

The section derives the gradients required in the hybrid Gaussian VB for the new model. 
First, we derive $\nabla_{\btheta_1} \log p(\by|\btheta_1,\btheta_2)$. 

\begin{align*}
    \dfrac{\partial}{\partial \balph_j} \log p(\by|\btheta_1,\btheta_2) &= \dfrac{\partial}{\partial \balph_j} \log p(\by_j|\balph_j,\bbeta,\bX_j) \\
    &= \sum\limits_{i=1}^{n_j} \dfrac{\partial}{\partial \balph_j} \log p(y_{ij}|\balph_j,\bbeta,X_{ij})\\ 
    &= \sum\limits_{i=1}^{n_j} \left( \dfrac{\partial \widetilde{\balph}_{ij}}{\partial \balph^\top_j} \right)^\top \dfrac{\partial}{\partial \widetilde{\balph}_{ij}} \log p(y_{ij}|\widetilde{\balph}_{ij})\\
    &= \sum\limits_{i=1}^{n_j} \left( \dfrac{\partial \widetilde{\balph}_{ij}}{\partial \balph^\top_j} \right)^\top \left( \dfrac{\partial \bpsi_{ij}}{\partial \widetilde{\balph}_{ij}^\top} \right)^\top \dfrac{\partial}{\partial \bpsi_{ij}} \log p(y_{ij}|\bpsi_{ij})\\
    &= \sum\limits_{i=1}^{n_j} \left( \dfrac{\partial \bpsi_{ij}}{\partial \widetilde{\balph}_{ij}^\top} \right)^\top \dfrac{\partial}{\partial \bpsi_{ij}} \log p(y_{ij}|\bpsi_{ij}); 
\end{align*}
$\bpsi_{ij}$ denotes the model parameters in the natural scale. The derivatives $\dfrac{\partial}{\partial \bpsi_{ij}} p(y_{ij}|\bpsi_{ij})$ for LBA and DDM densities are given below.
Note that $\widetilde{\balph}_{ij} = \balph_j + \bbeta X_{ij},$ so $ \dfrac{\partial \widetilde{\balph}_{ij}}{\partial \balph^\top_j} = I$. 
\begin{align*}
    \dfrac{\partial}{\partial \bbeta} \log p(\by|\btheta_1,\btheta_2) &= \sum\limits_{j=1}^{J} \dfrac{\partial}{\partial \bbeta} \log p(\by_j|\balph_j,\bbeta,\bX_j)\\
    &= \sum\limits_{j=1}^{J} \sum\limits_{i=1}^{n_j} \dfrac{\partial}{\partial \bbeta} \log p(y_{ij}|\widetilde{\balph}_{ij}).
\end{align*}
To derive the derivative with respect to the matrix $\bbeta$, we derive the derivative with respect to its rows. Let $\bbeta_k$ be the $k$th row in the coefficient matrix $\bbeta,k = 1,\dots, \Da$. We have 
\begin{align*}
    \dfrac{\partial}{\partial \bbeta_k} \log p(y_{ij}|\widetilde{\balph}_{ij}) &=  \dfrac{\partial}{\partial \alpha_{ij;k}} \log p(y_{ij}|\widetilde{\balph}_{ij}) \dfrac{\partial \alpha_{ij;k}}{\partial \bbeta_k}\\
    &=  \dfrac{\partial}{\partial \alpha_{ij;k}} \log p(y_{ij}|\widetilde{\balph}_{ij}) \dfrac{\partial \alpha_{ij;k}}{\partial \bbeta_k},
\end{align*}
where $\widetilde{\alpha}_{ij;k},\alpha_{j;k}$ are the $k$th element of vectors $\widetilde{\balph}_{ij}$ and $\balph_j$, respectively. Since $\alpha_{ij;k} = \alpha_{j;k} + \bbeta_k X_{ij}^\top,$ we have $\dfrac{\partial \alpha_{ij;k}}{\partial \bbeta_k} = X_{ij}$ and therefore,
\begin{equation*}
    \dfrac{\partial}{\partial \bbeta} \log p(y_{ij}|\widetilde{\balph}_{ij}) = \begin{bmatrix}  \dfrac{\partial}{\partial \bbeta_1^\top} \log p(y_{ij}|\widetilde{\balph}_{ij}) \\ \vdots \\  \dfrac{\partial}{\partial \bbeta_{\Da}^\top} \log p(y_{ij}|\widetilde{\balph}_{ij}) \end{bmatrix} = \begin{bmatrix} \dfrac{\partial}{\partial \alpha_{ij;1}} \log p(y_{ij}|\widetilde{\balph}_{ij})  X_{ij}^\top \\ \vdots \\  \dfrac{\partial}{\partial \alpha_{ij;\Da}} \log p(y_{ij}|\widetilde{\balph}_{ij})  X_{ij}^\top \end{bmatrix},
\end{equation*}
and
\begin{equation*}
    \dfrac{\partial}{\partial \bbeta} \log p(\by|\btheta_1,\btheta_2) = \sum\limits_{j=1}^{J} \sum\limits_{i=1}^{n_j} \dfrac{\partial}{\partial \bbeta} \log p(y_{ij}|\widetilde{\balph}_{ij}).
\end{equation*}

\subsubsubsection{LBA density}
\citet{brown2008simplest} show the LBA density (the joint density over response time $RT=t$ and response choice $RE=c$) is
$$ \textrm{LBA} (c,t|b,A,\bv,s,\tau) = f_c(t)\times \prod\limits_{k\neq c}(1-F_k(t)),$$
where 
$$ F_c(t) = 1+\uone \xzero - \utwo \zzero+\dfrac{1}{\wone} \xone - \dfrac{1}{\wtwo}\zone$$
and
$$ f_c(t) = \dfrac{1}{A}\left[ -v^c\xzero + s\xone +v^c\zzero - s\zone \right]$$
with $\wone = \dfrac{b-(t-\tau)v^c}{(t-\tau)s},\,\wtwo = \dfrac{A}{(t-\tau)s}$, where $\Phi(\cdot)$ and $\phi(\cdot)$ denote the distribution function and density of the standard normal distribution, respectively.
Hence,
\begin{equation*}
    \dfrac{\partial}{\partial \bpsi_{ij}} \textrm{LBA}(y_{ji}|\bpsi_{ij}) = \dfrac{\partial f_c(t)}{\partial \bpsi_{ij}}(1 - F_{k\neq c}(t)) - \dfrac{\partial F_{k\neq c}(t)}{\partial \bpsi_{ij}}f_c(t)
\end{equation*}
\vspace*{0.5cm}
\hspace*{-0.5cm}The partial derivatives of $f_c(t)$ with respect to $\bpsi$\footnote{We omit the subscript $ij$ for simplicity.} are\\

$\dfrac{\partial}{\partial b}f_c(t) = \dfrac{1}{A}\bigg[  -v^c\xone + s\xtwo + v^c\zone - s\ztwo \bigg]\dfrac{\partial \wone}{\partial b};$\\

$\dfrac{\partial}{\partial A}f_c(t) = -\dfrac{1}{A}\bigg[  f_c(t) - v^c\xone + s\xtwo\bigg]\dfrac{\partial \wone}{\partial A};$\\

$\dfrac{\partial}{\partial v^c}f_c(t) = \dfrac{1}{A}\bigg[  - \xzero +\zzero\bigg] + \dfrac{1}{A}\bigg[  - v^c\xone+s\xtwo+v^c\zone$
\hspace*{2.2cm}	$-s\ztwo\bigg]\dfrac{\partial \wone}{\partial v^c};$\\

$\dfrac{\partial}{\partial \tau}f_c(t) = \dfrac{1}{A}\bigg[   - v^c\xone +s\xtwo \bigg]\left( \dfrac{\partial \wone}{\partial \tau} -\dfrac{\partial \wtwo}{\partial \tau} \right) + \dfrac{1}{A}\bigg[    v^c\zone-s\ztwo\bigg]\dfrac{\partial \wone}{\partial \tau}.$\\

\vspace*{0.5cm}
\hspace*{-0.5cm}The partial derivatives of $F_c(t)$ with respect to $\bpsi$\footnote{We omit the subscript $ij$ for simplicity. } are:\\
$\begin{array}{ll}
	
	\dfrac{\partial}{\partial b}F_c(t) &= \bigg[\dfrac{1}{A}\xzero +\uone\xone\dfrac{\partial \wone}{\partial b} \bigg]\\
	&+\dfrac{1}{\wtwo} \bigg[\xtwo-\ztwo \bigg]\dfrac{\partial \wone}{\partial b}+ \bigg[ -\dfrac{1}{A}\zzero - \utwo\zone \dfrac{\partial \wone}{\partial b} \bigg];\\
	
	\dfrac{\partial}{\partial A}F_c(t) &= \dfrac{1}{A}\bigg[-\xzero +\xone(b-A-(t-\tau)v^c)\dfrac{\partial \wtwo}{\partial A} \bigg]\\
	&+\dfrac{1}{A^2}\bigg[ (b-(t-\tau)v^c)\zzero-(b-A-(t-\tau)v^c)\xzero \bigg]\\
	&+ \dfrac{\xtwo\dfrac{\partial \wtwo}{\partial A}\wtwo - \dfrac{\partial \wtwo}{\partial A}\xone }{\wtwo^2}+\zone\dfrac{\partial \wtwo}{\partial A};\\
	
	\dfrac{\partial}{\partial v^c}F_c(t) &= -\dfrac{t-\tau}{A}\xzero+\uone\xone\dfrac{\partial \wone}{\partial v^c}\\
	&+\dfrac{t-\tau}{A}\zzero - \zone\utwo\dfrac{\partial \wone}{\partial v^c}\\
	&+\dfrac{1}{\wtwo}\left( \xtwo-\ztwo\right)\dfrac{\partial \wone}{\partial v^c};\\
	
	\dfrac{\partial}{\partial \tau}F_c(t) &=\dfrac{v^c}{\tau}\xzero+\uone\xone\left( \dfrac{\partial \wone}{\partial \tau} -\dfrac{\partial \wtwo}{\partial \tau} \right) \\
	&-\dfrac{v^c}{\tau}\zzero-\utwo\zone\dfrac{\partial \wone}{\partial \tau} \\
	&+\bigg[\xtwo\wtwo\left( \dfrac{\partial \wone}{\partial \tau} -\dfrac{\partial \wtwo}{\partial \tau} \right)-\xone\dfrac{\partial \wtwo}{\partial \tau} \bigg]/(\wtwo^2)\\
	&-\bigg[\ztwo\wtwo\dfrac{\partial \wone}{\partial \tau}-\zone\dfrac{\partial \wtwo}{\partial \tau} \bigg]/(\wtwo^2).
\end{array}$

\subsubsubsection{The DDM density}
\paragraph{Deriving gradients of the diffusion density (large-time representation)}\label{appendix:gradients-large-time}

From Section~\ref{sec:ddm_density}, the density of the diffusion model (large-time representation) is

\begin{align*}
\dfpdflowerlarge &= \dfrac{1}{\sz\stau}\int\limits_{\muz - \frac{\sz}{2}}^{\muz + \frac{\sz}{2}}\int\limits_{\mutau - \frac{\stau}{2}}^{\mutau + \frac{\stau}{2}} \dfrac{\pi}{a^2\sqrt{1+t_d\sv^2}} \exp\left(-\dfrac{\muv^2 t_d + 2\muv z - z^2\sv^2}{2(t_d\sv^2+1)} \right)\\
&\times \sumWkl d \tau dz .
\end{align*}

Define the integrand as  
$$\glarge := \dfrac{\pi}{a^2\sqrt{1+t_d\sv^2}} \exp\left(-\dfrac{\muv^2 t_d + 2\muv z - z^2\sv^2}{2(t_d\sv^2+1)} \right)\sumWkl. $$
We now derive the partial derivatives of the integrand with respect to $\muv,\sv^2,a,z$, and $\tau$. The partial derivatives with respect to $\muz,\sz,\mutau,$ and $\stau$, as well as the transformed parameters can then be derived straightforwardly using the chain rule.\\

$\dfrac{\partial}{\partial \muv} \glarge = -\glarge\left[ \dfrac{t_d\muv + z}{t_d \sv^2 + 1} \right],$

$\dfrac{\partial}{\partial \sv^2} \glarge =\dfrac{1}{2} \glarge\left[ \dfrac{1}{2}\dfrac{(z+\muv t_d)^2- t_d(1+t_d\sv^2)}{(1+t_d\sv^2)^2} \right],$

\begin{align*}
    \dfrac{\partial}{\partial a}\glarge &= -\glarge \dfrac{2}{a} + \dfrac{\pi}{a^2\sqrt{t_d\sv^2+1}} \exp\left(-\dfrac{\muv^2 t_d + 2\muv z - z^2\sv^2}{2(t_d\sv2+1)} \right)\\ &\times\sumWkl\left( -\dfrac{\pi kz}{a^2}\cot\left(\dfrac{\pi kz}{a} + \dfrac{\pi^2 k^2t_d}{a^3}\right) \right) ,
\end{align*}

\begin{align*}
    \dfrac{\partial}{\partial z}\glarge &= \glarge \left[ \dfrac{z\sv^2-\muv}{t_d\sv^2+1}\right] + \dfrac{\pi}{a^2\sqrt{t_d\sv^2+1}} \exp\left(-\dfrac{\muv^2 t_d + 2\muv z - z^2\sv^2}{2(t_d\sv^2+1)} \right)\\ &\times \sumWkl\dfrac{\pi k}{a}\cot\left( \dfrac{\pi kz}{a} \right),
\end{align*}

\begin{align*}
    \dfrac{\partial}{\partial \tau}\glarge &= \glarge \left[ \dfrac{\muv^2(t_d\sv^2+1) - \sv^2\muv^2t_d}{2(t_d\sv^2+1)^2} +\dfrac{\sv^2}{2(t_d\sv^2+1)} - \dfrac{2\muv z-z^2\sv^2}{2(t_d\sv^2+1)^2}\sv^2\right] \\
    &+\dfrac{\pi^3}{2a^4\sqrt{t_d\sv^2+1}} \exp\left(-\dfrac{\muv^2 t_d + 2\muv z - z^2\sv^2}{2(t_d\sv^2+1)} \right) \sum\limits_{k=1}^{\infty} k^3\sin\left( \dfrac{\pi k z}{a}\right) \exp\left( -\dfrac{\pi^2k^2t_d}{a^2}\right). 
\end{align*}


\paragraph{Deriving gradients of the diffusion density (small-time representation)}\label{appendix:gradient-small-time}
From Section~\ref{sec:ddm_density},  the density of the diffusion model (small-time representation) is 
\begin{align*}
\dfpdflowersmall&= \dfrac{1}{\sz\stau}\int\limits_{\muz - \frac{\sz}{2}}^{\muz + \frac{\sz}{2}}\int\limits_{\mutau - \frac{\stau}{2}}^{\mutau + \frac{\stau}{2}} \dfrac{at_d^{-3/2}}{\sqrt{2\pi(1+t_d\sv^2)}} \exp\left(-\dfrac{\muv^2}{2\sv^2}+ \dfrac{(\muv - z\sv^2)^2}{\sv^2(1+t_d\sv^2)}\right)\\
&\times\sumWks d \tau dz .
\end{align*}
Define the integrand as 
$$\gsmall = \dfrac{at_d^{-3/2}}{\sqrt{2\pi(1+t_d\sv^2)}} \exp\left(-\dfrac{\muv^2}{2\sv^2}+ \dfrac{(\muv - z\sv^2)^2}{\sv^2(1+t_d\sv^2)}\right)\times\sumWks. $$
We now derive the partial derivatives of the integrand with respect to $\muv,\sv^2,a,z$, and $\tau$. The partial derivatives with respect to $\muz,\sz,\mutau,$ and $\stau$, as well as the transformed parameters can then be derived straightforwardly using the chain rule.

\begin{align*}
    \dfrac{\partial}{\partial \sv^2} \gsmall &= \gsmall \left[ -\dfrac{t_d}{2(1+t_d\sv^2)} + \dfrac{\muv^2}{2s^4_{v}} - \dfrac{z(\muv-z\sv^2)}{(1+t_d\sv^2) \sv^2}  \right]\\
    & -\gsmall \dfrac{ (1+2t_d\sv^2)(\muv-z\sv^2)^2}{2(1+t_d\sv^2)^2 s^4_{v}}, 
\end{align*}

\begin{align*}
    \dfrac{\partial}{\partial \sv^2} \gsmall &= \gsmall \left[ -\dfrac{t_d}{2(1+t_d\sv^2)} + \dfrac{\muv^2}{2s^4_{v}} \right]\\
    &- \gsmall \dfrac{2z(\muv-z\sv^2)(1+t_d\sv^2)\sv^2 + (1+2t_d\sv^2)(\muv-z\sv^2)^2}{2(1+t_d\sv^2)^2 s^4_{v}},  
\end{align*}

\begin{align*}
    \dfrac{\partial}{\partial a}\gsmall &= \gsmall \dfrac{1}{a} + \dfrac{at_d^{-3/2}}{\sqrt{2\pi(1+t_d\sv^2)}} \exp\left(-\dfrac{\muv^2}{2\sv^2}+ \dfrac{(\muv - z\sv^2)^2}{\sv^2(1+t_d\sv^2)}\right)\\ &\times \sum\limits_{k=-\infty}^{\infty} \exp\left( -\dfrac{(z+2ka)^2}{2t_d}\right) \left( -\dfrac{zt_d + 2ka(z+2ka)^2}{a^2t_d} \right),
\end{align*}


\begin{align*}
    \dfrac{\partial}{\partial z}\gsmall &= \gsmall \left[ \dfrac{z\sv^2-\muv}{t_d\sv^2+1} \right] + \dfrac{at_d^{-3/2}}{\sqrt{2\pi(1+t_d\sv^2)}} \exp\left(-\dfrac{\muv^2}{2\sv^2}+ \dfrac{(\muv - z\sv^2)^2}{\sv^2(1+t_d\sv^2)}\right)\\
    &\times \sum\limits_{k=-\infty}^{\infty} \exp\left( -\dfrac{(z+2ka)^2}{2t_d}\right) \left( \dfrac{t_d - (z+2ka)^2}{at_d} \right),
\end{align*}


\begin{align*}
   \dfrac{\partial}{\partial \tau}\gsmall &= \gsmall \left[  \dfrac{\sv^2(t_d\sv^2+1) + (\muv-z\sv^2)^2}{2(t_d\sv^2+1)^2} +\dfrac{3}{2t_d}\right]\\  
   &- \dfrac{at_d^{-3/2}}{\sqrt{2\pi(1+t_d\sv^2)}} \exp\left(-\dfrac{\muv^2}{2\sv^2}+ \dfrac{(\muv - z\sv^2)^2}{\sv^2(1+t_d\sv^2)}\right)\\
   &\times \sumWks \dfrac{ (z+2ka)^2}{2t_d^2}. 
\end{align*}

\subsubsection{Deriving $\nabla_{\btheta_1}\log p(\btheta_1,\btheta_2)$ }
From above,  that $\btheta_1 = (\balph_{1:J},\bbeta,\bmualph,\log \ba)$ and $\btheta_2 = \bSigalph$ and
$$p(\btheta_1,\bSigalph) = N(\bbeta|\mubeta,\Sbeta)\prod\limits_{j=1}^J N(\balph_j|\bmualph,\bSigalph)N(\bmualph|\mumu,\Sigmu)\textrm{IW}(\bSigalph|\nu,\bPsi) \prod\limits_{d=1}^{\Da}\textrm{IG}(a_d|\alpha_d,\beta_d)\left| J_{a_d\rightarrow\log a_d}\right|;$$
hence,
\begin{align*}
	\log p(\btheta_1,\bSigalph) &= \log N(\bbeta|\mubeta,\Sbeta) + \sum\limits_{j=1}^J \log N(\balph_j|\bmualph,\bSigalph)+ \log N(\bmualph|\mumu,\Sigmu)+ \log\textrm{IW}(\bSigalph|\nu,\bPsi)\\
	&+ \sum\limits_{d=1}^{\Da}\Big(\log \textrm{IG}(a_d|\alpha_d,\beta_d) + \log a_d\Big)\\
	&= - \dfrac{1}{2}(\bbeta -\mubeta)^\top\Sbeta^{-1} (\bbeta -\mubeta) -\dfrac{1}{2}\sum\limits_{j=1}^J(\balph_j-\bmualph)^\top\bSigalphinv(\balph_j-\bmualph) \\ 
	&- \dfrac{1}{2}(\bmualph -\mumu)^\top\Sigmu^{-1} (\bmualph -\mumu) +\dfrac{\nu}{2}\logdet(\bPsi) -\dfrac{1}{2}\tr (\bPsi\bSigalphinv)\\ &+\sum\limits_{d=1}^{\Da}\left\{-(\alpha_d+1)\log a_d - \dfrac{\beta_d}{a_d} + \log a_d\right\}+\text{constant}.
\end{align*}
Hence, 
\begin{equation*}
    \begin{array}{l}
    \dfrac{\partial}{\partial \balph_j}\log p(\btheta_1,\bSigalph) = -\bSigalphinv(\balph_j-\bmualph),\\
        \dfrac{\partial}{\partial \betavec} \log p(\btheta_1,\bSigalph) = \dfrac{\partial}{\partial \betavec} \log N(\betavec|\mubeta,\Sbeta) = -\Sbeta^{-1}\left( \betavec - \mubeta \right),\\
        \dfrac{\partial}{\partial \bmualph}\log p(\btheta_1,\bSigalph) = \sum\limits_{j=1}^J\bSigalphinv(\balph_j-\bmualph) - \Sigmu^{-1}\bmualph,\\
        \dfrac{\partial}{\partial \log a_d}\log p(\btheta_1,\bSigalph) = -\dfrac{\nu}{2} - \alpha_d + \dfrac{\beta_d}{a_d} +\nu_{\alpha}\dfrac{(\bSigalphinv)_{dd}}{a_d} \textrm{, for }d=1,\dots,\Da.\\
        \dfrac{\partial}{\partial \betavec} \log p(\btheta_2|\btheta_1,\by) = \boldsymbol{0}.
    \end{array}
\end{equation*}


\subsubsection{Deriving $\nabla_{\btheta_1}\log p(\btheta_2|\btheta_1,\by)$}
Obviously $\nabla_{\btheta_1}\log p(\btheta_2|\btheta_1,\by)\equiv \nabla_{\btheta_1}\log p(\bSigalph|\btheta_1,\by)\equiv \nabla_{\btheta_1} \log \textrm{IW}(\bSigalph|\nu,\bPsi')$. We now have 
\begin{align*}
	&\log \textrm{IW}(\bSigalph|\nu,\bPsi') = \dfrac{\nu}{2}\logdet(\bPsi')-\dfrac{1}{2}\tr (\bPsi'\bSigalphinv)+\text{constant}.\\
	&\propto \dfrac{\nu}{2}\logdet\left( \sum\limits_{j=1}^J (\balph_j-\bmualph)(\balph_j-\bmualph)^\top + 4\diag\left( a^{-1}_1,\dots,a^{-1}_{\Da}\right) \right)\\
	&-\dfrac{1}{2}\tr \left( \left( \sum\limits_{j=1}^J (\balph_j-\bmualph)(\balph_j-\bmualph)^\top + 4\diag\left( a^{-1}_1,\dots,a^{-1}_{\Da}\right) \right)\bSigalphinv \right).
\end{align*}
Hence, 


\begin{equation*}
    \begin{array}{l}
        \dfrac{\partial}{\partial \balph_j}\log \textrm{IW}(\bSigalph|\nu,\bPsi') = \left( \nu\left(\bPsi'\right)^{-1}- \bSigalphinv\right) (\balph_j-\bmualph), \textrm{ for }j = 1,\dots,J,\\
        \dfrac{\partial}{\partial \betavec} \log p(\btheta_2|\btheta_1,\by) = \boldsymbol{0},\\
        
        \dfrac{\partial}{\partial \bmualph}\log \textrm{IW}(\bSigalph|\nu,\bPsi')  =\left(\bSigalphinv -\nu\left(\bPsi'\right)^{-1}\right) \sum\limits_{j=1}^J(\balph_j-\bmualph),\\
        \dfrac{\partial}{\partial \log a_d}\log \textrm{IW}(\bSigalph|\nu,\bPsi') =
-\dfrac{\va}{a_d}\left(\nu \left(\left(\bPsi'\right)^{-1}\right)_{dd} +  \left(\bSigalphinv \right)_{dd} \right),\textrm{ for }d=1,\dots,\Da.\\
        
    \end{array}
\end{equation*}	


\subsection{VB for DDM \label{supp:sec-VB-HDDM}}

We now consider VB for the DDM. \citet{dao2022efficient} consider LBA models having the same hierarchical structure and show that the posterior distribution can be factored as
\begin{equation*}
    p(\balph_{1:J},\bmualph,\bSigalph,\ba|\by) = p(\balph_{1:J},\bmualph,\ba|\by)p(\bSigalph|\balph_{1:J},\bmualph,\ba,\by),
\end{equation*}
where  $p(\bSigalph|\balph_{1:J},\bmualph,\ba,\by)$ is the density of $\textrm{IW}(\bSigalph|\nu,\bPsi')$
with $\nu = 2\Da + J + 1$ and $\bPsi' = \sum_{j=1}^J (\balph_j - \bmualph)(\balph_j - \bmualph)^\top + 4\diag\left(a^{-1}_1,\dots,a^{-1}_{\Da}\right) $. Denote $\btheta_1 = (\bmualph,\log \ba)$. To exploit the structure of the posterior distribution, \citet{dao2022efficient} propose the Hybrid Gaussian VB of the form
\begin{equation*}\label{eq:lba-hybrid-vb}
q_{\blamb}(\balphJ,\btheta_1,\bSigalph) = N(\balphJ,\btheta_1|\bmulamb ,\Blamb \Blamb^\top + \Dlamb^2) \textrm{IW}(\bSigalph|\nu,\bPsi').
\end{equation*}

The LBA likelihood is equal to 0 if the observed response time $RT$ is less than the non-decision time $\tau$. Hence, the support\footnote{The support of a density is understood as a region where the density is positive. The rigorous definition is omitted as it requires more advanced mathematical concepts which are beyond the scope of this paper.} of the posterior density should not cover any regions where the likelihood is 0. However, the proposed Hybrid Gaussian VB does not capture this important feature. Instead, due to the tractability and simplicity of LBA density, \citeauthor{dao2022efficient} handles the issue by numerically truncating the likelihood (set extreme values equal to some predetermined threshold). Compared to LBA models, DDM is more mathematically intractable, hence a better approach is needed. We improve the variational density to tackle the irregularity issue of the diffusion density as follows.
Given the data for subject $j$,  the lower limit of the non-decision time $\Ltauj:= \mutauj - \dfrac{\stauj}{2}$ must be smaller than any observed response time, i.e., 
\begin{equation}\label{eqs:extra_constraints}
    \Ltauj:= \mutauj - \dfrac{\stauj}{2}< \minRT_j:= \min\limits_{i \in \{ 1,\dots, n_j\} } \text{RT}_{ij}, \textrm{ for all } j=1,\dots,J.
\end{equation}
Because these extra constraints in equation (\ref{eqs:extra_constraints}) only arise after the data becomes available, they are not captured in the transformed random effects $\balphJ$ (see section \ref{supp:sec-HDDMtransformation} of the online supplement), meaning that the choice $$q_{\blamb}(\balphJ,\btheta_1)=N(\balphJ,\btheta_1|\bmulamb,\Blamb\Blamb^\top + \Dlamb^2)$$ 
is now unsuitable. To derive a better parametric form for $q_{\blamb}(\balphJ,\btheta_1)$, we need to introduce some new transformations on the non-decision time parameters. Instead of using $\alpha_6 = \log(\mutauj - 0.5\stau)$, we replace it with the new transformation $$\alphtilde_6 = \log\left(\dfrac{\minRT_j - (\mutauj - 0.5\stauj)}{(\mutauj - 0.5\stauj)}\right).$$ Notice that we only need to change the transformation on the non-decision time parameters ($\alpha_6$), with all the other transformations unchanged. The new transformed random effects, denoted by $\balphtildeJ$, are now unrestricted, so a multivariate Gaussian approximation can be used.

\begin{equation*}
    q_{\blamb}(\balphtildeJ,\btheta_1)=N(\balphtildeJ,\btheta_1|\bmu,BB^\top + D^2).
\end{equation*}
A better choice for $q_{\blamb}(\balphJ,\btheta_1)$ is the one induced from $q_{\blamb}(\balphtildeJ,\btheta_1),$ i.e.

\begin{equation*}
    q_{\blamb}(\balphJ,\btheta_1) = N\left(T_{\by}(\balphJ),\btheta_1|\bmu,BB^\top + D^2\right) \left|J_{\bthetatilde_{\by}\rightarrow\btheta} \right|,
\end{equation*}
where
\begin{equation*}
    \left|J_{\bthetatilde_{\by}\rightarrow\btheta} \right| = \prod\limits_{j=1}^J\dfrac{\partial \alphtilde_{j;6} }{\partial \alpha_{j;6}}= \prod\limits_{j=1}^J \dfrac{\minRT_j}{e^{\alpha_{j;6}} - \minRT_j},
\end{equation*}
is the Jacobian of the function transforming $\bthetatilde_{\by} = (\balphtildeJ,\btheta_1) $ to $\btheta = (\balphJ,\btheta_1)$. To see this, note that
\begin{equation*}
    \alphtilde_{j;6} = \log\left(\dfrac{\minRT_j - (\mutauj - 0.5\stauj)}{(\mutauj - 0.5\stauj)}\right) = \log\left(\dfrac{\minRT_j - e^{\alpha_{j;6}} }{e^{\alpha_{j;6}}}\right),\textrm{ for }j = 1,\dots,J.
\end{equation*}
We obtain
\begin{equation*}
    \dfrac{\partial \alphtilde_{j;6} }{\partial \alpha_{j;6}} = \dfrac{\minRT_j}{e^{\alpha_{j;6}} - \minRT_j},\textrm{ for }j = 1,\dots,J.
\end{equation*}


\section{Particle Metropolis within Gibbs (PMwG)}\label{supp:PMwGalgorithm}
This section gives further details of the PMwG algorithm. Let $\btheta$ be all the model parameters, except the random effects, i.e., $\btheta=(\bbeta,\bmualph,\bSigalph,\ba)$. For convenience, we denote by  $\btheta_{-\bbeta}$ all the model parameters except $\bbeta$, i.e., $\btheta_{-\bbeta} = (\bmualph,\bSigalph,\ba)$. We similarly define  $\btheta_{-\bmualph},\btheta_{-\bSigalph},$ and $\btheta_{-\ba}$. With the model structure and the choice of prior distribution introduced in Section \ref{sec:reg-eams-model-specification}, the full conditional densities of $\bmualph,\bSigalph,$ and $\ba$ (hyperparameters) are known, hence sampling from these distributions is straightforward. However, the full conditional densities $p(\balph_j|\btheta,\by_j), j = 1,\dots,J,$ and $p(\bbeta|\balphJ,\btheta_{-\bbeta},\by) $ are 
unavailable in closed form, so sampling directly from them is difficult.

Algorithm~\ref{algo:PMwG-eam-reg} is the PMwG sampler for estimating the EAMs with  covariates. The algorithm starts by initialising the values for $\btheta,\bbeta,$ and random effects $\balphJ$. The sampler then runs for a predetermined number of iterations $T_{iter}$. In each iteration, we sequentially sample the group-level parameters $\bmualph$ (Step 2a), $\bSigalph$ (Step 2b) and the hyperparameters $\ba$ (Step 2c) using Gibbs step conditional on the selected particles $\balphJk$ from the previous iteration. In Step 2d, $R-1$ new particles are generated using the conditional MC method given in algorithm~\ref{algo:CMC_alpha} in section \ref{supp:PMwGalgorithm}. Note that we keep the particles $\balphJk$ fixed and set the first set of particles $\balphJ^1 = \balphJk$. The conditional MC algorithm gives the particles $\balphJR$ and the normalised weights $W_j^{(r)},\textrm{ for } j=1,\dots,J \textrm{ and } r= 1,\dots,R.$ Step 2e samples the new index vector $\bk = (k_1,\dots,k_J)$ and the random effects are selected according to the index vector $\bk$, i.e., we discard the rest of the particles $\balphJ^{(-\bk)}$ and only keep $\balphJk$. 
Similar steps (2f and 2g)  are applied to sample $\bbeta$. The PMwG sampler in Algorithm \ref{algo:PMwG-eam-reg} is for the hierarchical EAM with covariates. When there are no covariates, steps (2f) and (2g) can be omitted.

\begin{algorithm}
\caption{Particle Metropolis within Gibbs (PMwG) for the hierarchical EAM with covariates}\label{algo:PMwG-eam-reg}
\begin{enumerate}
    \item Initialisation: select initial values for $\balphJ,\btheta$ and set $\balph^1_j = \balph_j^{k_j} , \bbeta^1 = \bbeta^{\kb}$.
    \item For $t=1,\dots,T_{iter}$:
    \begin{itemize}
        \item[(a)] Sample $\bmualph|\bk,\balphJ^{\bk},\btheta_{-\bmualph},\by$ from $N(\overline{\bmu},\overline{\bSig})$, where
        $$\begin{array}{ll}
             \overline{\bmu} &= \overline{\bSig}\left(\bSigalphinv\sum\limits_{j = 1}^J\balph_j^{k_j} \right), \\
             \overline{\bSig} &=\left( J\bSigalphinv + I\right)^{-1}.
        \end{array}$$
        \item[(b)] Sample $\bSigalph|\bk,\balphJ^{\bk},\btheta_{-\bSigalph}$ from $\textrm{IW}(k_{\alpha},B_{\alpha}),$ where
         $$\begin{array}{ll}
             k_{\alpha} &= \va + \Da -1 +J,  \\
             B_{\alpha} &= 2\va \diag\left( a_1^{-1},\dots,a_1^{-1} \right) + \sum\limits_{j=1}^J(\balph_j^{k_j} - \bmualph)(\balph_j^{k_j} - \bmualph)^\top.
         \end{array}$$
         \item[(c)] Sample $a_d|\bk,\balphJ^{\bk},\btheta_{-a_d}$ from $\textrm{IG}(\overline{\alpha},\overline{\beta}), $ where
         $$\begin{array}{ll}
             \overline{\alpha} = & \dfrac{\va + \Da}{2}, \\
             \overline{\beta} = & \va \left( \bSigalphinv \right)_{dd} + \dfrac{1}{\mathcal{A}^2_d}, \textrm{ for }d=1,\dots,\Da.
         \end{array}$$
         \item[(d)] Sample $\balphJ^{-\bk}$ using Algorithm \ref{algo:CMC_alpha} in section \ref{supp:PMwGalgorithm} of the online supplement
         \item[(e)] Sample the index vector $\bk = (k_1,\dots,k_J)$ with probability given by
         $$P(k_1=l_1,\dots,k_J = l_J|\balph_{1:J},\bbeta,\bmualph,\bSigalph) = \prod\limits_{j = 1}^J W_j^{l_j}.$$
         \item[(f)] Sample $\bbeta^{-\kb}$ using Algorithm \ref{algo:CMC_beta} in section \ref{supp:PMwGalgorithm} of the online supplement
         \item[(g)] Sample the index $\kb$ with probability given by
         $$P(\kb=l|\balph_{1:J},\bbeta,\bmualph,\bSigalph) = W_{\beta}^{l}.$$
    \end{itemize}
\end{enumerate}
\end{algorithm}

\subsection{Conditional Monte Carlo Algorithm \label{CMCalgorithm}} 
This section presents the conditional Monte Carlo (CMC) algorithm used in the PMwG algorithm. The CMC is required to sample the random effects $\balphJ$ (Algorithm \ref{algo:CMC_alpha}) and the parameter $\bbeta$ (Algorithm \ref{algo:CMC_beta}).

\begin{algorithm}
\caption{Conditional Monte Carlo for sampling $\balph_{1:J}$}\label{algo:CMC_alpha}
\textbf{Require:} proposal densities $\malpha,j=1,\dots,J.$
\begin{enumerate}
	\item Fix $\balphJ^{(1)} = \balphJ^{(\bk)}.$\\
	
	\item For $j = 1,\dots,J$
	\begin{itemize}
	    \item Sample $R-1$ particles $\balph_j^{(r)},r=2,\dots,R$ from the proposal $\malpha$.
	    \item Compute the weights
    	$$ \omega_j^{(r)} = \dfrac{p(\by_j|\balph^{(r)}_j,\bbeta,\bX_j)p(\balph^{(r)}_j|\bmualph,\bSigalph)}{m_j(\balph_j^{(r)}|\by_j,\bX_j,\bbeta,\bmualph,\bSigalph) },\textrm{ for } r = 1,\dots,R.$$
    	\item Normalize the weights
    	$$ W_j^{(r)}=\dfrac{\omega_j^{(r)}}{\sum\limits_{r=1}^R \omega_j^{(r)}}, \textrm{ for } r = 1,\dots,R.$$
	\end{itemize}
\end{enumerate}
\end{algorithm}

\begin{algorithm}
\caption{Conditional Monte Carlo for sampling $\bbeta$}\label{algo:CMC_beta}
\textbf{Require:} proposal density $m(\bbeta|\by,\balphJ,\bX).$
\begin{enumerate}
	\item Fix $\bbeta^{(1)} = \bbeta^{\kb}.$\\
	
	\item Sample $R-1$ particles $\bbeta^{(r)},r=2,\dots,R$ from the proposal $m(\bbeta|\by,\balphJ,\bX)$.
	\item Compute the weights
    $$ \omega_{\beta}^{(r)} = \dfrac{ \prod\limits_{j=1}^J p(\by_j|\balph_j,\bbeta^{(r)},\bX_j)p(\bbeta^{(r)}|\mubeta,\Sbeta)}{m(\bbeta^{(r)}|\by,\balphJ,\bX)}, \textrm{ for } r = 1,\dots,R.$$
    \item Normalise the weights
    $$ W_{\beta}^{(r)}=\dfrac{\omega_{\beta}^{(r)}}{\sum\limits_{r=1}^R \omega_{\beta}^{(r)}}, \textrm{ for } r = 1,\dots,R.$$
\end{enumerate}
\end{algorithm}

\subsection{Proposal distributions and tuning parameters}
The efficiency of the PMwG sampler depends on a careful choice of the proposal densities $m_j(\balph_j|\by_j,\bmualph,\bSigalph),j = 1,\dots,J$ for the random effects and $m(\bbeta_{vec}|\by_j,\balph_j),j = 1,\dots,J$ for the parameters $\bbeta$. This section discusses how we select and tune these distributions. As recommended by \citeauthor{gunawan2020new}, the proposal densities are adaptively improved over time. 
At the first stage, the proposal density for subject $j$ at iteration $t$ is chosen to be a mixture of the prior $N(\balph_j|\bmualph^{(t)},\bSigalph^{(t)})$ and a normal distribution centred on the previous sample for the random effect, $N(\balph_j^{(t-1)},\epsilon\boldsymbol{\Sigma}^{(t)}_{\balph} )$, i.e.,
\begin{equation*}
    \malpha = \omega N(
\balph_j|\balph_j^{(t-1)},\epsilon\boldsymbol{\Sigma}^{(t)}_{\balph}  ) + (1-\omega) N(\balph_j|\boldsymbol{\mu}^{(t)}_{\balph} ,\boldsymbol{\Sigma}^{(t)}_{\balph}  ).
\end{equation*}
The scale parameter $\epsilon$ ($0<\epsilon<1$) is used to control the "step-size" and is determined by trial-and-error. The scale parameter should be small when the number of random effects is large. We set $\epsilon = 0.5$. \citet{wall2021identifying} set  $\epsilon = 0.1$ for estimating a hierarchical LBA with 110 subjects, each having 30 random effects.
Similarly, the proposal density for sampling $\bbeta$ is the mixture of a Gaussian distribution centred at the previous draw $\bbeta^{(t-1)}$ with the block diagonal covariance matrix $\Szero$  having $\Da$ blocks with the matrix $(\bX^\top \bX)^{-1}$ on each diagonal block and the prior

\begin{equation*}
    \mbeta = \omega N(
\betavec|\betavec^{(t-1)},\epsilon\Szero ) + (1-\omega) N(\betavec|\mubeta,\Sbeta ).
\end{equation*}

After the burn-in and initial adaptation stages, at iteration $t (t>T_{burn} + T_{adapt})$, more efficient proposal densities for subject $j$ are obtained by first approximating the joint posterior distribution $p(\balph_j,\bmualph,\bSigalph,\bbeta|\by)$ by fitting a multivariate Gaussian distribution using the draws from the adaptation stage
$\left\{ \balph_j^{(i)},\bmualph^{(i)},\bSigalph^{(i)},\bbeta^{(i)} \right\}_{i = T_\textrm{burn} +1 }^{t-1}$. From this multivariate Gaussian, we obtain the conditional distribution of $\balph_j$ given $\bmualph=\boldsymbol{\mu}^{(t)}_{\balph},\bSigalph=\boldsymbol{\Sigma}^{(t)}_{\balph}, \bbeta = \bbeta^{(t)}$ and $\by,$ which we denote by $N(\widehat{\bmu}_j,\widehat{\bSig}_j)$.
The process of updating better proposal distributions can be done at each iteration. However, for computational efficiency, we update new proposals after every 20 iterations and stop updating when the proposal distributions stablise (after 5,000 iterations).

The proposal density for subject $j$ is taken as a`  mixture of the conditional distribution $N(\hat{\bmu}_j,\hat{\bSig}_j)$, the Gaussian distribution centred at the previous draws with the covariance matrix $\hat{\bSig}_j$ and the prior group-level distribution for the random effects, $N(\boldsymbol{\mu}^{(t)}_{\balph},\boldsymbol{\Sigma}^{(t)}_{\balph}  )$,
\begin{equation}
\label{Eq:prop_alpha}
    \malpha = \omega_1 N(
\balph_j|\hat{\bmu}_{j},\hat{\Sigma}_j ) + \omega_2 N(
\balph_j|\balph_j^{(t)},\hat{\Sigma}_j ) + \omega_3 N(\balph_j|\boldsymbol{\mu}^{(t)}_{\balph},\boldsymbol{\Sigma}^{(t)}_{\balph}  ).
\end{equation}
We set $\omega_1 = 0.65,\, \omega_2 = 0.3$ and $\omega_3=0.05$. Following \citet{hesterberg1995weighted}, including the prior density $p(\boldsymbol{\alpha}_j|\boldsymbol{\mu_\alpha},\boldsymbol{\Sigma_\alpha})$
in Equation \eqref{Eq:prop_alpha} with small weight $\omega_3=0.05$ ensures that the importance weights are bounded. We put a larger weight $\omega_1 = 0.65$ on the most efficient proposal $N(\balph_j|\hat{\bmu}_{j},\hat{\Sigma}_j )$, and a smaller weight $\omega_2 = 0.3$ for the random walk type proposal.

More efficient proposal distributions for $\bbeta$ are obtained similarly. Since the full conditional density $p(\bbeta|\by,\balphJ,\btheta_{-\bbeta})=p(\bbeta|\by,\balphJ)$, we use $\left\{ \balphJ^{(i)},\bbeta^{(i)} \right\}_{i = T_\textrm{burn}+ 1 }^{t-1}$ to fit a multivariate Gaussian. From the fitted Gaussian, obtain the conditional distribution of $\bbeta$ given $\balphJ=\balphJ^{(t)}$ and $\by,$ which we denote by $N(\widehat{\bmu}_{\beta},\widehat{\bSig}_{\beta})$. The better proposal density for $\bbeta$ is now taken to be the mixture of the conditional distribution $N(\widehat{\bmu}_{\beta},\widehat{\bSig}_{\beta})$, the Gaussian distribution centred at the previous draws with the covariance matrix $\widehat{\bSig}_{\beta}$ and the prior,
\begin{equation*}
    \mbeta = \omega_1 N(
\bbeta|\widehat{\bmu}_{\beta},\widehat{\bSig}_{\beta} ) + \omega_2 N(\betavec|\betavec^{(t-1)},\widehat{\bSig}_{\beta} ) + \omega_3 N(\betavec|\mubeta,\Sbeta).
\end{equation*}

Again, we set $\omega_1 = 0.65,\, \omega_2 = 0.3$ and $\omega_3=0.05$.

\section{Integrated Autocorrelation Time (IACT)} \label{supp:IACT}

The \emph{integrated autocorrelated time} (IACT) for a scalar parameter $\theta$ is defined as \citep{liu2001monte}
\begin{equation*}
    \text{IACT}_{\theta} := 1 + 2\sum\limits_{j=1}^{\infty}\rho_{j,\theta},
\end{equation*}
where $\rho_{j,\theta}$ is the correlation of the iterates of $\theta$ in the MCMC after the chain has converged.
The $\text{IACT}_{\theta}$ measures the inefficiency of the sampling scheme with respect to that parameter and can be interpreted as follows: if the $\text{IACT}_{\theta}$ = 100, then we need 100 times as many iterates as an independent scheme to obtain the same  variance of the estimator. Hence,  the larger the value of the $\text{IACT}_{\theta}$, the poorer the performance.

We estimate $\textrm{IACT}_{\theta}$ based on $M$ MCMC iterates $\theta^{\left[1\right]},...,\theta^{\left[M\right]}$ (after convergence) as
\begin{align*}
{\widehat{\rm IACT}}_{\theta,M} &=1+2\sum_{j=1}^{L_{M}}\widehat{\rho}_{j,\theta},
\end{align*}
where $\widehat{\rho}_{j,\theta}$ is the estimate of $\rho_{j,\theta}$, $L_{M}=\min(1000,L)$ and $L=\min_{j\leq M}|\widehat{\rho}_{j,\theta}|<2/\sqrt{M}$ because $1/\sqrt M$
is approximately the standard error of the autocorrelation estimates when the series is white noise.

\section{Initialisation Methods\label{supp:sec-vb-initialisation1}}
The selection of starting values for the experiment in Section \ref{sec:vb-initialization} for the DDM using the domain knowledge is now discussed. 
	
\begin{enumerate}
    \item First, we initialise $\bmulambmu$ (the mean of $\bmualph$) in the natural scale: $v^{(h)} \sim U(-3,-2), v^{(m)} \sim U(1,2), v^{(e)} \sim U(5,6), \sv \sim U(0,2), a\sim U(0.5,2), \muz = a/2, \sz \sim U(0,0.5), \stau \sim U(0,0.1),$ and $\mutau = \minRT + 0.5\stau + 0.1u,$ where $u\sim U(0,0.1)$ and $\minRT$ is the smallest observed response time in the sample.
	To obtain $\bmulambmu$, apply the transformations defined in ~Equation \eqref{eq:eeg-hddm-transform} on these values. 
	
	We now explain how domain knowledge is used in specifying implausible regions for each parameter. Assuming the upper boundary of the diffusion process corresponds to correct responses, it is reasonable to expect, under `hard' conditions, subjects tend to make incorrect decisions, hence, the drift rate $v^{(h)}$ is expected to be negative. For a similar reason, the drift rate under `easy' conditions $v^{(e)}$ is expected to be positive and is greater than the drift rate under `medium' conditions $v^{(m)}$. The plausible intervals for the variability parameters $\sv,\sz,\stau,$ and the boundary separation $a$ are chosen based on the typical range (see ``rtdists'' R package). We set $\muz = a/2$ to exclude bias. Finally, the non-decision time $\tau$ is chosen such as the upper bound for non-decision time $\Ltau:= \mutau - \dfrac{\stau}{2}$ must be greater than any observed response time in the data, hence $\mutau = \minRT + 0.5\stau + 0.1u,$ where the term $0.1u$ is added to represent some random noise. 
    \item Next, we initialise $\bmulambalpha$ (means of the random random effects) by sampling from a prior given the chosen value of the group-level mean $\bmulambmu$ and the group-level covariance which is set to be a diagonal matrix with the diagonal elements equal 0.1, i.e., sample $\balph^0_j \sim N(\bmualph,0.1I), j = 1,\dots, J$ and set $\bmulambalpha = (\balph^0_1,\dots,\balph^0_{\Da})$.
    \item Finally, we set $\bmulambbeta$ (the mean of the coefficients $\bbeta$) and $\bmulambhyper$ (the mean of $\log \ba$) to be zeros. 
\end{enumerate}

\section{Simulation study for DDM}\label{sec:supplement-hddm-simulation}

The true values in the simulation study (both settings) are given by 
\begin{align*}
    \bmualph = (3.31,    2.12,    1.33,   -2.17,   -0.48,   -0.29,   -1.07,  -0.28,   -1.19,   -1.28,   -1.52),\\
    \diag (\bSigalph) = (1.44 , 0.62 , 0.41 , 0.72 , 0.38 , 0.13 , 0.09 , 0.2, 0.11 , 0.15 , 0.08 , 0.19).
\end{align*}
We provide the true values for the group-level mean and variances. The true values for the covariances are available upon request. Note that the same true values are used in both simulation settings $(J=12 \textrm{ and } J = 50)$.

\section{Simulation study for DDM with regressors}\label{supp:sec-reg-eams-simulation}
\subsection{Data simulation}
We study the performance of the proposed methods using simulation. The data is simulated from the DDM with neural covariates. There are 20 subjects, each subject performs 600 trials under three conditions (\lq easy \rq, \lq medium \rq, and \lq hard \rq; there are 200 trials for each condition). In each trial, 5 neural covariates are independently drawn from a Normal distribution with mean $0$ and standard deviation $0.001$. The fixed-effect coefficients are generated independently from a Normal distribution with mean $0$ and standard deviation $2$, i.e., $\beta_{ij}\sim N(0,2^2),$ for $i=1,\dots,\Da; j = 1,\dots,d.$  
\begin{equation*}
 \bbeta=   \begin{bmatrix}
    -0.24 & 3.46 & 2.96 & -0.71 & 0.03\\
    1.1 & -2.16 & 3.03 & -0.19 & 0.1\\
    0.7 & -0.55 & -1.88 & 2.2 & 0.03\\
    0.72 & 0.36 & -0.37 & -3.93 & -0.02\\
    1.8 & 3.02 & -2.2 & -2.9 & 0.13\\
    -3.85 & 3.21 & 2.42 & 2.04 & 0.12\\
    0.52 & -3.68 & -3.25 & -2.84 & 0.02\\
    1.83 & 3.25 & 0.21 & -1.21 & -0.02\\
    0.03 & 0.26 & -2.91 & -3.17 & -0.12\\
    \end{bmatrix}
\end{equation*}
Finally, the group-level parameters $\bmualph$ and $\bSigalph$ are chosen as follows.
\begin{align*}
    \bmualph = (0.65 , 1.63 , 3.53 , 0.09 , -0.41 , -0.46 , -0.59 , -1.21 , -2.14),\\
    \diag (\bSigalph) = (0.24 , 1.39 , 3.83 , 0.26 , 0.13 , 0.19 , 0.18 , 0.03 , 0.45).
\end{align*}
We provide the true values for the group-level mean and variances. The true values for the covariances are available upon request. 

\subsection{Model specification}\label{sec:simulation-model}
We consider fitting the DDM having the mixture form \eqref{eq:reg_mixture} with the weight $\overline{\omega} = 0.0001$. Note that the true data generating process is DDM with no contaminant reaction times ($\overline{\omega} = 0$). To capture the trial difficulty, different values for the drift rate means are considered: $\muv^{(h)}$ (drift rate mean under the hard condition), $\muv^{(m)}$ (drift rate mean under the moderate condition), and $\muv^{(e)}$ (drift rate mean under the easy condition). The remaining model parameters are held constant over all conditions. Each subject has 9 random effects
$$ \bldeta := (\muv^{(h)},\muv^{(m)},\muv^{(e)},\sv,a,\muz,\sz,\mutau,\stau). $$

To capture the natural constraints of the random effects $\bldeta$, we apply proper transformations\footnote{The transformations are in Section~\ref{supp:sec-regeam-simulation-transform}.} and denote the transformed random effects by $\balph$.
For demonstration purposes, we link all diffusion parameters to the neural covariates.

\subsection{Results}

The performance of the PMwG sampler is assessed based on the IACT estimates (efficiency) and the true parameter recovery (accuracy). Tables~\ref{tab:PMwG_HDDM_mimicEEGdata_medium_nonzero_beta1} and \ref{tab:PMwG_HDDM_mimicEEGdata_medium_nonzero_beta2} show the IACT estimates for the fixed-effect coefficients and the group-level parameters, respectively. Overall, the PMwG sampling scheme is efficient as most of the IACT estimates are small.

\begin{table}[H]
    \centering
    \begin{tabular}{c|c||c|c||c|c||c|c||c|c}
     & IACT &  & IACT &  & IACT &  & IACT &  & IACT \\
     \hline
     $\beta_{1  1}$&1.01&$\beta_{1  2}$&1.00&$\beta_{1  3}$&1.03&$\beta_{1  4}$&1.00&$\beta_{1  5}$&1.04\\
$\beta_{2  1}$&1.00&$\beta_{2  2}$&1.03&$\beta_{2  3}$&1.02&$\beta_{2  4}$&1.02&$\beta_{2  5}$&1.03\\
$\beta_{3  1}$&1.03&$\beta_{3  2}$&1.00&$\beta_{3  3}$&1.02&$\beta_{3  4}$&1.02&$\beta_{3  5}$&1.02\\
$\beta_{4  1}$&1.02&$\beta_{4  2}$&1.00&$\beta_{4  3}$&1.02&$\beta_{4  4}$&1.02&$\beta_{4  5}$&1.01\\
$\beta_{5  1}$&1.02&$\beta_{5  2}$&1.05&$\beta_{5  3}$&1.01&$\beta_{5  4}$&1.04&$\beta_{5  5}$&1.02\\
$\beta_{6  1}$&1.01&$\beta_{6  2}$&1.00&$\beta_{6  3}$&1.04&$\beta_{6  4}$&1.00&$\beta_{6  5}$&1.01\\
$\beta_{7  1}$&1.00&$\beta_{7  2}$&1.02&$\beta_{7  3}$&1.01&$\beta_{7  4}$&1.02&$\beta_{7  5}$&1.04\\
$\beta_{8  1}$&1.11&$\beta_{8  2}$&1.12&$\beta_{8  3}$&1.03&$\beta_{8  4}$&1.04&$\beta_{8  5}$&1.06\\
$\beta_{9  1}$&1.03&$\beta_{9  2}$&1.03&$\beta_{9  3}$&1.00&$\beta_{9  4}$&1.01&$\beta_{9  5}$&1.01\\
    \end{tabular}
    \caption[DDM Regression - Simulation Study. IACT estimates of the coefficients $\bbeta$.]{IACT estimates of the fixed effect coefficients $\bbeta$ using the PMwG draws.}
    \label{tab:PMwG_HDDM_mimicEEGdata_medium_nonzero_beta1}
\end{table}

\begin{table}[H]
    \centering
    \begin{tabular}{c|c||c|c}
    Parameter & IACT & Parameter & IACT\\
    
     \hline
     
     $\bmualph_{1}$&3.33&$\bSigalph_{1}$&5.28\\
$\bmualph_{2}$&1.78&$\bSigalph_{2}$&2.27\\
$\bmualph_{3}$&2.19&$\bSigalph_{3}$&3.11\\
$\bmualph_{4}$&8.03&$\bSigalph_{4}$&7.99\\
$\bmualph_{5}$&14.31&$\bSigalph_{5}$&14.99\\
$\bmualph_{6}$&7.25&$\bSigalph_{6}$&6.88\\
$\bmualph_{7}$&11.29&$\bSigalph_{7}$&14.42\\
$\bmualph_{8}$&2.20&$\bSigalph_{8}$&2.14\\
$\bmualph_{9}$&2.80&$\bSigalph_{9}$&3.43\\
    \end{tabular}
    \caption[DDM Regression - Simulation Study. IACT estimates of the group-level parameters.]{IACT estimates for the group-level mean $\bmualph$ and variances (diagonal of $\bSigalph$) using the PMwG draws.}
    \label{tab:PMwG_HDDM_mimicEEGdata_medium_nonzero_beta2}
\end{table}

\begin{figure}[H]
	\hspace{-0.5cm}\includegraphics[scale=0.25]{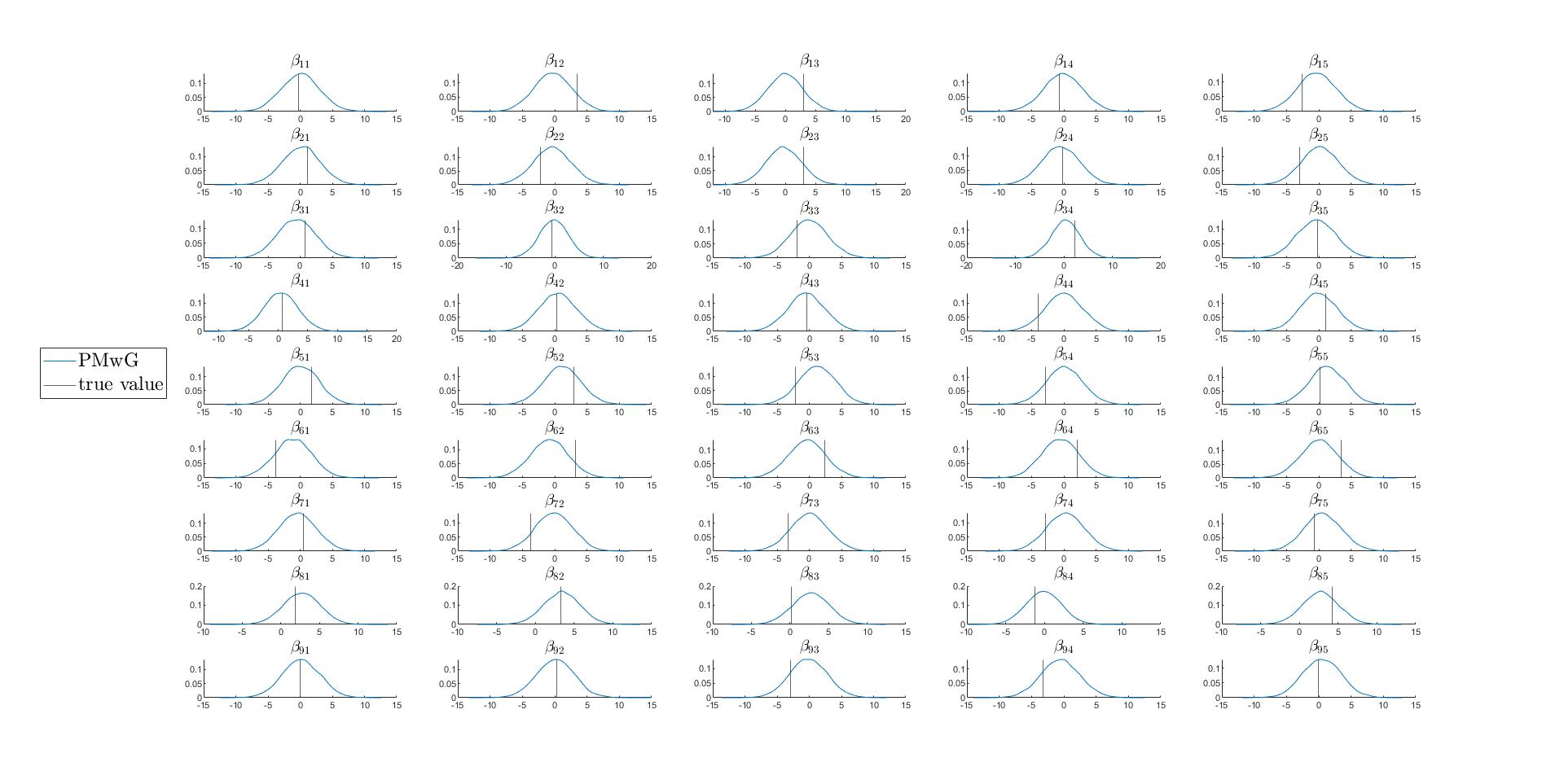}
	\caption[DDM Regression - Simulation Study. PMwG posterior density estimates of the coefficients.]{Posterior marginal densities of $\bbeta$ estimated from PMwG (blue) and the true values indicated by the black vertical lines.} \label{fig:HDDM_Reg_EEGdatabased_nonzero_beta_medium_PMwG_density_beta}
\end{figure}

\begin{figure}[H]
	\hspace{-0.5cm}\includegraphics[scale=0.25]{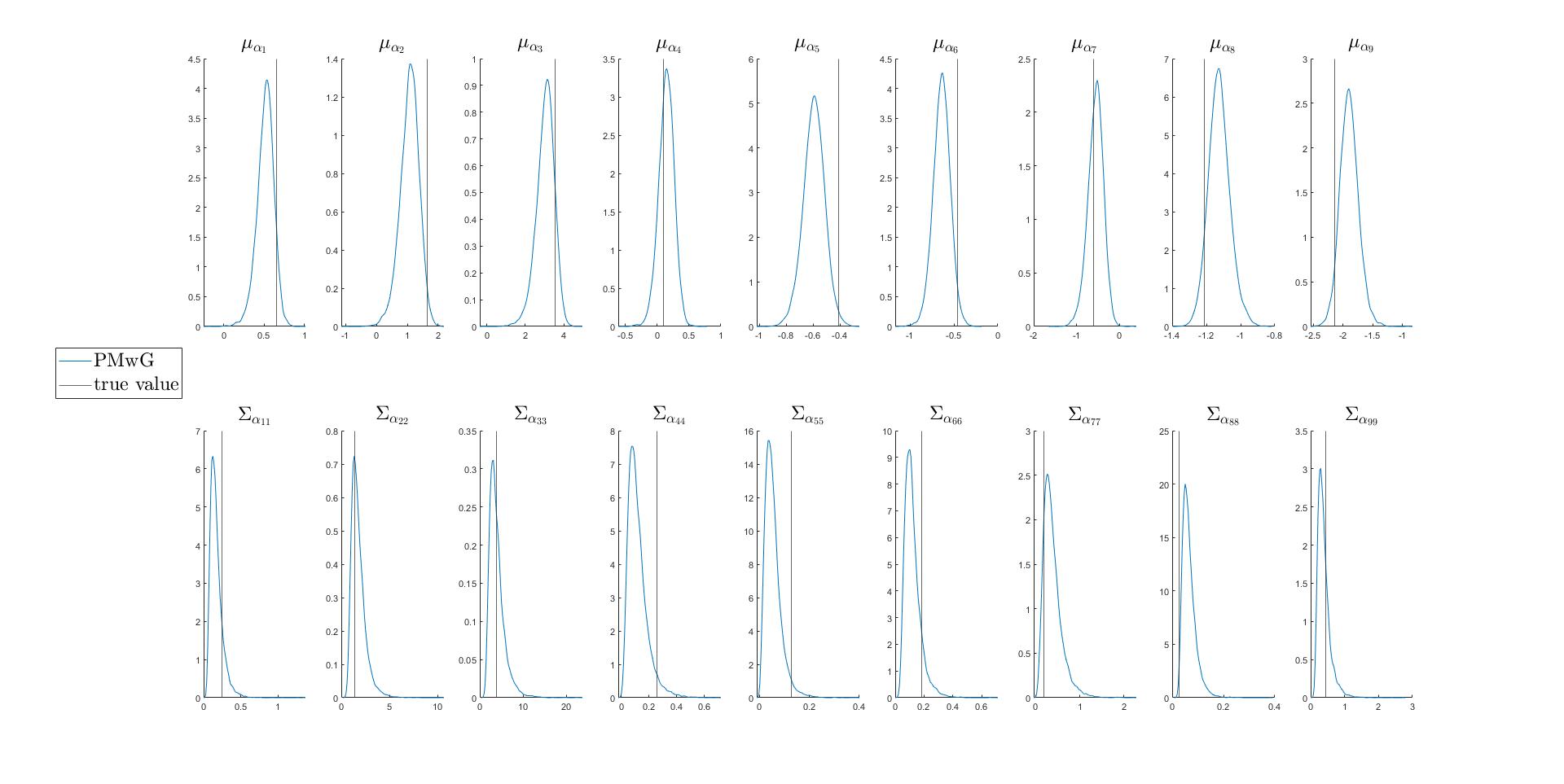}
	\caption[DDM Regression - Simulation Study. PMwG posterior density estimates of the group-level parameters.]{Posterior marginal densities estimated from PMwG of the group-level mean $\bmualph$ (top panels) and variance $\diag(\bSigalph)$ (bottom panels); the true values represented by the black vertical lines.} \label{fig:HDDM_Reg_EEGdatabased_nonzero_beta_medium_PMwG_density_group_level}
\end{figure}

Figures~\ref{fig:HDDM_Reg_EEGdatabased_nonzero_beta_medium_PMwG_density_beta} and \ref{fig:HDDM_Reg_EEGdatabased_nonzero_beta_medium_PMwG_density_group_level} display the marginal posterior densities estimated by the PMwG sampler of the fixed-effect coefficients $\bbeta$ and the group-level parameters, respectively. The figures show that the true values (represented by vertical lines) are recovered accurately.

\begin{figure}[H]
		\centering
		
        \hspace*{-2cm} \includegraphics[scale=0.22]{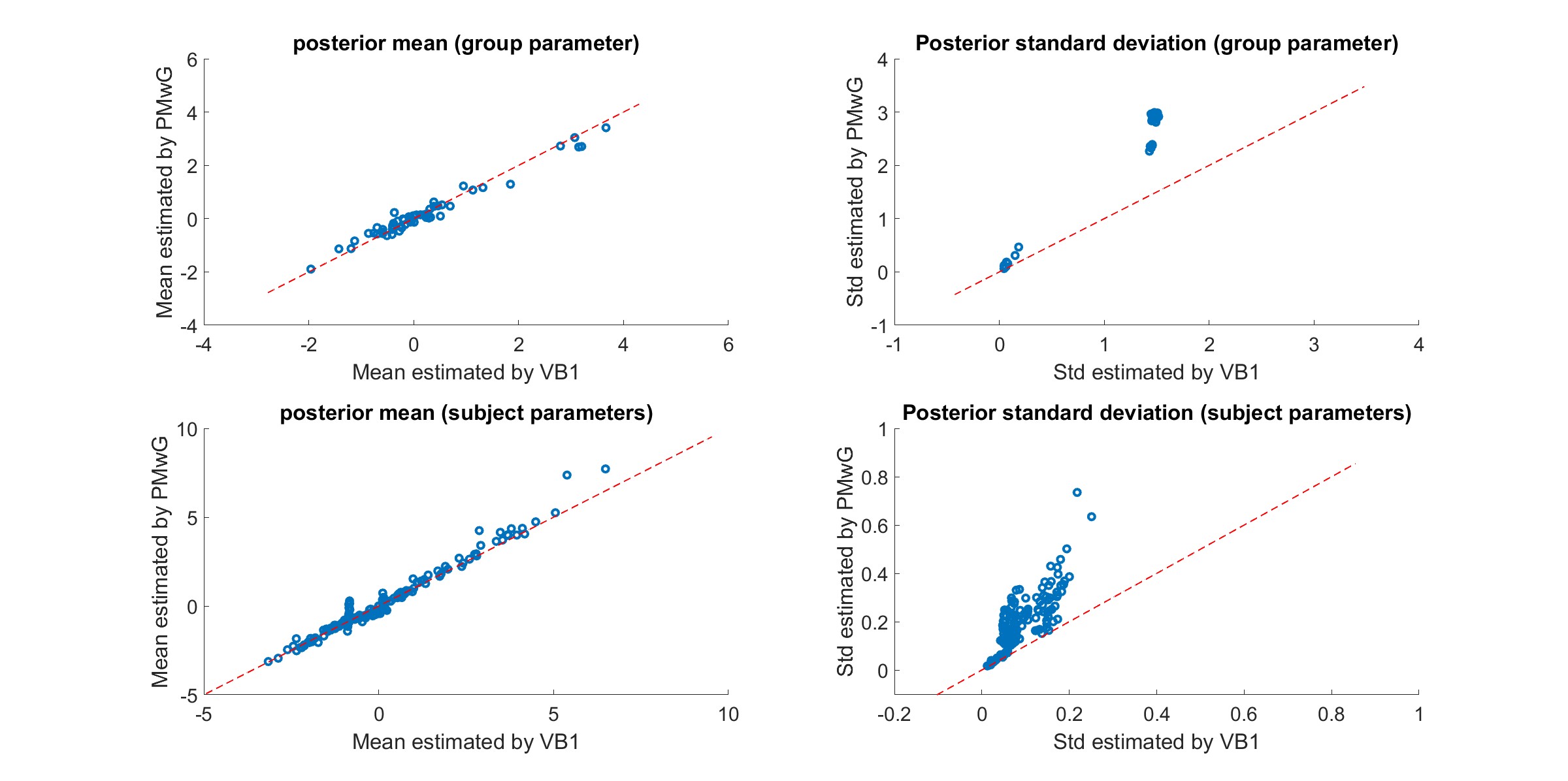}
		\caption[DDM Regression - Simulation Study. Comparing PMwG and VB using different starting points.]{Comparing the means and standard deviations of the marginal posterior distributions estimated by VB (horizontal axis) against the exact values calculated using PMwG (vertical axis). The top panels show the means and standard deviations of the group-level parameters. The bottom panels show the means and standard deviations of the random effects.}\label{fig:HDDM_Reg_EEGdatabased_nonzero_beta_medium_moment_plot}
\end{figure}

We now study the performance of the VB approximations. The overall accuracy is assessed by comparing the posterior means and the posterior standard deviations, estimated by PMwG and VB, of all the model parameters and the subject parameters as shown in Figure~\ref{fig:HDDM_Reg_EEGdatabased_nonzero_beta_medium_moment_plot}. The main benefit of using VB is its computational efficiency, which is shown in  Table~\ref{tab:HDDM_Reg_EEGdatabased_nonzero_beta_medium}. VB is 14 times faster than PMwG, and requires much less computational resources. 
\begin{table}[H]
    \centering
    \begin{tabular}{c|l|c}
   \multirow{2}{*}{\textbf{Method}} & \multicolumn{1}{|c|}{\textbf{Running time}} & \textbf{Number of} \\
    & \multicolumn{1}{|c|}{\textbf{(hour:minute)}}  & \textbf{cpu-cores} \\
    \hline
    \hline
    PMwG & \multicolumn{1}{|c|}{41:47} & 48 \\
    \hline
    VB & \multicolumn{1}{|c|}{3:05} & 8\\
     \hline
    \end{tabular}
    \caption[DDM Regression - Simulation Study. Computation time comparison between PMwG and VB.]{A comparison between PMwG and VB in terms of running time and computational resources in the simulation study.}
    \label{tab:HDDM_Reg_EEGdatabased_nonzero_beta_medium}
\end{table}

\section{Simulating predicted data from EAMs}\label{sec:simulate-pred-data}
This section discusses how to simulate predicted data from a RegEAM and estimate the predictive posterior distributions for the summary statistics considered in Section~\ref{sec:mental-rotation}. Suppose we obtain draws $\{ \balphJ^{(t)},\bbeta^{(t)},\bmualph^{(t)},\bSigalph^{(t)}\}_{t=1}^T$ from the posterior $p(\balphJ,\bbeta,\bmualph,\bSigalph|\by,\bX)$ by using PMwG. For each posterior draw, we simulate the predicted data according to the model and calculate the summary statistics. Algorithm~\ref{supp:algo-pred-reg-eam} gives a step-by-step explanation.

\begin{algorithm}
\caption{Predicted data simulation for EAMs}\label{supp:algo-pred-reg-eam}
\begin{enumerate}
    \item Input: posterior draws $\{\balphJ^{(t)},\bbeta^{(t)},\bmualph^{(t)},\bSigalph^{(t)}\}_{t=1}^T \sim p(\balphJ,\bbeta,\bmualph,\bSigalph|\by,\bX)$.
    \item For $t=1,\dots,T$:
    \begin{enumerate}
        \item  Generate  predicted data for subject $j (j=1,\dots,J)$ as 
        \begin{equation*}
            \tilde{y}^{(t)}_{ij}:= (\predRTij, \predREij) \sim p(y_{ij}|\widetilde{\balph}^{(t)}_{ij}),  \widetilde{\balph}^{(t)}_{ij} = \balph^{(t)}_j + \bbeta^{(t)} X_{ij}, \textrm{ for } i = 1,\dots,n_j.
        \end{equation*}
        \item From the predicted data $\left\{ \predRTij, \predREij \right\}$, we calculate the summary statistics. 
        \begin{enumerate}
            \item The median response time statistics corresponding to stimulus type $S=s$ ($s\in \{\textrm{``same'', ``mirror''} \})$ and rotation condition $E=e$ ($e\in \{0,45,90,135,180 \})$, denoted by $\medse$, is calculated as follows.
            \begin{equation*}
                \medse=\dfrac{1}{J}\sum\limits_{j=1}^J\medsej,
            \end{equation*}
            with
            \begin{equation*}
                \medsej = \textrm{Med} \left( \left\{ \predRTjse \right\} \right),
            \end{equation*}
            where $\textrm{Med}(\{ U \})$ denotes the sample median of a set $U$, and $\predRTjse$ represents all the predicted response time data of subject $j$ from the trials in which the stimulus type $S=s$ and the rotation condition $E=e$. 

            \item The mean accuracy statistics corresponding to stimulus type $S=s$ ($s\in \{\textrm{``same'', ``mirror''} \})$ and rotation condition $E=e$ ($e\in \{0,45,90,135,180 \})$, denoted by $\accuracyse$, is calculated as follows.
            \begin{equation*}
                \accuracyse = \dfrac{1}{J}\sum\limits_{j=1}^J\accuracysej,
            \end{equation*}
            where $\accuracysej$ is the proportion of correct response of subject $j$ calculated based on trials with stimulus type $S=s$ and rotation condition $E=e$.
        \end{enumerate}
    \end{enumerate}
    \item Construct the box-plot of the summary statistics based on the predicted values $\left\{ \medse, \accuracyse  \right\}_{t=1}^T$.
\end{enumerate}
\end{algorithm}

\section{Further results}\label{supp:sec-further-results}
This section represents further estimation results of the simulated and real data.

\subsection{Simulation Study \label{additionalresultssimstudy}}

\begin{table}[H]
    \centering
    \begin{tabular}{c|c|c|c|c|c}
    Parameter & \multicolumn{2}{|c|}{IACT} & Parameter & \multicolumn{2}{|c}{IACT} \\
    &$J=12$&$J=50$&  &$J=12$&$J=50$\\
    \hline
$\bmualph_{1}$&3.56&1.19&$\bSigalph_{1}$&3.82&1.79\\
$\bmualph_{2}$&4.04&1.34&$\bSigalph_{2}$&4.72&2.04\\
$\bmualph_{3}$&3.15&1.23&$\bSigalph_{3}$&3.33&1.83\\
$\bmualph_{4}$&3.65&1.13&$\bSigalph_{4}$&3.86&1.41\\
$\bmualph_{5}$&3.47&3.85&$\bSigalph_{5}$&4.24&5.37\\
$\bmualph_{6}$&4.55&1.93&$\bSigalph_{6}$&4.95&1.89\\
$\bmualph_{7}$&10.95&4.59&$\bSigalph_{7}$&4.90&4.39\\
$\bmualph_{8}$&2.63&1.70&$\bSigalph_{8}$&2.85&1.78\\
$\bmualph_{9}$&8.06&3.14&$\bSigalph_{9}$&5.02&3.09\\
$\bmualph_{10}$&69.28&16.05&$\bSigalph_{10}$&33.43&9.12\\
$\bmualph_{11}$&2.89&1.08&$\bSigalph_{11}$&3.14&1.18\\
$\bmualph_{12}$&3.43&1.12&$\bSigalph_{12}$&3.30&1.23\\
    \end{tabular}
    \caption[DDM - Simulation Study. Estimation results for the group-level parameters.]{The efficiency (IACT) of the group-level means and variances estimated using the PMwG sampler. $J$ is the number of participants.}
    \label{tab:HDDM_PMwG_mimic_Lexical}
\end{table}

\begin{figure}[H]
    \centering
    \includegraphics[scale=0.2]{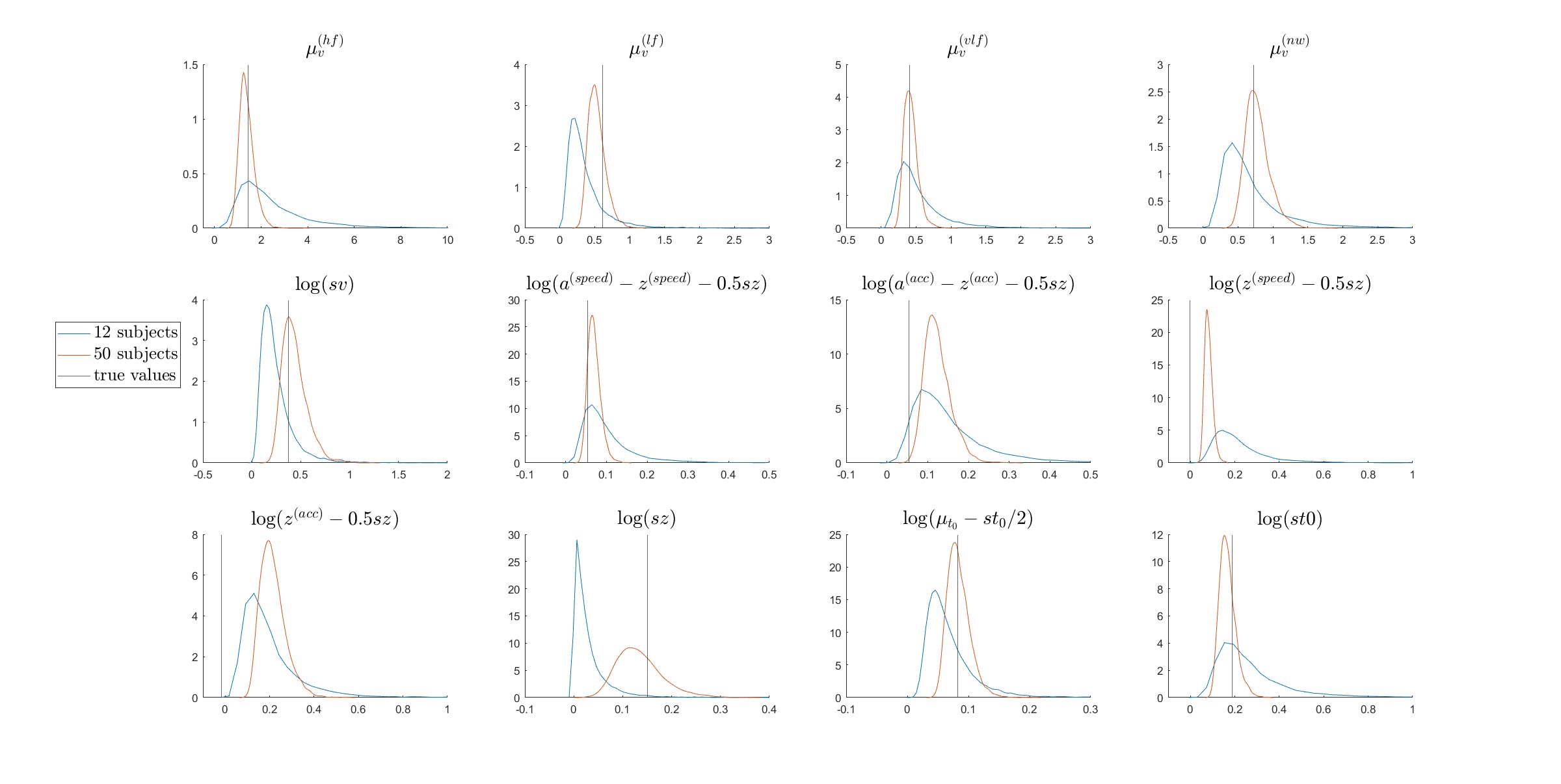} 
    \caption[DDM - Simulation Study. Kernel density estimates of the group-level variances]{Kernel density estimates of the group-level variance ($\diag (\bSigalph)$) estimated using the PMwG method in the simulation setting 1 (12 subjects, blue color) and setting 2 (50 subjects, red color). The vertical lines show the true values.}
    \label{fig:hddm_mimic_Lexical_density_var}
\end{figure}

\begin{figure}[H]
    \centering
   
    \hspace*{-2cm} \includegraphics[scale=0.22]{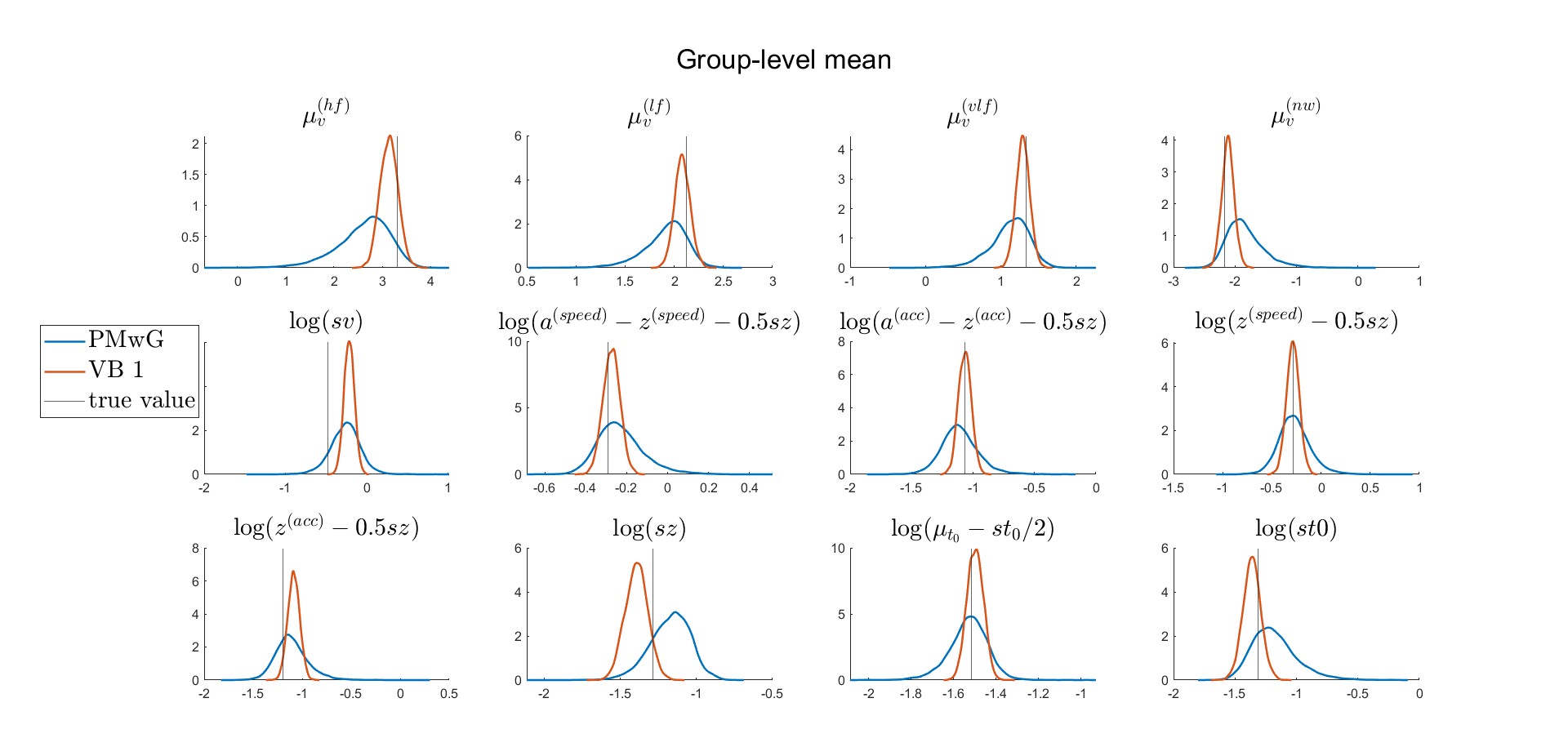}
    \caption[DDM - Simulation Study (12 subjects). Comparing the posterior densities estimated using PMwG against using VB.]{
    Kernel density estimates of marginal posterior densities of the group level means in the simulation study with $J=12$ subjects estimated using PMwG and VB. The vertical lines show the true values.}
    \label{fig:hddm_simdata_medium_vb_pmwg_density}
\end{figure}
\begin{figure}[H]
    \centering
    
    \hspace*{-2cm} \includegraphics[scale=0.22]{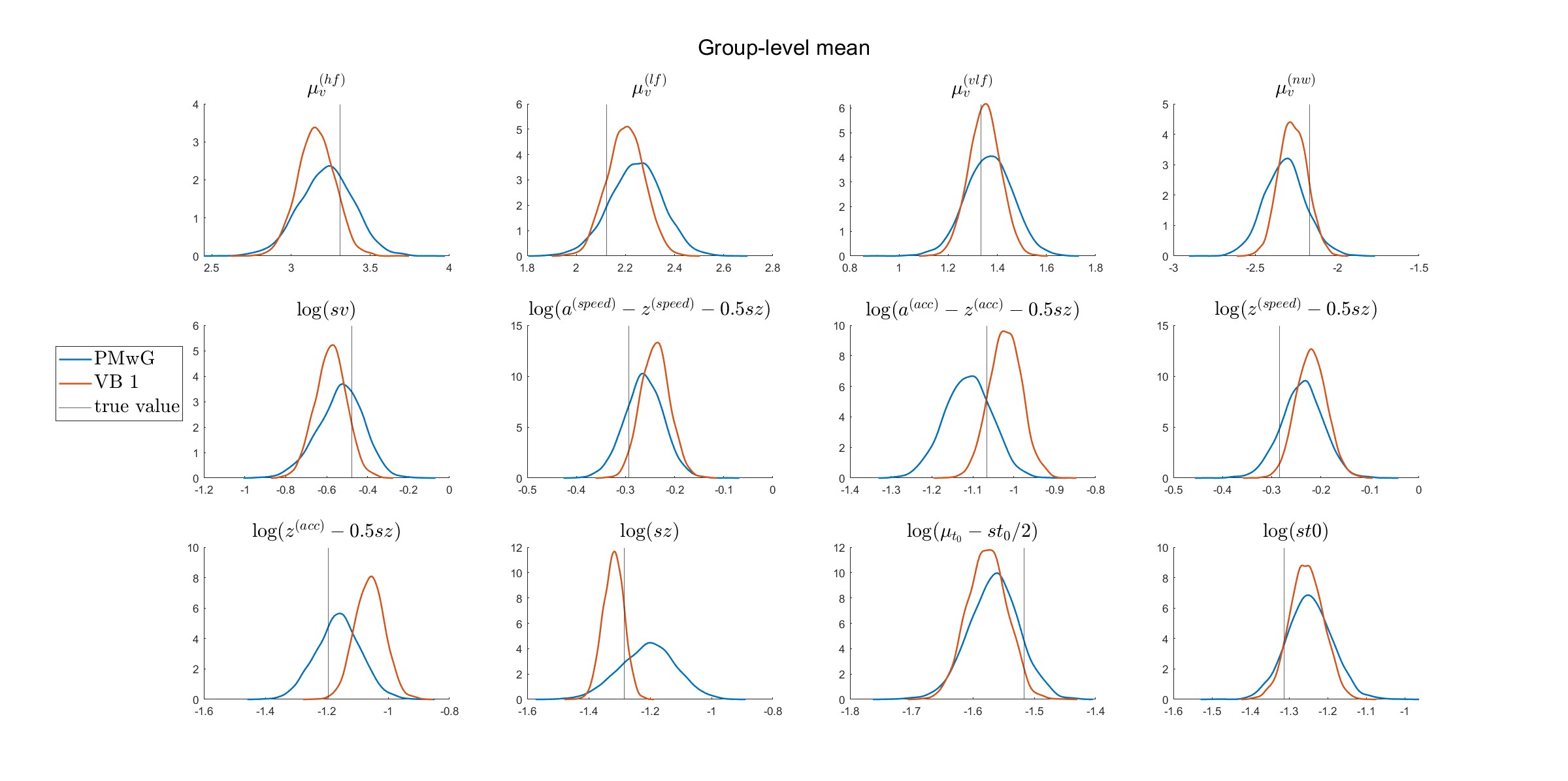}
    \caption[DDM - Simulation Study (50 subjects). Comparing the posterior densities estimated using PMwG against using VB.]{Kernel density estimates of marginal posterior densities of the group level means in the simulation study with $J=50$ subjects estimated using PMwG and VB. The vertical lines show the true values.}
    \label{fig:hddm_simdata_large_vb_pmwg_density}
\end{figure}

\subsection{Data set 2}

\begin{figure}[H]
	\centering
	\hspace*{-0cm}\includegraphics[scale=0.2]{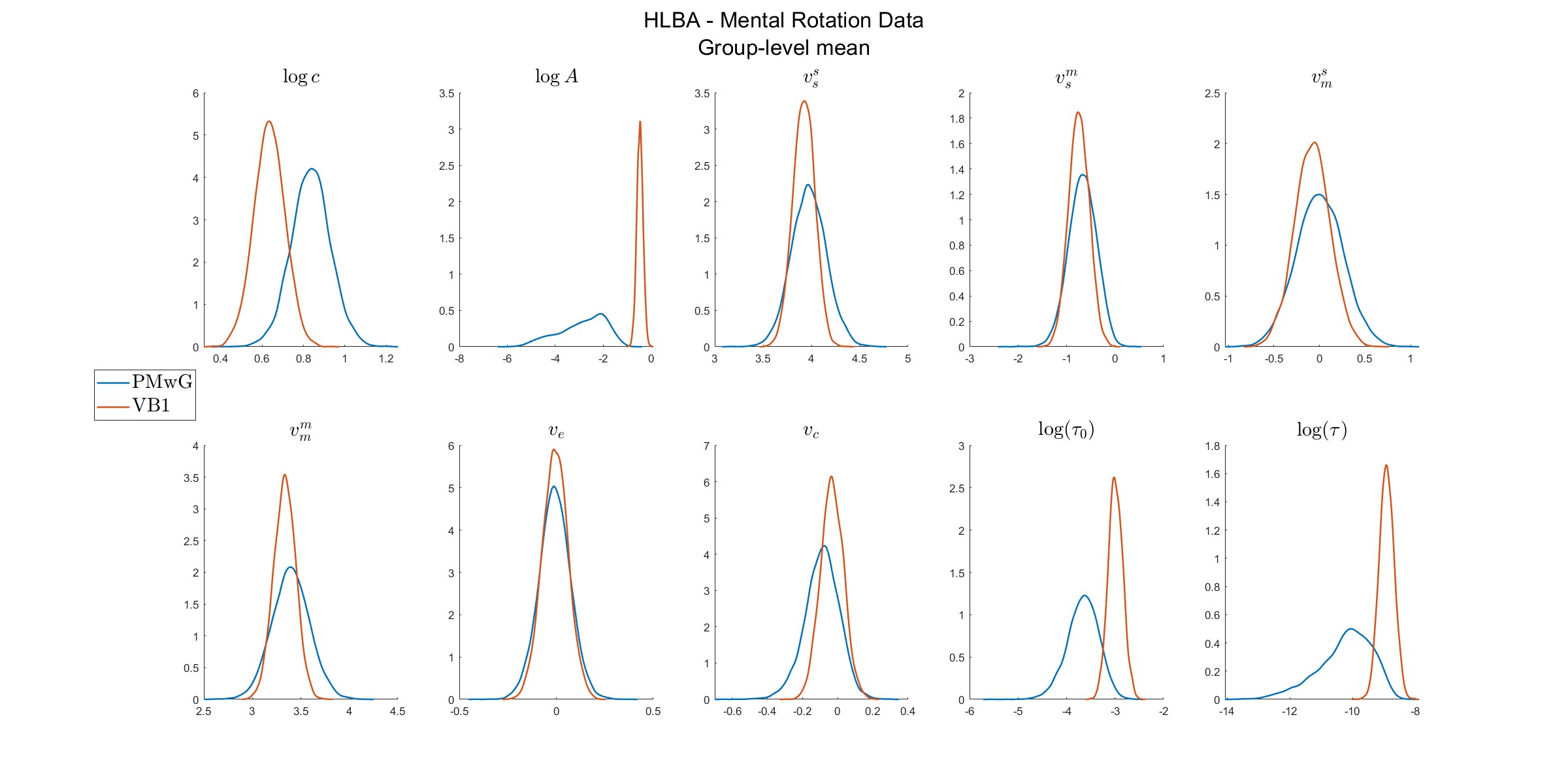}
	\caption[LBA - Mental Rotation. Comparing the posterior densities estimated using PMwG against using VB.]{
    LBA - Kernel density estimates of marginal posterior densities of the group level means estimated using PMwG and VB.}
    \label{fig:reglba_mental_rotation_vb_pmwg_density_mu}
\end{figure}

\begin{figure}[H]
    \centering
    
    \hspace*{-2cm} \includegraphics[scale=0.22]{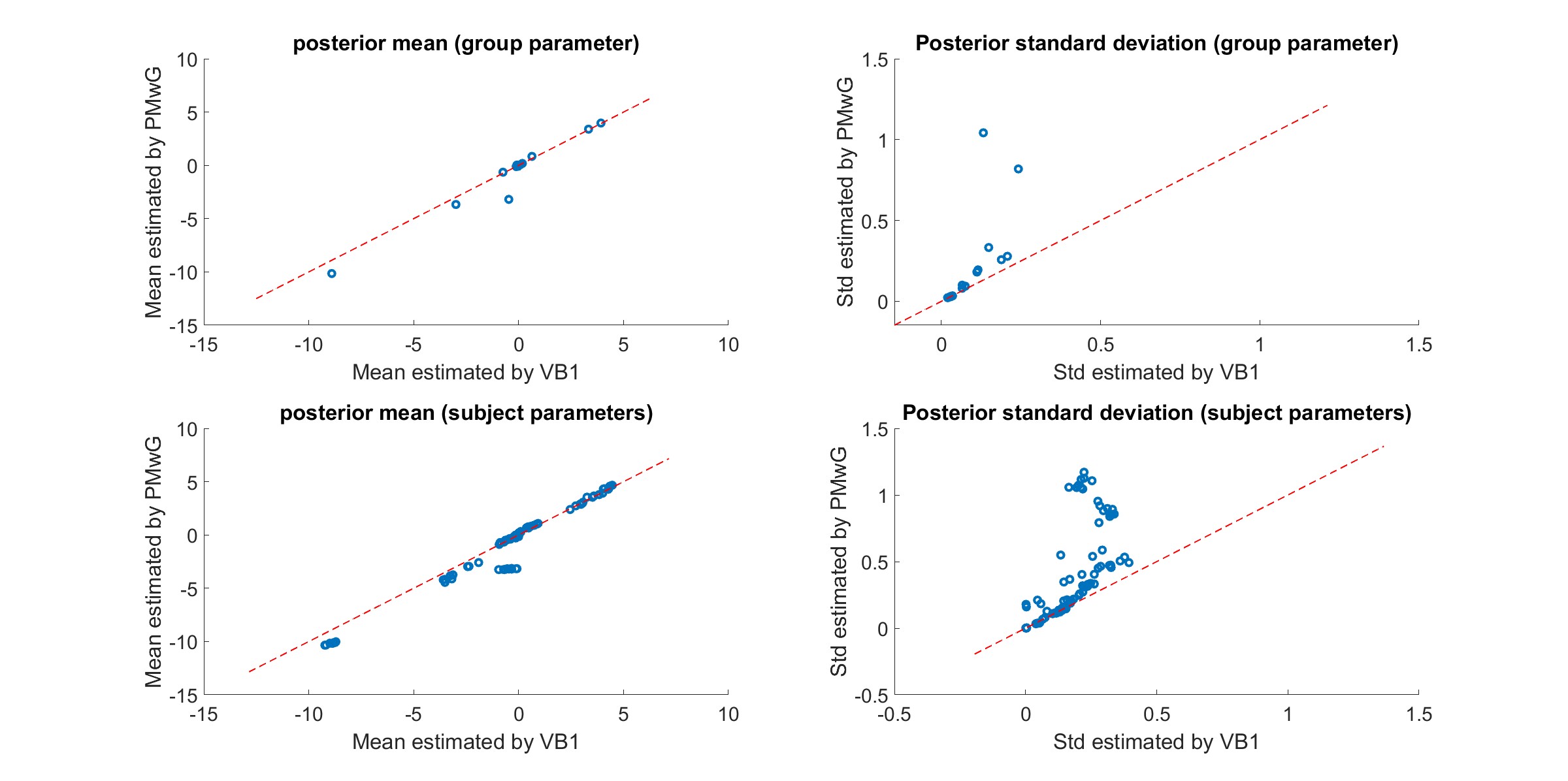}
    \caption[LBA - Mental Rotation data. Comparing the posterior moments estimated using PMwG against using VB .]{LBA - Comparing the means and standard deviations of the marginal posterior distributions estimated by VB (horizontal axis) against the exact values calculated using PMwG (vertical axis) for the simulation study with $J=50$ subjects. The top panels give the means and standard deviations of the group-level parameters. The bottom panels show the means and standard deviations of the subject parameters.}
    \label{fig:reglba_mental_rotation_vb_pmwg_moment}
\end{figure}

\begin{figure}[H]
	\centering
	\hspace*{-0cm}\includegraphics[scale=0.2]{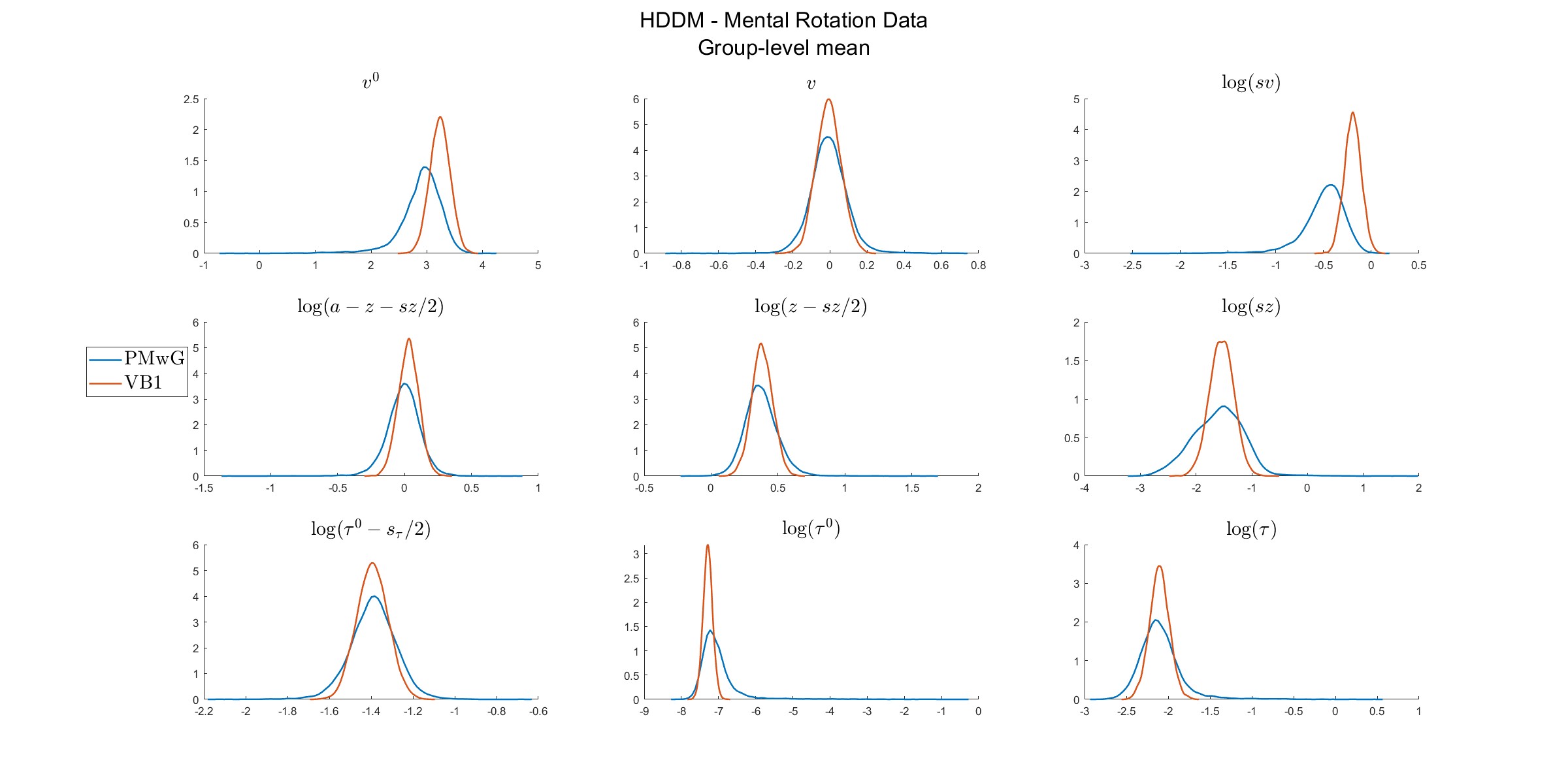}
	\caption[DDM - Mental Rotation. Comparing the posterior densities estimated using PMwG against using VB.]{
    DDM - Kernel density estimates of marginal posterior densities of the group level means estimated using PMwG and VB.}
    \label{fig:reghddm_mental_rotation_vb_pmwg_density_mu}
	\end{figure}

\begin{figure}[H]
    \centering
    
    \hspace*{-2cm} \includegraphics[scale=0.22]{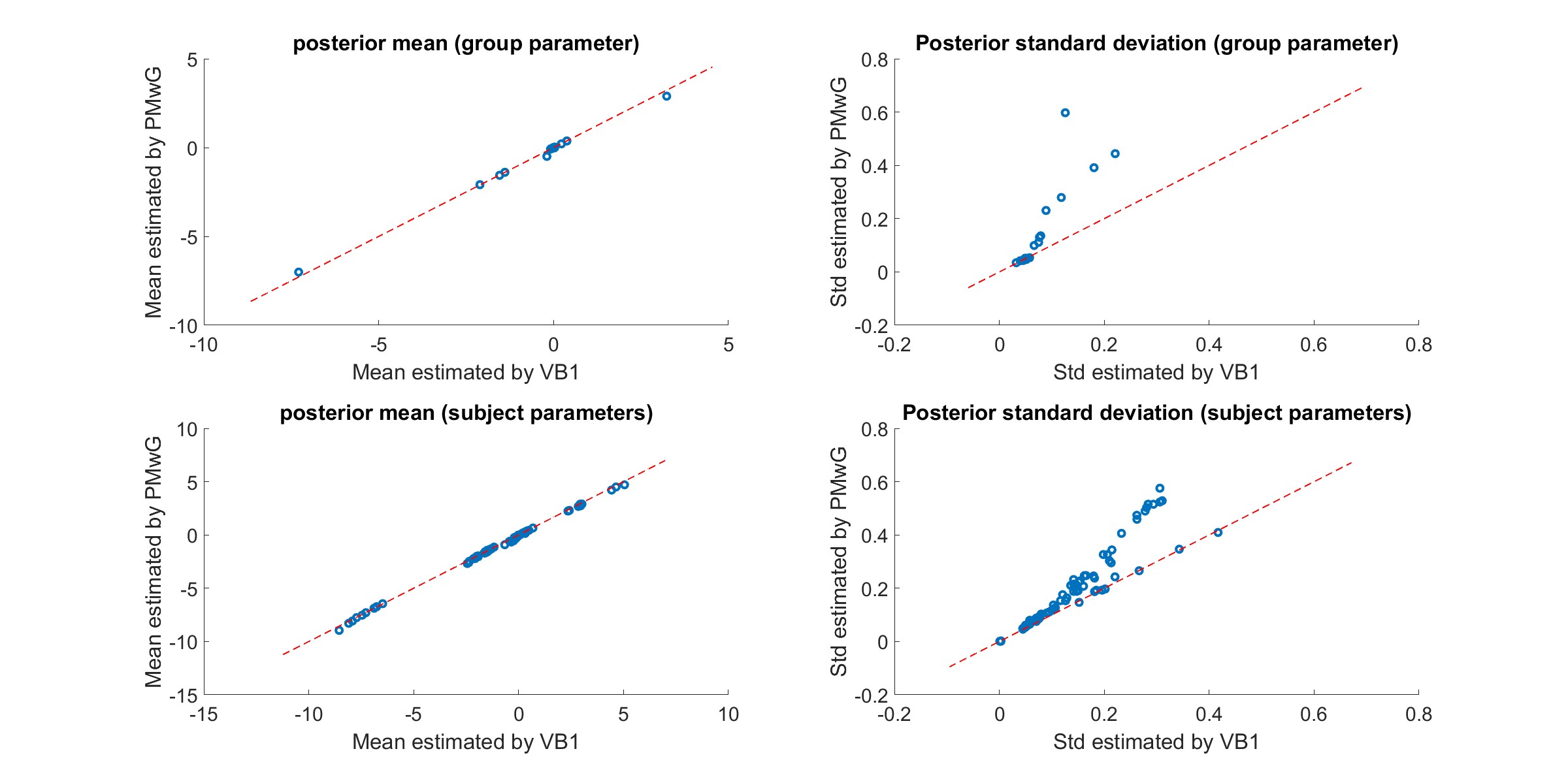}
    \caption[DDM - Mental Rotation data. Comparing the posterior moments estimated using PMwG against using VB .]{DDM - Comparing the means and standard deviations of the marginal posterior distributions estimated by VB (horizontal axis) against the exact values calculated using PMwG (vertical axis) for the simulation study with $J=50$ subjects. The top panels give the means and standard deviations of the group-level parameters. The bottom panels show the means and standard deviations of the subject parameters.}
    \label{fig:reghddm_mental_rotation_vb_pmwg_moment}
\end{figure}

Figure~\ref{fig:reglba_mental_rotation_vb_pmwg_density_mu} and \ref{fig:reghddm_mental_rotation_vb_pmwg_density_mu} show the kernel density estimates of the group-level mean $\bmualph$ from the LBA and DDM model, respectively. VB captures relatively well the marginal posterior distributions of the group-level mean, except for some components in the LBA. For other group-level parameters as well as the random effects, we assess the accuracy of VB by comparing the mean and standard deviations as shown in figures ~\ref{fig:reglba_mental_rotation_vb_pmwg_moment} (LBA) and 
 \ref{fig:reghddm_mental_rotation_vb_pmwg_moment} (DDM). Overall, VB can approximate the posterior means relatively precisely. In terms of computational efficiency, table \ref{tab:ERPdata_computational_comparison} displays the running time of the two methods. Again, VB is about 10 times faster than PMwG when estimating both models.

\begin{table}[H]
    \centering
    \begin{tabular}{c|c|c}
     & \textbf{LBA} & \textbf{DDM}\\
    \hline
    \hline
    PMwG & 1:15 & 30:56 \\
    \hline
    VB & 0:07 & 5:58 \\
     \hline
    \end{tabular}
    \caption[RegHEAM- Real data. Running time comparison between PMwG and VB.]{Mental Rotation. A comparison between PMwG and VB in terms of running time (hour:minute).}
    \label{tab:ERPdata_computational_comparison}
\end{table}

\subsection{Data set 3}
\begin{figure}
  \hspace*{-1cm}\includegraphics[scale=0.20]{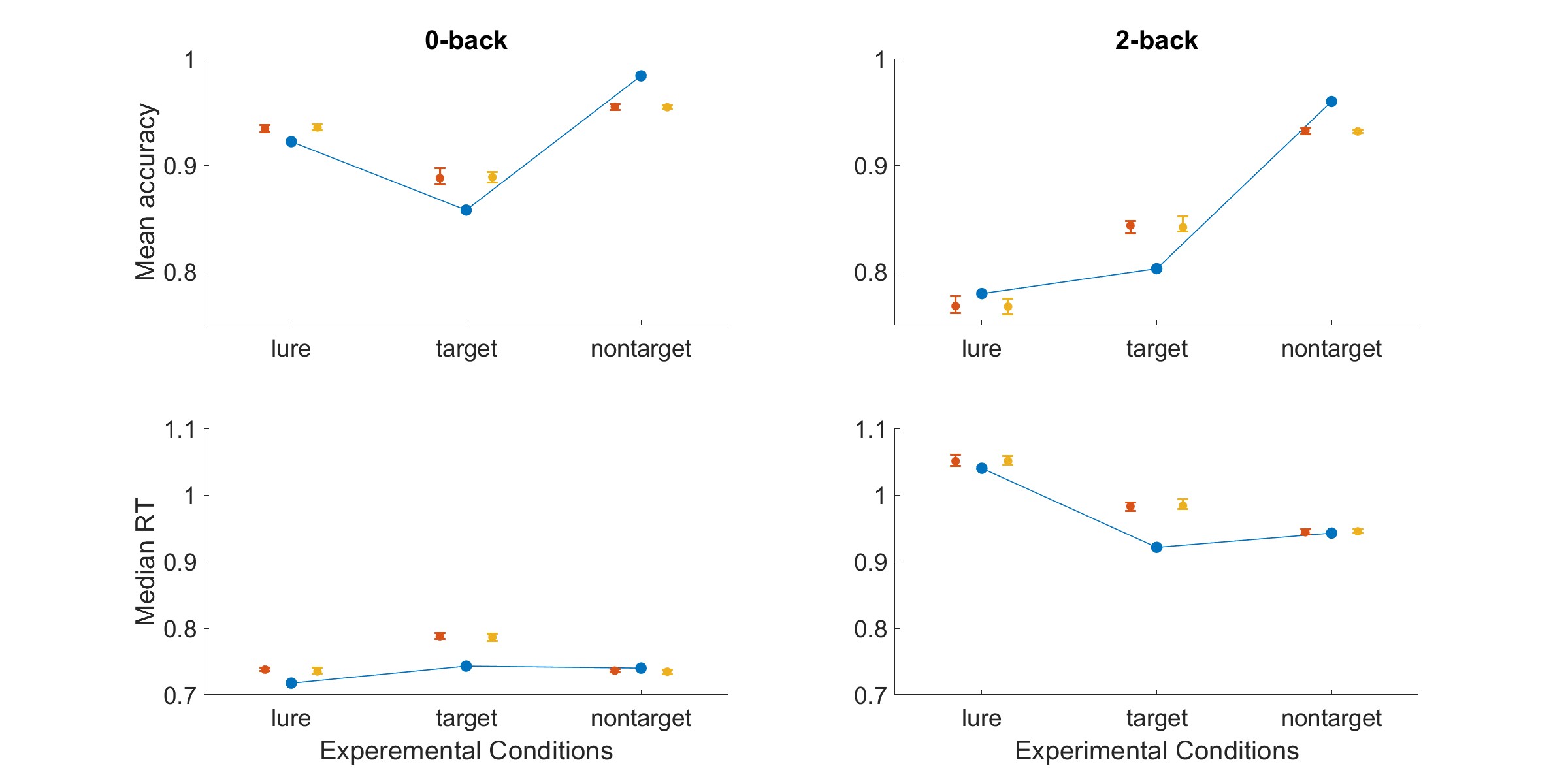}	
	\caption{Comparing the estimates of posterior predictive summary statistics for the LBA model using two methods: PMwG (represented by red vertical bars) and VBL (represented by yellow vertical bars), against the observed data (represented by blue dots). Within each vertical bar, the upper and lower tails, along with the dot in between, represent the $2.5\%$, $97.5\%$, and $50\%$ quantiles of the predictive distribution, respectively. The top panels show the posterior predictive mean accuracy, while the bottom panels show median RT. Left panels correspond to the 0-back conditions while the right panels correspond to the 2-back conditions.}
	\label{fig:hcp-data-hlba-no-covariates-boxplot-sumstats}	
\end{figure}

\begin{figure}
	\hspace*{-1cm}\includegraphics[scale=0.20]{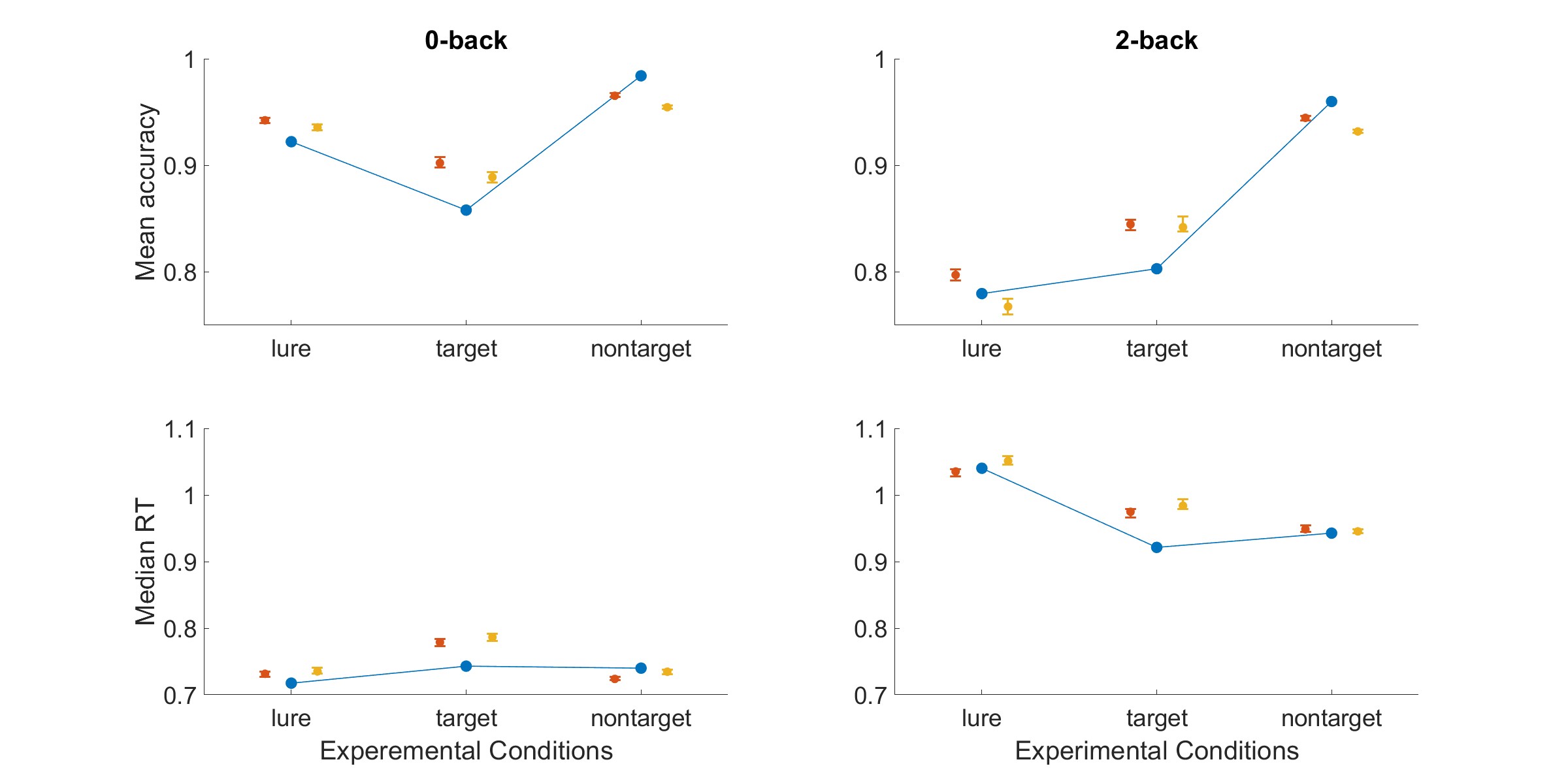}	
\caption{Comparing the estimates of posterior predictive summary statistics for the DDM  (represented by red vertical bars) and LBA models (represented by yellow vertical bars) using VBL, against the observed data (represented by blue dots). Within each vertical bar, the upper and lower tails, along with the dot in between, represent the $2.5\%$, $97.5\%$, and $50\%$ quantiles of the predictive distribution, respectively. The top panels show the posterior predictive mean accuracy, while the bottom panels show median RT. Left panels correspond to the 0-back conditions while the right panels correspond to the 2-back conditions.}
	\label{fig:hcp-data-lba-vs-hddm-without-covariates-boxplot-sumstats}	
\end{figure}

\begin{figure}
   
    \hspace*{-1.2cm}\includegraphics[scale=0.2]{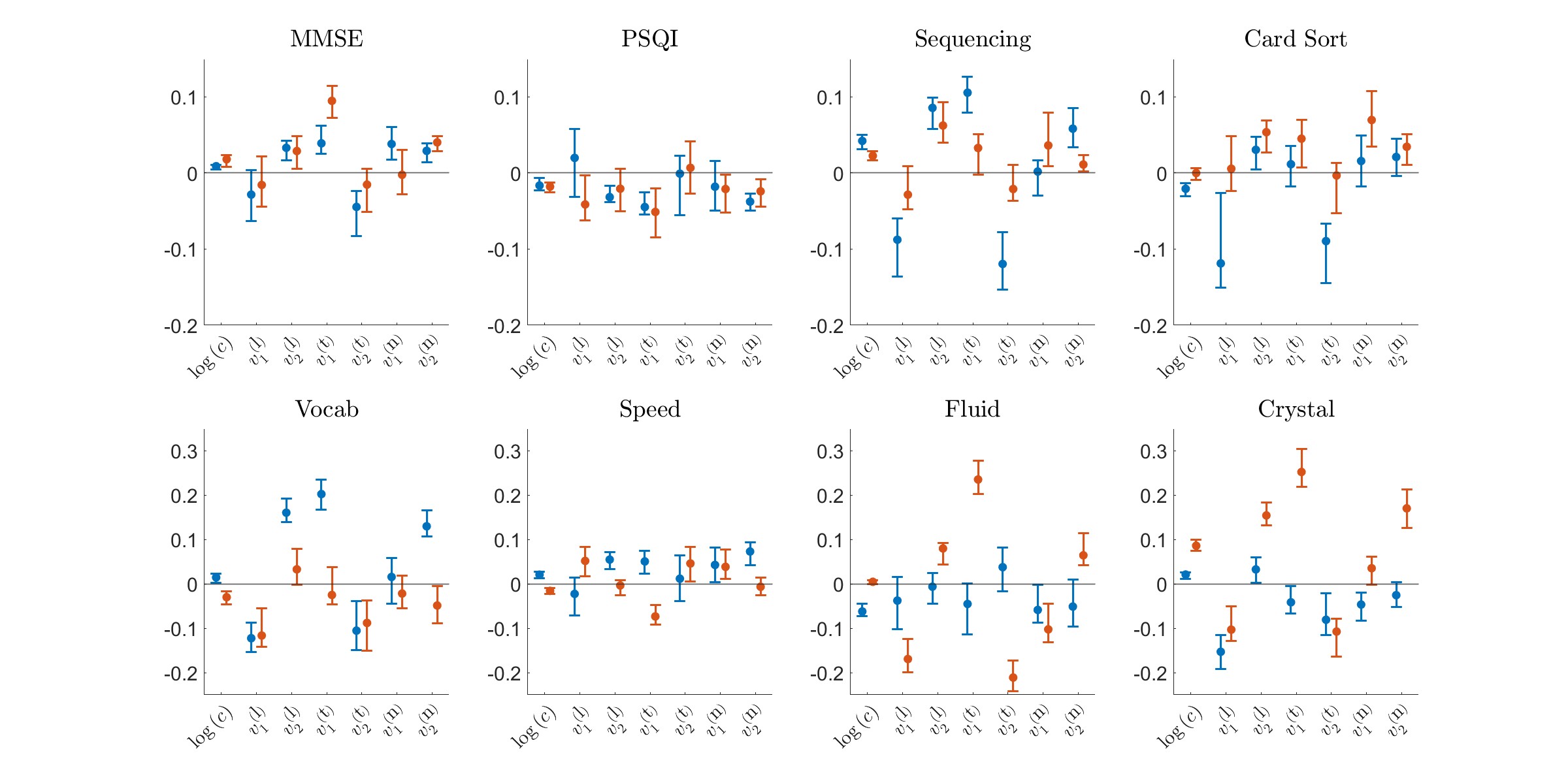}	
	\caption{HCP 0-back and 2-back data. Posterior distribution of the coefficients $\bbeta$ estimated using the VBL method for the LBA, with vertical bars representing the 0-back (blue) and 2-back (red) conditions. Within each vertical bar, the upper and lower tails, along with the dot in between, represent the $2.5\%$, $97.5\%$, and $50\%$ quantiles of the posterior distribution, respectively. Each panel corresponds to a covariate. For example, the top left panel shows the posterior distribution of all the coefficients of MMSE. The second top left panel displays the posterior distribution of all coefficients corresponding to PSQI, and so on and for other panels.}
	\label{fig:hcp-data-reglba-boxplot-beta-covariates}	
\end{figure}

Figure~\ref{fig:hcp-data-hlba-no-covariates-boxplot-sumstats} compares the posterior predictive distribution obtained from LBA (without covariates) with the observed data. Figure~\ref{fig:hcp-data-lba-vs-hddm-without-covariates-boxplot-sumstats} compares the posterior predictive distribution obtained from DDM (without covariates) with the one obtained from LBA (without covariates). The (marginal) posterior distributions of the coefficients of the covariates for the LBA are shown in Figure~\ref{fig:hcp-data-reglba-boxplot-beta-covariates}.

\bibliography{all_references,uoncoglabshort}	


\end{document}